%% file: Hauptdatei.tex
\documentclass[a4paper,12pt,headsepline]{scrartcl}[]
\input{latex_einstellungen/variablen}

\usepackage{graphicx}

\usepackage[ngerman]{babel}

\usepackage[right]{eurosym}

\usepackage[utf8]{inputenc}

\usepackage[T1]{fontenc}

\usepackage{lmodern}
\usepackage{fix-cm}

\usepackage{float}
\restylefloat{figure}

\usepackage{longtable}

\usepackage{csquotes}

\usepackage{geometry}
\usepackage{fancybox}

\usepackage[hyphens,obeyspaces,spaces]{url}

\usepackage{color}

\usepackage{amssymb}

\usepackage[bookmarksnumbered,pdftitle={\titleDocument},hyperfootnotes=false]{hyperref} 

\usepackage{fancyhdr} 
\usepackage{array}

\usepackage{setspace}

\usepackage{capt-of}

\usepackage{makeidx}

\usepackage{listings}
\makeindex

\usepackage[german]{nomencl}

\makenomenclature

\begin{document}
\input{latex_einstellungen/trennung}



\include{latex_einstellungen/deckblatt}


\onehalfspacing

\newpage
\thispagestyle{empty} 
\section*{ }
\include{abstract}

\newpage
\thispagestyle{empty} 
\section*{ }
\singlespacing

\newpage
\tableofcontents

\newpage
\listoffigures
\newpage
\thispagestyle{empty} 
\section*{ }



\setlength{\columnsep}{25pt}

\newpage
\fancyhead[L]{\nouppercase{\leftmark}} 

\onehalfspacing

\input{1_einleitung}

\newpage
\thispagestyle{empty} 
\section*{ }
\input{2_kap2}
\newpage
\thispagestyle{empty} 
\section*{ }
\input{3_kap3}

\input{4_kap4}

\newpage
\thispagestyle{empty} 
\section*{ }
\input{5_kap5}

\input{6_kap6}

\input{7_kap7}

\onecolumn
\singlespacing
\newpage
\addcontentsline{toc}{section}{Literaturverzeichnis}
\nocite{Literatur}
\renewcommand\refname{Literaturverzeichnis}

\bibliography{Literatur}


\onehalfspacing
\newpage

\addcontentsline{toc}{section}{Eidesstattliche Erklärung}
\include{erklaerung}

\newpage
\thispagestyle{empty} 
\section*{ }

\end{document}

%% file: latex_einstellungen/variablen.tex
\newcommand{\titleDocument}{Bachelorarbeit}

%% file: latex_einstellungen/trennung.tex
\hyphenation{
Film-pro-du-zen-ten
Lux-em-burg
Soft-ware-bau-steins
zeit-in-ten-
Instru-men-tarien
}

%% file: latex_einstellungen/deckblatt.tex




\thispagestyle{empty}


\begin{figure}[t]
 \centering
 \includegraphics[width=0.6\textwidth]{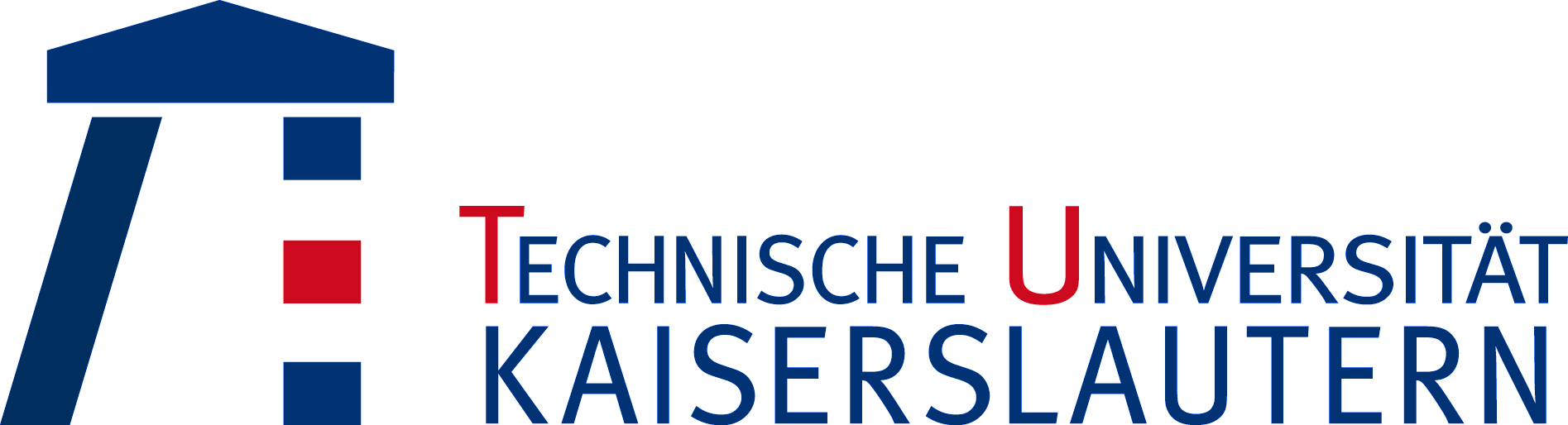}
~~~~~~~~~~
 \includegraphics[width=0.3\textwidth]{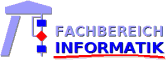}
\end{figure}

\begin{verbatim}

\end{verbatim}

\begin{center}
\Large{Technische Universität }\\
\Large{Kaiserslautern}\\
\end{center}

\begin{center}
\Large{}
\end{center}
\begin{verbatim}

\end{verbatim}
\begin{center}
\doublespacing
\textbf{\LARGE{\titleDocument}}\\
\singlespacing
\begin{verbatim}

\end{verbatim}
\textbf{}
\end{center}
\begin{verbatim}

\end{verbatim}
\begin{center}

\end{center}
\begin{verbatim}

\end{verbatim}
\begin{center}
\textbf{zur Erlangung des akademischen Grades \\ Bachelor of Science}
\end{center}
\begin{verbatim}

\end{verbatim}
\begin{flushleft}
\begin{tabular}{llll}
\textbf{Thema:} & & Vorstellung eines sozioinformatischen Analyseansatzes\\
&& zur Technikfolgenabschätzung in Anlehnung an Vesters\\
&& Sensitivitätsmodell am Beispiel des Unternehmens\\
&& ‚Uber‘ als sozio-technisches System\\
& & \\
\textbf{Autor:} & & Tobias Krafft & \\
& & MatNr. 382531 & \\
& & \\
\textbf{Version vom:} & & 15. Juni 2015 &\\
& & \\
\textbf{1. Betreuerin:} & & Prof. Dr. Katharina Zweig &\\
\textbf{2. Betreuer:} & & Prof. Dr. Paul Lukowicz &\\
\end{tabular}
\end{flushleft}

%% file: abstract.tex
\section*{Kurzzusammenfassung}
\begin{center} 
\fbox{
 \begin{minipage}{\textwidth}
	\begin{itshape}
	\centering 
	\textsf{\\} 
	Wer will was Lebendiges erkennen und beschreiben,\\
Sucht erst den Geist herauszutreiben,\\
Dann hat er die Teile in seiner Hand,\\
Fehlt leider! nur das geistige Band.\\
		
		\cite[S.56 Vers 1936ff]{Goethe}
		\label{Goethe}
    
	\end{itshape}
	
\end{minipage}
}
\end{center} 

\begin{verbatim}

Zunehmende technische Fortschritte sowie ihre Verflechtung mit der 
Gesellschaft stellen die Forschung vor neue Problemstellungen.
Technikfolgenabschätzung für Systeme, in welche sowohl soziale als
auch technische Komponenten integriert sind, bildet eines dieser
neuen Themenfelder. 

Ziel dieser Arbeit war eine kritische Auseinandersetzung mit dem
biokybernetischen Denkansatz anhand des ‚Sensitivitätsmodells‘
von Frederic Vester, das in die sozioinformatische Analyse eines
sozio-technischen Systems integriert und hinsichtlich der
Fragestellung geprüft wurde, inwieweit sich dieses Konzept
zur Beurteilung von Planungsvorhaben bei sozio-technischen 
Systemen eignet. Dazu wurden F. Vesters 
‚Biokybernetischenn Grundregeln‘ zur Feststellung der 
Überlebensfähigkeit eines Systems, auf ihre Anwendbarkeit 
in Bezug auf sozio-technische Systeme untersucht. Die
Funktionalität der vorgestellten Analysemethode wurde auf das 
nach der Definition von Andrea Kienle und Gabriele Kunau als 
sozio-technisches System charakterisierte Startup-Unternehmen 
‚Uber‘ angewendet. 

Das Sensitivitätsmodell erwies sich mit gewissen Einschränkungen
auf sozio-technische Systeme übertragbar, allerdings waren nur 50% 
der Regeln auf solche anwendbar. Im Gegensatz zu unsystemischen
Analysen besteht der Vorteil des Sensitivitätsmodells in der 
Vermittlung von Erkenntnissen über innere Systemzusammenhänge. 
Abschließend wurde der vorgestellte Analyseansatz als
transdisziplinäre Weiterentwicklung des Sensitivitätsmodells
charakterisiert.   
\end{verbatim}
\newpage
\section*{Abstract}
What fits the human brain does not necessarily mean whatever is invented is meant for human brain. This is exactly what is implied by the definition of algorithm: \textit{Reducing and Simplifying}. Our cognitive apparatus, although endowed with certain parameters, ceases to comprehend chains of mathematical and logical steps used for algorithmic or mathematical purposes. Such a cognitive limit allows us to pay attention to the symbolic regularities which are meaningful. The more meaningless these symbols become, the fuzzier is our mental representation for it. Needless to say, that we actually tend to interact with activities that are meaningful, and discard those that are meaningless. By activities we mean steps, steps that form an algorithm. Could it be just an assumption? Are we really aversing from algorithms in modern day life? \\

%% file: 1_einleitung.tex
\section{Einführung und Fragestellung}\label{kapitel1}
\subsection{ Problemstellung }\label{kapitel1.1}
Im Jahr 1969 nannte der damalige Generalsekretär der Vereinten Nationen Sithu U Thant „Wettrüsten, Umweltverschmutzung, Bevölkerungsexplosion und wirtschaftliche Stagnation“ \cite[S.11]{Meadows} als wichtigste Probleme, die die Menschheit zu lösen hätte. Ob Bankenkrisen oder Zunahme von Umweltkatastrophen infolge der globalen Erwärmung – nach über 45 Jahren muss man leider konstatieren, dass sich die Gesamtsituation trotz intensiver Bemühungen nicht wesentlich gebessert hat \cite {fr1}. Wurde ein Problem gelöst, taucht schon wieder ein neues auf \cite{aaa}. \\
Frederic Vester, der so genannte Vater des ‚vernetzten Denkens‘ (vgl. \ref{kapitel2.2}) sieht die Problematik der heutigen Zeit vor allem darin, dass wir beim Versuch, die aktuellen Probleme „in unserer immer komplexeren Welt“ \cite[S.18]{KunstVester} zu lösen, weiterhin nach dem tradierten linear-kausalen Denkmuster \cite[vgl. S.145]{KunstVester} verfahren.\\
Durch das cartesianische Denken der abendländischen Kultur geprägt, zerlegt man die Wirklichkeit in immer kleinere Teile und versucht durch die Analyse dieser Einzelteile zu einem Verständnis des Ganzen zu kommen. Die daraus resultierende, immer weitere Spezialisierung der Wissenschaften führt aber trotz oder besser wegen der Fülle von gesammelten Daten in den einzelnen, voneinander oft  isoliert arbeitenden Spezialgebieten zu keinem umfassenderen Verständnis der Welt denn „das Ganze mehr ist als die Summe seiner Teile“ \cite[S.32]{KunstVester},  wie Frederic Vester, schon Ende des letzten Jahrhunderts zutreffend feststellt hat. Der gravierende Nachteil dieser im westlichen Kulturkreis fest verankerten Methode liegt dabei in der vorwiegend kurzfristig ausgelegten, problemorientierten Vorgehensweise, bei welcher der Fokus einseitig auf der Verbesserung von Einzelaspekten liegt und „die vernetzten Zusammenhänge unserer Welt“  ignoriert.\\
Daraus ergeben sich bei Projekten aller Art immer wieder gravierende Planungsfehler, deren Folgen aber oft  zeitverzögert eintreten, sodass sie dann nicht immer mehr zu korrigieren sind. 
So machte Vester schon 1978 während der von ihm initiierten Wander-Ausstellung „Unsere Welt – ein vernetztes System“ am Beispiel des Assuan-Staudamms auf die fatalen Spätfolgen des Baus für die Ökosysteme in Ägypten aufmerksam \cite[vgl. S.105ff.]{Welt}. Diese Problematik ist heute im Bereich der  Technikfolgenabschätzung unter dem Begriff ‚Collingridge Dilemma‘ geläufig, auf das im zweiten Kapitel noch näher eingegangen wird.  \\
Es ist also zumindest im wissenschaftlichen Bereich schon länger bekannt, dass ein Umdenken und eine neue wissenschaftliche Betrachtungsweise zwingend erforderlich sind, da eine einseitig ausgerichtete Betrachtung von linearen Wirkbeziehungen unzureichend ist. So kritisierte Ludwig von Bertalanffy (1901-1972)  bereits 1948  in seinem Aufsatz ‚Zu einer allgemeinen Systemlehre‘, der 1969 mit anderen seiner Schriften in  „General System Theory: Foundations, Development, Applications“ \cite{Bertalanffy} herausgegeben wurde, die isolierte Betrachtung von Einzelphänomenen in der klassischen Physik und führte mit seinem Entwurf der ‚Allgemeinen Systemlehre‘ ein neues wissenschaftliches Paradigma ein \cite[vgl. S.32ff]{Bertalanffy}.\\
 Weitere entscheidende Impulse kamen dann von Frederic Vester, der in seinen Veröffentlichungen immer wieder versuchte, das systemische Denken einer breiteren Öffentlichkeit zugänglich zu machen. In seinem 1999 erschienenen Werk ‚Die Kunst vernetzt zu denken, Ideen und Werkzeuge für einen neuen Umgang mit Komplexität‘ stellt er ausführlich das von ihm entwickelte ‚Sensitivitätsmodell‘ vor, ein computergestütztes Instrumentarium für den Umgang mit komplexen Systemen, mit dem relevante Einflussgrößen und bestehende Interdependenzen erfasst und eine gezielte Steuerung und Planung möglich werden sollen. Sein Modell basiert zusätzlich auf acht biokybernetischen Regeln, die er als entscheidend für die Beurteilung der Lebensfähigkeit aller komplexen Systeme  erachtet.  Dieses auch für ökologische Fragestellungen bedeutsame Buch wurde von Vester danach noch um einige Kapitel erweitert, unter anderem mit Veröffentlichungen aus den Jahren vor der Nuklearkatastrophe von Tschernobyl, in denen er bereits auf konkrete Missstände aufmerksam machte und vor einer möglichen Nuklear-Katastrophe warnte, was jedoch mit einer Hetzkampagne durch die „Gesellschaft für Reaktorsicherheit“  bekämpft wurde \cite[vgl. S. 124ff]{Neuland}. Die erweiterte Auflage erschien unter dem gleichnamigen Titel mit dem Zusatz ‚Bericht an den Club of Rome‘  dann im Frühjahr 2002 erneut \cite{Neuland}.\\ 	
Trotz der Notwendigkeit eines systemischen Denkens findet es insgesamt immer noch wenig Beachtung und dient demzufolge auch selten als Grundlage bei Problemlösungen. Auch die  dringend erforderliche interdisziplinäre Zusammenarbeit zwischen den verschiedenen Fachbereichen setzt sich nur sehr zögerlich durch, im aktuellen Bildungssystem hat das fächerübergreifende Lernen in den Bildungsstandards der Bundesländer  ebenfalls bis jetzt kaum Eingang gefunden.\footnote{ Vgl. Pahde, Martina (2002): Aussagen zum fächerübergreifenden Unterricht in den Richtlinien und Lehrplänen der Sekundarstufe II in den einzelnen Bundesländern, S.4. In: Weiterentwicklung der Prinzipien der gymnasialen Oberstufe und des Abiturs – Abschlussbericht der von der Kultusministerkonferenz eingesetzten Expertenkommission. Kiel 1995}
Offenbar mangelt es an für das Erfassen komplexer Systeme geeigneter Instrumentarien, die Voraussetzung für die Entwicklung effizienter Lösungsstrategien. 
\newpage
\subsection{Zielsetzung der Arbeit}\label{kapitel1.2}
Die vorliegende Arbeit beabsichtigt daher, Vesters Sensitivitätsmodell im Rahmen eines  sozioinformatischen Analyseansatz‘ zur Technikfolgenabschätzung vorzustellen.\\
Dabei soll das Modell hinsichtlich der Fragestellung untersucht werden, inwieweit sich dieses Konzept für eine Beurteilung von Planungsvorhaben bei sozio-technischen Systemen eignet. Die Funktionalität der vorgestellten  Analysemethode soll auf das 2009 in den USA gegründete Startup-Unternehmen ‚Uber‘ angewendet werden, wobei das Unternehmen ‚Uber‘ als sozio-technisches System definiert wird.\\
Weiterhin sollen die von Vester aufgestellten acht biokybernetischen Regeln, für die Vester eine allgemeine Gültigkeit postuliert, auf ihre Übertragbarkeit auf sozio-technische Systeme kontrolliert werden

%% file: 2_kap2.tex
\newpage
\section{Die Notwendigkeit des Umdenkens zu systemorientierten Problemlösungsstrategien}\label{kapitel2}

Das folgende Kapitel verfolgt das Ziel, die allgemeine Notwendigkeit eines Umdenkens zu systemorientierten Problemlösungen aufzuzeigen, wobei neben Vesters Ausführungen noch zwei weitere, für die Thematik relevante Ansätze vorgestellt werden.
Zu Beginn wird der ‚Tanaland‘ - Versuch aus dem Jahr 1975 von Dietrich Dörner erläutert, bei dem es sich um „das früheste publizierte deutsche Simulationsprogramm“ \cite[S.5]{Funke} für komplexe Problemlösungsprozesse handelt und auf den sich die einschlägige Literatur noch heute beruft \cite[vgl.]{Hettlage}. \\
Wie Vester gehört auch Dörner zu den ‚Pionieren‘ bei der Erforschung komplexer Systeme und erkannte gleichfalls das systemische Denken als einzige Möglichkeit, die Probleme unserer komplexen Wirklichkeit angemessen erfassen und lösen zu können. Während Vester biologische Systeme als Ausgangspunkt für seine Theorien verwendet, untersuchte der 2006 emeritierte Professor für Kognitive Psychologie den Umgang mit Komplexität im computersimulierten Szenario von ‚Tanaland‘. Dörner identifizierte dabei typische Planungs- und Denkfehler, die ihm zufolge für das „beklagenswerte Schicksal von Tanaland“ \cite[S.22]{Doerner} verantwortlich waren.   
\\\\
Anschließend befasst sich Abschnitt \ref{kapitel2.2} mit den wesentlichen Aspekten der Ausführungen Vesters bezüglich des Scheiterns monokausaler Denkstrukturen beim Umgang mit komplexen Systemen. Zu Beginn wird in \ref{kapitel2.2.1} die Biographie von Frederic Vester kurz vorgestellt, der das ‚vernetzte Denken‘ sowie die daraus resultierende Interdisziplinarität in seinen Veröffentlichungen und Büchern zu einer Zeit anmahnte, als dies noch als „komplett ketzerisch“ angesehen wurde \cite{Streich}. 
Anschließend werden die von Vester postulierten Fehler bei unserem Umgang mit komplexen Systemen aufgezeigt, die er in Anlehnung an Dörners Ergebnisse des Tanaland-Versuchs entwickelt hat. Vester betrachtet diese ‚zentralen Fehler‘ als Beweis für die Unzulänglichkeit  linear-kausaler Denkstrukturen des so genannten ‚Klassifizierungs-Universums‘, in dem wir seiner Ansicht nach aktuell leben. Es empfiehlt sich also  für die sozioinformatischen Analyse darauf zurück zugegriffen. Eine Betrachtung des 'Collingridge Dilemma‘ , welches für den Bereich der Technikfolgenabschätzung ein wichtiges Problem darstellt,  bildet den Abschluss dieses Kapitels.
\newpage
\subsection{Der Untergang von ‚Tanaland‘  als Beispiel für die Schwierigkeiten im Umgang mit komplexen Systemen }\label{kapitel2.1}
Dietrich Dörner, inzwischen emeritierter Professor für Psychologie und langjähriger Direktor des Instituts für Theoretische Psychologie der Universität Bamberg, gibt mit dem eingangs schon erwähnten ‚Tanaland‘ - Versuch ein anschauliches Beispiel, wie schwer das ‚vernetzte Denken‘ den meisten Menschen fällt.  Auch wenn es sich bei nur 12 Versuchspersonen um kein wissenschaftlich repräsentatives Experiment handelt, ist das Ergebnis derart bezeichnend, dass es kurz dargestellt werden soll. \\
Probanden aus unterschiedlichen Fachrichtungen bekamen in einem Computersimulationsspiel den Auftrag, in Tanaland, einer fiktiven Region Ostafrikas, für eine Zeitdauer von zehn Jahren für bessere Lebensbedingungen zu sorgen. Für diese Aufgabe wurden sie mit allen notwendigen Handlungskompetenzen ausgestattet und konnten über alle erforderlichen Kreditsummen verfügen. Bei sechs Sitzungen, deren Zeitpunkt frei gewählt werden konnte, bestand die Möglichkeit, Informationen zu sammeln und Maßnahmen zu planen bzw. einzuleiten. Dabei war die Anzahl der ergriffenen Maßnahmen bei allen sechs Eingriffspunkten unbegrenzt \cite[vgl. S.22]{Doerner}.  \\
Trotz dieser, vermeintlich, optimalen Bedingungen endeten alle Versuche desaströs. Eine Rückzahlung der Kredite wurde bei allen Probanden unmöglich, da nach kurzfristigen Verbesserungen die verschiedensten Katastrophen auftraten: Stieg die Lebenserwartung, kam es zu Hungersnöten; wenn die Zahl der Mäuse, Ratten und Affen deziemiert werden, um den Ertrag der Felder und Gärten zu erhöhen, vermehrten sich die Insekten ungehemmt oder die größeren Raubkatzen begannen die Viehbestände zu reduzieren, weil ihre übliche Beute, wie Kleinsäuger oder Affen, nicht mehr ausreichend zur Verfügung stand \cite[vgl. S.27]{Doerner}.  Es ist anzunehmen, dass den ausgewählten Versuchspersonene das nötige interdisziplinäre Denken fehlte, um Tanaland als komplexes Wirkungsgefüge samt seiner fächerübergreifenden Abhängigkeiten zu erfassen, denn „Tanaland enthielt keine Probleme, die nur mit einem speziellen Fachwissen zu bewältigen waren“ \cite[S.27]{Doerner}.\\
Dörner beschreibt vorliegende Problematik metaphorisch folgendermaßen:\\
\emph{ „So können wir sagen, dass ein Akteur in einer komplexen Handlungssituation einem Schachspieler gleicht, der mit einem Schachspiel spielen muss, welches sehr viele (etwa: einige Dutzend) Figuren aufweist, die mit Gummifäden aneinander hängen, so dass es ihm unmöglich ist, nur eine Figur zu bewegen. Außerdem bewegen sich seine und des Gegners Figuren auch von allein, nach Regeln, die er nicht genau kennt oder über die er falsche Annahmen hat. Und obendrein befindet sich ein Teil der eigenen und der fremden Figuren im Nebel und ist nicht oder nur ungenau zu erkennen“ \cite[S.66]{Doerner}.}\\
Mit diesem metaphorischen Vergleich veranschaulicht Dörner das wahrscheinliche 
Scheitern der nach den linearen Denkstrukturen des traditionellen Bildungssystems  ausgebildeten Experten beim Versuch, langfristige  Problemlösungen für komplexe Systeme zu finden.\\
Dörner fasst abschließend folgende Hauptursachen zusammen, die seiner Meinung nach mit verantwortlich für den Untergang Tanalands sind, wenn auch bei den einzelnen Probanden in unterschiedlicher Ausprägung. NAch Dörner, wird zum einen „die Ablaufgestalt von Prozessen“ \cite[S.32]{Doerner} nicht genügend beachtet und zum anderen werden Maßnahmen oft ohne ausreichende vorherige Situationsanalyse eingeleitet, sodass Fern- oder Nebenwirkungen nicht berücksichtigt werden, wie das oben skizzierte Beispiel der Vernetzung zwischen Jagd auf \emph{Kleinsäuger – Ernteertrag –  Großkatzen und Viehbestand} gezeigt hat. \\
Neben diesen Faktoren, die auf ein ungenügendes systemisches Denken zurückzuführen sind, 
stellte Dörner, seines ZeichenoPsychologe, während der sechs Sitzungen im Verlauf des Versuchs bei allen Probanden eine ähnliche Veränderung des Verhaltensspektrums fest. Die Anzahl der Protokolleinheiten zur Situationsanalyse, also Reflexionen oder Fragen, nahm in den letzten Sitzungen drastisch ab, weil,- so die nachvollziehbare Interpretation Dörners,-  die Probanden dann wohl glaubten „ein genügend genaues Bild von der Situation bekommen zu haben, welches keiner Korrektur (…) mehr bedurfte.“ \cite[S.29]{Doerner}. Ein Irrtum, wie der Versuch gezeigt hat.\\
Die offensichtlichen Parallelen der Ergebnisse des Tanaland-Versuchs mit der Realität veranlassten Dörner zu dem Folgeprojekt ‚Lohhausen‘ \cite[vgl. S.32ff]{Doerner}, bei dem 48 Probanden in Lohhausen, einer fiktiven Kleinstadt mit 3.700 Einwohnern, jeweils für zehn Jahre zum Bürgermeister mit umfassenden „ Kontroll - und Eingriffsmöglichkeiten“ \cite[S.34]{Doerner} ernannt wurden. \\
Die während der beiden Computersimulationen ermittelten psychologischen Faktoren überprüfte Dörner anhand der Reaktorkatastrophe vom 26. April 1986 in Tschernobyl auf ihre Übertragbarkeit auf die Realität. Eine genaue Analyse der  Umstände, die schließlich zum Super Gau führten \cite[vgl.„Tschernobyl in Tanaland“, S.47ff]{Doerner}, brachte einen überzeugenden Nachweis, dass der Unfall „zu hundert Prozent auf psychologische Faktoren zurückzuführen“ \cite[S.48]{Doerner} sei und nicht auf die veraltete Technologie.\newpage
 Die im Tanaland- und und Lohhausen-Szenario ermittelten Ergebnisse sind also durchaus auf die Realität übertragbar, was sich  auch in den beobachteten Verhaltensweisen widerspiegelt, denn sowohl die „Reaktorfahrer“  in Tschernobyl als auch die Versuchspersonen von Tanaland hatten übereinstimmend folgende Schwierigkeiten \cite[vgl. S.57]{Doerner}: 
 \begin{itemize}
 \item beim Umgang mit der Zeit
 \item bei der Einschätzung exponentieller Entwicklungen
 \item beim Umgang mit Neben- oder Fernwirkungen.
 \end{itemize}
Dörner schloss daraus „die Tendenz zu einem isolierten Ursache-Wirkungs-Denken“ \cite[vgl. S.57]{Doerner}, was angesichts der 93 Atomkraftwerke, die derzeit in 18 europäischen Ländern am Netz sind \cite{Atom},  beunruhigend wirken kann. 
\newpage
\subsection{Scheitern monokausaler Denkstrukturen beim Umgang mit komplexen Systemen nach Frederic Vester}\label{kapitel2.2}

Frederic Vester stützt die These Dörners, dass bei der Beurteilung heutiger Probleme die gängigen Erklärungsansätze und lineare Lösungsstrategien nicht mehr genügen. Während Dörner bezüglich der Umsetzung seiner Forderung: „Wir müssen lernen in Systemen zu denken“ \cite[S.326]{Doerner},  eher allgemein auf den „gesunden Menschenverstand“ \cite[S.325]{Doerner} setzt und als „Belehrungsmittel“ \cite[S.323]{Doerner} das Durchspielen von  Simulationsszenarios empfiehlt, versucht Vester mit seinem Konzept vom ‚vernetzten Denken‘  und dem von ihm entwickelten Sensitivitätsmodell dem Menschen konkrete Handlungshilfen anzubieten.\\
Das vorliegende Teilkapitel stellt hier zunächst den Werdegang Frederic Vesters vor, anschließend werden die Hauptaspekte in seinem Argumentationsgang bezüglich des Scheiterns monokausaler Denkstrukturen beim Umgang mit komplexen Systemen ausgeführt. 

\subsubsection{Kurzbiographie Frederic Vester}\label{kapitel2.2.1}
Die folgenden Ausführungen zum Leben von Frederic Vester stützen sich im Wesentlichen auf Informationen, die auf seiner Homepage \cite{Vester} zu finden sind und auf biographische Angaben von Jürgen Streich, der von Vester noch zu seinen Lebzeiten zum Verfassen seiner Biographie autorisiert worden war. Die Biographie erschien ursprünglich 1997 in Streichs Buch ‚30 Jahre Club of Rome‘ und wurde von ihm anlässlich von Vesters Tod im November 2003 mit geringfügigen Modifikationen im Internet veröffentlicht \cite{Streich}. \\
Frederic Vester, geboren 1925 in Saarbrücken, lässt sich kaum mit wenigen Begriffen charakterisieren und ein Überblick über seinen Werdegang verdeutlicht seine Vielseitigkeit und Universalität.\\
Nach dem Studium der Chemie in Hamburg und an der Pariser Sorbonne promovierte Vester in Hamburg in Biochemie und arbeitete dann am Heidelberger Institut für experimentelle Krebsforschung. Ausschlaggebend hierfür war die Krebserkrankung seiner Mutter. Es folgten verschiedene Forschungsaufenthalte in den USA und in Cambridge, England, bis er 1958 Dozent für Biochemie an der Universität Saarbrücken wurde. Dort wurde ihm,- laut der Biographie von Jürgen Streich-, die Habilitation unter dubiosen Umständen verweigert \cite[vgl.]{Streich}.  Seiner Darstellung zufolge wurde die Habilitationsschrift „Untersuchungen aus verschiedenen Disziplinen bis hin zu philosophischen Überlegungen“ \cite{Streich}, trotz überwiegend positiven Votums von elf bei insgesamt zwölf auswärtigen Gutachtern aber abgelehnt. 1969  habilitierte sich Vester jedoch erfolgreich in Konstanz.\\
Seit 1966 leitete er am Max-Planck-Institut in München eine eigene Arbeitsgruppe für Krebsforschung und stand dort nach eigener Aussage vor der Entscheidung, \enquote{den Systemgedanken zu vergessen und wissenschaftlich 'exakt', das heißt in einem spezifischen Fachgebiet, zu arbeiten oder aber ständig Vorwürfe zu ernten, die sich auf mein Vorhaben, unterschiedliche Fachbereiche zu verbinden, bezogen} \cite{Streich}.  Vester folgte jedoch seinen Überzeugungen, verließ 1970 die etablierte Forschung und gründete die ‚Studiengruppe für Biologie und Umwelt, Frederic Vester GmbH für interdisziplinäre Forschung, Publizistik und Beratung‘ in München, die er zusammen mit seiner Frau Anne aus eigenen finanziellen Mitteln aufbaute. Dort konnte er sein Ziel verfolgen, das \enquote{vernetzte Denken} zu vermitteln, welches er als Vordenker der Umweltbewegung als unerlässlich für den weiteren  Fortbestand der Menschheit erachtete. \\
Durch die Verwendung des damals weitgehend unbekannten Adjektivs ‚vernetzt‘ im Titel seiner internationalen Wanderausstellung \enquote{Unsere Welt - ein vernetztes System}  hat Vester den Begriff maßgeblich geprägt. Durch den großen Erfolg der 1978 als Jubiläumsausstellung zum 75-jährigen Bestehen des Deutschen Museums in Berlin eröffneten Wanderausstellung, die schließlich auf eine bemerkenswert lange ‚Wanderzeit‘ von achtzehn Jahren kam, machte er die Notwendigkeit einer Umorientierung zur ‚vernetzten‘ Denkweise einer breiten Öffentlichkeit zugänglich. Wie dem Bericht von Christian Baumann „Was hat die Ausstellung Wanderausstellung ‚Unsere Welt - ein vernetztes System‘ bewirkt?“ \cite[S.161ff]{Welt} zu entnehmen ist, steht der mittlerweile in den allgemeinen Sprachgebrauch übergegangenen Ausdruck ‚vernetzt‘ seit 1981 im Großen Duden.\\
Vester hatte „ein unbeirrbares Vertrauen in die Macht der Information“ \cite[S.12]{Neuland}, sodass er alle verfügbaren Möglichkeiten ausschöpfte, um sein neues Denkmodell des ‚vernetzten Denkens‘ in der Öffentlichkeit publik zu machen. Er ist Autor von insgesamt 17 Sachbüchern, entwarf das kybernetische Umweltsimulationsspiel ‚Ökolopoly‘, das im Jahr 2000 unter dem Namen ‚Ecopolicy‘ als Windows-Version 2.0 auf CD-ROM im Schulbuchverlag Westermann erhältlich war und entwickelte das Sensitivitätsmodell so weit, dass es seit ca. 1980 in zum Teil umfangreichen Studien eingesetzt worden ist \cite[vgl. S.359ff]{Kunst2005}. \\
Auch die Vielzahl von Ämtern in den unterschiedlichsten Bereichen, die er im Laufe seines Lebens bekleidete, spricht für Vesters Universalität: In den Jahren 1974-1978 war er Präsident des Bayerischen Volkshochschulverbandes, 1982 Gründungspräsident der Deutschen Energiegesellschaft, und, obwohl seines Zeichens Pazifist, von 1981 bis 1989 Ordinarius für Interdependenz von technischem und sozialem Wandel an der Universität der Bundeswehr in München. Zwischen 1989 und 1991 lehrte er als Gastprofessor für Betriebswirtschaft an der Hochschule St. Gallen und war seit 1993 Mitglied des Club of Rome. \\\\
Frederic Vester, ein Vorreiter auch in bildungstheoretischen Fragen, erkannte schon früh die Defizite im aktuell bestehenden Bildungssystem und forderte eine erweiterte Ausbildung in ‚Systemkunde‘, in der Spezialgebiete immer in den dazugehörigen Lebens- und Wirkungsraum eingebettet und die jeweiligen Wechselbeziehungen zwischen den existierenden Bereichen erarbeitet werden \cite[vgl. S.18]{KunstVester}. 
Als ob er schon im Vorfeld geahnt hätte, wie lange es dauern würde, bis das vernetzte Denken Einzug in die aktuellen Lehrpläne hält, stellte er bereits 1978 das Planspiel \enquote{Ökolopoly} als Computer-Exponat auf seiner  Wanderausstellung \enquote{Unsere Welt, ein vernetztes System}  vor, durch das er Kinder mit dieser Denkweise vertraut machen wollte.\\
Bei diesem Spiel werden  Lösungen für eine tragfähige Zukunft in drei fiktiven Ländern, nämlich einem  Industrie-, Schwellen- und Entwicklungsland gesucht, wodurch sich spielerisch in den Umgang mit Komplexität einarbeitet werden kann. Es gilt über 12 Jahre erfolgreich zu regieren und die Länder möglichst in den \enquote{Paradieszustand} zu führen. Bei jeder Entscheidung müssen wesentliche Bereiche, wie Politik, Produktion, Umwelt, Lebensqualität, Sanierung, Aufklärung und Bevölkerungsentwicklung berücksichtigt werden. Indem das Spiel die Folgen simuliert, die sich aus den jeweiligen Spielentscheidungen ergeben, verschafft es ein Verständnis für die ökologischen, wirtschaftlichen und sozialen Wirkungszusammenhänge.

\subsubsection{Die  zentralen Fehler in unserem Umgang mit komplexen Systemen }\label{kapitel2.2.2}
Unsere vorwiegend linearen Denkstrukturen haben uns Menschen zwar jahrhundertelang erfolgreich das Bestehen auf diesem Planeten ermöglicht, versagen allerdings jetzt im Zeitalter hochkomplexer, miteinander vernetzter Strukturen, sodass wir nach Vester im Umgang mit komplexen Systemen zu typischen Fehlern neigen. Seine Darstellung der sechs zentralen Fehler folgt, wie er selbst betont \cite[vgl. S.36f]{KunstVester}, den Ergebnissen Dörners vom ‚Tanaland-Versuch‘ [vgl.\ref{kapitel2.1}].   \\
Als ersten zentralen Fehler nennt Vester die „Falsche Zielbeschreibung“ \cite[vgl. S.36f]{KunstVester}.  Demzufolge beschränkt man sich ohne Blick auf die Gesamtsituation auf die Lösung von Einzelproblemen, wodurch eine Art von „Reparaturdienstverhalten“ \cite[S.36]{KunstVester} entsteht. Dabei beseitige man einen Missstand nach dem anderen, erkennt aber aufgrund des  fehlenden Überblicks nicht, wenn beispielsweise die letzte Korrektur gar kein neues Problem korrigiert, sondern nur die Folge einer vorherigen Korrektur behoben hat. Der zweite von Vester konstatierte Fehler betrifft die „Unvernetzte Situationsanalyse“. Es wird dazu tendiert, große Datenmengen zu erheben, ohne diese aber in Ermangelung geeigneter Ordnungsprinzipien sinnvoll auswerten zu können, sodass die Dynamik des Systems nicht erkannt wird \cite[vgl. S.37]{KunstVester}. Als nächsten Fehler identifiziert Vester die „Irreversible Schwerpunktbildung“ \cite[S.37]{KunstVester}. Er bezeichnet damit die Tendenz, sich nach anfänglichen Erfolgen weiterhin nur auf diesen Schwerpunkt zu fixieren, sodass wesentliche Probleme in anderen Bereichen nicht registriert werden. Mit dem vierten Fehler „Unbeachtete Nebenwirkungen“ \cite[S.37]{KunstVester} verweist Vester auf fehlende Nebenwirkungsanalyse in linear-kausalen Lösungsansätzen. Bei der „Tendenz zur Übersteuerung“ \cite[S.37]{KunstVester}, dem fünften Fehler, beruft sich Vester bei seiner Erklärung auf Dietrich Dörner. Es käme häufig zu falsch dosierten Eingriffen, da das Phänomen der Zeitverzögerung für das Eintreten etwaiger Fern- bzw. Nebenwirkungen nicht berücksichtigt würde. In Erwartung eines schnellen Feedbacks tendiere man dann zum  Übersteuern: Das erste Eingreifen in ein System erfolge eher  zaghaft, wenn dann aber der gewünschte Erfolg zu lange ausbleibt, würde der Eingriff sofort um einen großen Faktor erhöht, um dann alles bei ersten negativen Rückmeldungen wieder abzubremsen. \\
Zu welchen fatalen Folgen diese bisher genannten Fehler führen können, hat Dörner bei der Analyse der Umstände und der durchgeführten Maßnahmen vor dem Reaktorunglück  in Tschernobyl minutiös beschrieben \cite[vgl.„Tschernobyl in Tanaland“, S.47ff]{Doerner}.  Wichtig wäre es hier noch anzumerken, dass die zahlreichen Fehlentscheidungen nicht aus mangelnder Sachkenntnis getroffen wurden, sondern aufgrund der „Unfähigkeit zum nichtlinearen Denken in Kausalnetzen statt in Kausalketten“ \cite[S.54]{Doerner}.  Verantwortlich war zu dem Zeitpunkt „ein gut eingespieltes Team hoch angesehener Fachleute“ \cite[S.55]{Doerner}. \\
Der letzte bei Vester aufgezeigte Fehler betrifft die „Tendenz zu autoritärem Verhalten“ \cite[S.37]{KunstVester}. Wer in ein System einzugreifen vermag und meint, es verstanden zu haben, neigt Vester zufolge zu diktatorischem Verhalten, was für ein komplexes System völlig ungeeignet sei. Persönlichkeitsabhängig steige dann eventuell die Gefahr, Entscheidungen eher im Hinblick auf eigenen Prestigegewinn zu treffen und dabei das eigentliche Ziel, die Funktionsverbesserung des Systems zu vernachlässigen \cite[vgl. S.37]{KunstVester}.

\subsubsection{Die Unzulänglichkeit des Klassifizierungs-Universums nach Frederic Vester}\label{kapitel2.2.3}
Nach Vester identifiziert drei Ursachen für die oben genannten Fehler \cite[vgl. S.39ff]{KunstVester}: 
\begin{enumerate}
\item Auftrennung der Wirklichkeit
\item Mangel an kybernetischem (vgl Kapitel \ref{kapitel4}) Verständnis
\item Zu kurzer Planungshorizont
\end{enumerate}
Bei näherer Betrachtung resultieren jedoch alle der oben genannten Faktoren aus den Denkstrukturen, wie sie nach Vester für die Wirklichkeitswahrnehmung unseres Kulturkreises
aufgrund der üblichen Auftrennung der Wirklichkeit in immer kleinere Bereiche typisch sind. 
Vester unterscheidet in Anlehnung an Maruyama Masao (1914-1996), einem japanischen Politikwissenschaftler und Historiker, drei Möglichkeiten die Wirklichkeit zu erfassen \cite[S.143ff]{KunstVester}:
\begin{itemize}
\item als Klassifizierungs-Universum
\item als Relations-Universum
\item als Relevanz-Universum
\end{itemize}

Die abendländische Kultur geht bei der Betrachtung der Wirklichkeit von einem künstlich geschaffenen ‚Klassifizierungs-Universum‘ \cite[S.143f]{KunstVester} aus, das ein logisches Funktionieren der verschiedenen Systeme voraussetzt. Diese kategorische Einteilung der Welt in Fächer, Ressorts usw. des Klassifizierungs-Universum lässt den Menschen nach Vester das globale Wirkungsgefüge vergessen oder  \enquote{übersehen}.\\
Vester zufolge werden Kinder im Alter von sechs Jahren beim Schuleintritt aus einer ganzheitlichen Aneignung der Wirklichkeit, dem so genannten „Relations-Universum“ \cite[S.144f]{KunstVester} herausgeholt und mit einer in einzelne Disziplinen aufgefächerten Welt konfrontiert, also dem für unseren Kulturkreis typischen \enquote{Klassifizierungs-Universum} \cite[S.143]{KunstVester}. Nach seiner Auffassung erfasst ein Kleinkind die Wirklichkeit noch ganzheitlich, d.h. die Dinge der Umwelt werden nicht als Begriffe, sondern in ihrer Wirkung und Bedeutung innerhalb der Wirklichkeit betrachtet, es lebt demnach also in einem Relations-Universum. Während der ersten Schuljahre verliert sich jedoch diese Sichtweise. Man lernt die Wirklichkeit im Einzelnen zu definieren, analysieren und zu katalogisieren, wobei es allerdings an der Fähigkeit mangelt, das in den verschiedenen Unterrichtsfächern erworbene isolierte Faktenwissen fächerübergreifend in einen größeren Zusammenhang stellen zu können. Nach den Regeln des Klassifizierungs-Universums ausgebildet, denken und handeln die Menschen in Schubladen, was natürlich wenig hilfreich ist, wenn die Realität im ganzen durchschauen werden soll.  Die Realität funktioniert interdisziplinär und hält sich nicht an die künstlich erschaffenen Regeln eines Klassifizierungs-Universums.\\

Vester zufolge macht eine Ausbildung nach den Kategorien des ‚Klassifizierungs-Universums‘ den Menschen natürlich auch empfänglich für dessen Trugschlüsse und Fallen. Die Hinwendung zum  Rationalen begünstigt den allgemein vorherrschenden Glauben an den Wahrheitsgehalt von Statistiken und Hochrechnungen. Dadurch wird häufig auf zurückliegende Ereignisse zurück gegriffen und versucht diese in die Zukunft zu extrapolieren. Für statische Problemstellungen ist dies zwar ein durchaus erfolgsversprechendes Verfahren, jedoch bleibt der Systemzustand unberücksichtigt. Die einfachen Rückschlüsse durch die Aktion - Reaktion in einem Klassifizierungs-Universums lassen sich nicht so einfach postulieren, da in der Realität deutlich mehr Variablen auf die direkte Entscheidungsfindung Einfluss haben. \\
Der Glaube an eine prognostische Aussagekraft von solchen Voraussagen trägt nach Vester neben den oben genannten Problemen bezüglich der unsystemischen Herangehensweise einen großen Teil zu den Problemen bei, die durch falsche Entscheidungen entstehen, da dabei nicht berücksichtigt wird, dass sich komplexe Systeme nur in zwei Fällen linear wie Maschinen verhalten, die sich dann durch Hochrechnungen voraussagen lassen:\\
Zum einen kann eine solche Prognose nur für einen sehr kurzfristigen Moment der wirklichen Reaktion eines komplexen Systems entsprechen, wobei der jeweilige Zeithorizont je nach System variiert. Innerhalb dieses Zeitraums können meist alle direkten Abhängigkeiten überblickt werden, und einigermaßen exakte Vorhersagen über die möglichen Entwicklungen eines komplexen Systems sind machbar. Es ist allerdings ein Trugschluss, diesen Zeithorizont vergrößern zu können, indem  eine Prognose durch ein erhöhtes Datenaufkommen aus der Vergangenheit langfristig in die Zukunft erweitert wird. Diese Idee scheitert meist, was sich am Beispiel der Wettervorhersage gut veranschaulichen lässt. Der Zeithorizont liegt für Wetterprognosen im Bereich von Stunden. Daran ändert auch nichts, dass sich seit 1960 die Anzahl der Wettermessstationen etwa vervielfacht hat. Eine zielsichere Prognose  ist immer noch kaum 24 Stunden in die Zukunft möglich, da das komplexe System des Weltklimas von einer sehr hohen Anzahl an Variablen bestimmt wird, die alle direkt oder indirekt Einfluss nehmen \cite[S.93f]{KunstVester}.  
Zum anderen verhält sich ein System in seiner Wachstumsphase  häufig linear, weswegen gerne Prognosen auf
Grundlage der Vergangenheit/Historie zur Extrapolierung der Zukunft genutzt werden.\\
 Allerdings wird versucht, im Anschluss daran, die  Erfolge während einer Wachstumsphase, welche durch Hochrechnungen erhalten wurden, auch auf andere Phasen zu übertragen und bedenkt nicht, dass komplexe Systeme sich außerhalb der oben genannten Fälle auch völlig akausal verhalten können.\\

Das im Klassifizierungs-Universum dominierende cartesianische Denken favorisiert natürlich eine unsystemische Sichtweise der Wirklichkeit, sodass vorwiegend in  Kausalketten gedacht wird, was letztendlich nach Vester zu einem unzureichenden  kybernetischem Verständnis führt.
Durch die Aufsplitterung  der Wirklichkeit in verschiedene Fachgebiete werden die kybernetischen Steuerungs- und Regelungsvorgange nicht erkannt und das Prinzip der Selbstregulation verletzt, was in Kapitel \ref{kapitel4} noch näher ausgeführt wird.\\

Abschließend sei noch auf das von Vester konstatierte Problem des zu kurzen Planungshorizonts verwiesen. Vester zufolge leben wir alle, einschließlich Entscheidungsträger in Politik und Wirtschaft, leben nach Vester noch heute, also seit etwa 6.000 Jahren „nach wie vor in trautem Einklang mit der jährlichen Haushaltsplanung der ersten Pflanzer und Hirten“ \cite[S.75]{KunstVester}.  Insofern konzentriert man sich vorzugsweise auf kurzfristige Lösungsstrategien, die natürlich heutzutage angesichts der Komplexität der weltweit überall existierenden  Probleme, man denke nur an Staatsverschuldung, Arbeitslosigkeit oder die diversen Umweltprobleme, versagen müssen. Ein Jahr reicht nicht aus, um die Auswirkungen von solchen Entscheidungen bezüglich eines Systems zu beobachten und dann angemessen zu analysieren. Komplexen Systemen muss also die  Zeit gelassen werden,  alle seine direkten und indirekten Wirkungsbeziehungen zu entfalten.

\newpage
\subsection{Das ‚Collingridge Dilemma‘}\label{kapitel2.3}
Abschließend soll in diesem Kapitel noch auf das so genannte ‚Collingridge Dilemma‘ verwiesen werden, welches für den Bereich der Technikfolgenabschätzung nachweist, dass mit den gewohntem monokausalen Denkstrukturen keine überzeugenden Ergebnisse erreicht werden. \\
Der Begriff der Technikfolgenabschätzung  bezeichnet ein Teilgebiet der Technikphilosophie und –soziologie und versucht durch Beobachtung und Analysen von Technik, Wirtschaft und Gesellschaft Zusammenhänge zu erkennen und für eine zu analysierende Thematik Chancen und Risiken abzuwägen. Obwohl der Begriff seit den 1960er Jahren existiert,  gibt es noch keine einheitliche Definition \cite[vgl. S.14]{Grunwald}. Insofern besitzt die nachfolgende Definition von Armin Grunwald keine Allgemeingültigkeit, sondern wird hier lediglich als eine mögliche Begriffsklärung aufgeführt: „Die implizite Definition von Technikfolgenabschätzung als Antwort(en) auf gesellschaftliche Bedarfs- und Problemlagen erlaubt es, die vielfältigen Facetten der Technikfolgenabschätzung als verschiedene Antworten auf verschiedene Aspekte der Problemlagen aufzufassen und zuzuordnen“ \cite[S.16]{Grunwald}.\\

Das  nach David Collingridge, einem britischen Technikforscher des 20 Jahrhunderts, benannte ‚Collingridge Dilemma‘, auch Kontroll- oder  Steuerungsdilemma genannt, befasst sich mit der methodischen  Unstimmigkeit zwischen dem Informations- und dem Umsetzungsproblem während einer Technikfolgenabschätzung.\\
Bei dem Informationsproblem handelt es sich um dass Wirkungen eines Systems mit fortschreitender Entwicklung und Umsetzung immer besser zu erkennen und zu prognostizieren sind, wohingegen das Umsetzungsproblem sich mit der Einflussnahme von Veränderungswünschen befasst. Das Problem liegt darin, dass eine Gestaltung und Einflussnahme auf ein System mit fortschreitender Umsetzung zunehmend schwieriger wird, andererseits jedoch bei einer noch nicht umgesetzten Technologie im Vorfeld nicht alle eventuell auftretenden Probleme und Konsequenzen erkannt werden.\\
Anders ausgedrückt: Solange eine Technologie noch nicht eingeführt ist, können Systemänderungen zwar einfach durchgeführt werden, jedoch sind ihre Konsequenzen noch unklar. Werden jedoch die Folgen klarer, so ist das System bereits so weit umgesetzt und entwickelt, dass elementare Änderungen schwierig sind: „When change is easy, the need for it cannot be foreseen; when the need for change is apparent, change has become expensive, difficult and time consuming” \cite[S.11]{Collingridge}.\\
Am Beispiel der „Grünen Revolution“ welche in den 1960er Jahren begonnen hat und moderne landwirtschaftliche Geräte und Düngekulturen sowie Hochertragssorten in Entwicklungsländer zur Bekämpfung von Hungersnöten brachte, veranschaulicht Collingridge das nach ihm benannte Dilemma. Der ursprüngliche Nutzen, nämlich die Reduzierung des Hungerproblems zur Prävention gegenüber gewaltsamen Revolutionen zeigte erst bei fortgeschrittener Umsetzung große Missstände sowie gravierender Umweltschäden, wie folgende Beschreibung von Raj Patel, Eric Holt-Gimenez \& Annie Shattuck in dem Artikel \enquote{Das Ende von Afrikas Hunger} \cite{Hunger} in  der progressiv-linksliberale Wochenzeitschrift  „The Nation“ aus den Vereinigten Staaten im September 2009 zeigt:\\ „Außer der massiven Vertreibung von Bauern, brachte die Grüne Revolution noch weitere soziale Missstände mit sich – Bildung von Armenviertel für die vertriebenen Bauern um die Städte herum, Anstieg des Einsatzes von Schädlingsbekämpfungsmitteln, Absenkung des Grundwasserspiegels und Anstieg der Umweltbelastungen durch die industrielle Landwirtschaft“  \cite[S.2]{Hunger}. \\
Collingridge fasst diese sich immer wieder neu entwickelnde  Diskrepanz zwischen dem Informations- und dem Umsetzungsproblem in folgender Feststellung zusammen: \\
„\emph{It is this huge disparity between our technical competence and our understanding of how the fruits of this competence affect human society which has given rise of the widespread hostility to technology}“ \cite[S.11]{Collingridge}. \\

Es gilt also ein Analysemodell zu etablieren, das innerhalb eines komplexen Systems die vernetzten Beziehungsstrukturen in seiner Gesamtheit erfasst, damit im Voraus mögliche Spät- und Fernwirkungen erkannt und dementsprechend berücksichtigt werden.

%% file: 3_kap3.tex
\newpage
\section{Definition des ‚sozio-technisches System’ nach Kienle und Kunau}\label{kapitel3}
Mitte des 20. Jahrhunderts wurde erstmals eine systemische Betrachtungsweise der Beziehungen zwischen Arbeitern und technischen Bedingungen eines Betriebs durch „The Tavistock Institute“ etabliert, das sich in den 1950er Jahren beim britischen Steinkohlbergbau  mit der Frage befasst hat, warum trotz innovativer technischer Neuerungen der gewünschte Erfolg ausblieb. Während dieser Forschung zu den Ursachen für die ausbleibende Effizienzsteigerung wurde der Begriff des ‚sozio-technischen‘ Systems begründet, indem vorhandene Arbeitsstrukturen sowohl in technische als auch soziale Systeme untergliederte wurden, die in verschiedenster Weise miteinander kombinierbar sind und nur in Kombination verstanden bzw. gestaltet werden können \cite [S. 143ff]{Gertraude}.\\
Seit den Ergebnissen des Tavistock Instituts wurde der Begriff des sozio-technischen Systems sehr uneinheitlich verwendet, wie die folgende Bestimmung des sozio-technischen Systems aus der Habilitationsschrift des inzwischen emeritierten Professors für Allgemeine Technologie Günter Ropohl zeigt:\\\\
„\emph{Ein Computer wird erst wirklicher Computer, wenn er zum Teil einer Mensch-Maschine-Einheit geworden ist. Wenn Text geschrieben wird, tut das nicht allein der Mensch, aber es ist auch nicht allein der Computer, der den Text schreibt; erst die Arbeitseinheit von Mensch und Computer bringt die Textverarbeitung zuwege. Da freilich im benutzten Computer immer schon die Arbeit Verwendung anderer Menschen verkörpert ist, da also die Mensch-Maschine-Einheit nicht nur durch den einzelnen Nutzer gebildet, sondern auch von anderen Menschen mitgeprägt wird, bezeichne ich sie als soziotechnisches System}“ \cite[S.58]{Ropoh}.\\\\
Für den IT-Bereich wurde der Begriff ‚sozio-technisches System‘ im Wesentlichen durch Andrea Kienle und Gabriele Kunau systematisiert. In der vorliegenden Arbeit wird daher der Begriff ‚sozio-technisches System’ im Sinne der Definition von Kienle und Kunau verwendet,  sodass im Folgenden die zentralen Aspekte zur Bestimmung eines sozio-technischen Systems von Kienle/Kunau erläutert werden.\\ 
Ausgangspunkt von Kienle/Kunau ist ein allgemeiner Systembegriff, der weitgehend der auch im normalen Sprachgebrauch üblichen Auffassung von einem System entspricht: „Unter einem System im Allgemeinen versteht man eine Menge an Elementen, die in Beziehung zueinander stehen und so eine Einheit bilden“ \cite[S.83]{Kienle}. Als notwendigen Begleitfaktor eines jeden Systems fügen Kienle/Kunau  jedoch ergänzend die Umwelt hinzu, die alles enthält, „was nicht zum System selber gehört“ \cite[S.83]{Kienle}. Allerdings ist nach Kienle/Kunau \enquote{die Grenze zwischen Umwelt und System keineswegs immer objektiv eindeutig gegeben} \cite[S.83]{Kienle},  worauf  später noch näher eingegangen wird.\\
Die weiteren Ausführungen folgen der Vorgehensweise bei Kienle/Kunau, die zunächst  technische und soziale Systeme voneinander getrennt definieren. An dieser Stelle sei noch angemerkt, dass Kienle/Kunau dabei auch hinsichtlich der Terminologie auf die Systemtheorie von Niklas Luhmann  zurückgreifen \cite[vgl. S.83]{Kienle}, was dann aber jeweils im Einzelnen erläutert wird. \\\\
\textbf{Technische Systeme:}\\
Abgesehen von der allgemeinen Systemdefinition, die nach Kienle/Kunau auch für technische Systeme ihre Gültigkeit behält, konstatieren sie als Gemeinsamkeit bei allen technischen  Systemen, dass sie „das Ergebnis eines Konstruktions- und Produktionsprozesses sind“ \cite[S.84]{Kienle}. Demzufolge reagiert ein funktionierendes technisches System seinem Funktionscharakter entsprechend vorhersagbar, d.h. es entspricht den Regularien, nach denen es erstellt wurde, und zeigt klar definierte Verhaltensweisen sowie einen bestimmten funktionalen Katalog an Reaktionen, die bei entsprechendem Input ausgeführt werden. 
In Anlehnung an die Luhmann’sche Terminologie bezeichnen Kienle und Kunau ein solches System als ‚allopoietisch‘, entsprechend des altgriechischen Ausdrucks ‚allopoiesis‘ (‚allo‘ in der Bedeutung ‚fremd‘/‚verschieden“ und ‚poiein‘ für ‚machen‘/‚bauen‘), was frei übersetzt „von außen hergestellt“ \cite[S.84]{Kienle} bedeutet. 
Bei einem technischen System handelt es sich also um ein System, das sich nicht selbst reproduzieren kann.\\\\
\textbf{Soziale Systeme:}\\
Die Zusammenführung der beiden Systeme zum bei Kienle/Kunaus Konzept des sozio-technischen Systems basiert auf der engen Verbindung zwischen technischen und sozialen Systemen mit ihrer wechselseitigen Prägung. \\
Nach Kienle/Kunau ist die naheliegende Definition „Soziale Systeme bestünden aus Menschen“ zu eng gezogen, denn dadurch würden „Firmen wie Siemens oder (…) die römisch-katholische Kirche, politische Parteien“ \cite[S.86]{Kienle} nicht erfasst. Sie definieren daher in Anlehnung an Luhmann: „Soziale Systeme bestehen aus Kommunikationen, die sich sinnvoll aufeinander beziehen“ \cite[S.86]{Kienle}. Dies bedeutet bei Kienle und Kunau, jede Kommunikation in einem sozialen System müsse „in dem Sinne anschlussfähig sein, dass sie wieder neue Kommunikation im System erlaubt“ \cite[S.88]{Kienle}.  
Demzufolge sind für Kienle/Kunau soziale Systeme ‚autopoietisch‘, da diese von Menschen produziert werden und sich im Gegensatz zu technischen Systemen „von innen von selber“ \cite[S.87]{Kienle} entwickeln. Auch hier verweist der von Luhmann übernommene Begriff auf eine zentrale Eigenschaft sozialer Systeme: Das griechische Präfix ‚auto‘ bedeutet ‚selbst‘, dies kombiniert mit dem schon erläuterten ‚poiein‘ für ‚machen‘/‚bauen‘ heißt nach Kienle/Kunau, „dass sich soziale Systeme selbst erzeugen“ \cite[S.88]{Kienle}. Bezogen auf die oben erwähnte Definition heißt dies, dass soziale Systeme ihre Elemente (Kommunikationen) selber erzeugen \cite[vgl. S.88f]{Kienle}.\\
Weiterhin folgen Kienle/Kunau Niklas Luhmanns Anschauung ‚autopoietischer‘ Systeme als \enquote{geschlossene Systeme insofern, als sie das, was sie als Einheit zu ihrer eigenen Reproduktion verwenden (also: ihre Elemente, ihre Prozesse, sich selbst) nicht aus ihrer Umwelt beziehen können. Sie sind gleichwohl offene Systeme insofern, als sie diese Selbstreproduktion nur in einer Umwelt, nur in Differenz zu einer Umwelt vollziehen können}\cite[S.49]{Luhmann}.\\
Daraus resultierend, gelten für Kienle/Kunau soziale Systeme als „operativ geschlossen. Dies bedeutet, dass der Fortgang der Kommunikation ausschließlich innerhalb des sozialen Systems bestimmt wird“ \cite[vgl. S.88f]{Kienle}. Dadurch sind soziale Systeme zwar autonom, da sie sich „durch fortwährende Kommunikationen am Leben“ \cite[S.89]{Kienle} erhalten, jedoch nicht autark in dem Sinne, dass sie unabhängig von ihrer jeweiligen Umwelt wären: „Sie sind sehr wohl abhängig von den gegebenen Rahmenbedingungen und sie reagieren auch auf Impulse aus ihrer Umwelt“ \cite[S.8]{Kienle}. 
In dem Zusammenhang stellt sich für Kienle/Kunau für soziale Systeme jedoch die Frage nach der Abgrenzung: „Wo endet das System und wo beginnt seine Umwelt?“ \cite[S.90]{Kienle} Da soziale Systeme von außen nicht determiniert werden können, schlussfolgern Kienle/Kunau, dass soziale Systeme ihre Grenzen durch Selbstbeschreibungen bestimmen müssen, womit sie das letzte zentrale Merkmal eines sozialen Systems definiert haben: „Mit Selbstbeschreibungen beschreiben soziale Systeme ihre Grenzen. Sie regeln, welche Kommunikationen zum sozialen System gehören und welche nicht“ \cite[S.91]{Kienle}. \\\\
\textbf{Sozio-technische Systeme:}\\
Das grundlegend Neue im Ansatz von Kienle/Kunau besteht darin, dass sie die \\enge Verbindung zwischen technischen und sozialen Systemen mit ihrer wechselseitigen Prägung erkannten und beide im Konzept des sozio-technischen Systems zusammenführten. Ihrer Definition folgend wird aus einem sozialen ein sozio-technisches System, sobald es zur Unterstützung der Kommunikationsprozesse eine Beziehung zu mindestens einem technischen System eingeht, wobei sich beide Systeme wechselseitig prägen und das technische System in die Selbstbeschreibung des sozialen Systems einbezogen wird \cite[S.96f]{Kienle}.\\
In dieser Arbeit wird die normalerweise stetige Veränderung der Selbstbeschreibungen von sozio-technischen Systemen fixiert, um so aussagekräftig genug zu sein, dass Einflüsse und Komponenten sich mit Hilfe dieser in externe und interne Faktoren kategorisieren lassen, was für den später vorgestellten Analyseansatz wichtig ist.\\\\
\textbf{Das Internet als sozio-technisches System:}\label{ref_internet}\\
Am Beispiel des Internets als sozio-technisches System kann die enge Verbindung und wechselseitige Prägung der Systeme sehr deutlich aufgezeigt werden. Das Internet besteht unter anderem aus einer Vielzahl von einzelnen Computern, die als technische Komponenten dezentral, also ohne explizites Zentrum miteinander verbunden sind. Abschätzungen über die genauen Nutzerzahlen und Größen sind extrem schwierig zu treffen, jedoch lässt sich der öffentlich direkt zu erreichende Teil mit 4,71 Milliarden Seiten anhand der von der Suchmaschine Google veröffentlichten Zahlen \cite{Size} der indizierten Webseiten grob erfassen. Über die Interaktionen von menschlichen Akteuren kann demzufolge zwar auch keine genaue Zahl geboten werden, jedoch sind sowohl menschliche Akteure wie auch Unternehmen in dieses System integriert und interagieren täglich darüber. \\
Das Internet ist heutzutage eine wichtige Kommunikationsplattform. Anwendungen wie E-Mails oder der schnelle Austausch von Nachrichten bestätigen dies. Die Anfeorderungen Kienle/Kunaus nach Unterstützung der Kommunikationsprozesse der sozialen Komponente, also des Menschen, durch die technische Komponente, der aufgebauten Rechnerstruktur sind folglich erfüllt.\\
So hat sich in den letzten Jahren bei den Bereichen Informationsaustausch und Kommunikation eine wechselseitige Prägung ergeben, wie der Wandel um den Wissensbegriff und Wissensaneignung bei den menschlichen Akteuren eindrucksvoll zeigt: \enquote{An die Stelle des Buches bzw. der Typographie als gesellschaftlichem Leitmedium rückt zunehmend die vernetzte Welt des Cyberspace} \cite[S.413]{Pscheida} wodurch \enquote{eine Umorientierung in den Wahrnehmungsgewohnheiten und den Handlungsmustern notwendig} \cite[S.425]{Pscheida} geworden ist . 
In Korrelation hat sich die Wissenspräsentation im Internet stark an die Bedürfnisse der aktuellen Akteure angepasst, wie man zum Beispiel an dem 2001 gegründeten freien Online-Lexikon ‚Wikipedia‘ erkennen kann, welches bereits heute zu den sechs weltweit am meisten besuchten Internetseiten gehört \cite[S.280]{Weinberg}. Der Bezeichnung des Internets als sozio-technisches System steht somit nichts im Weg. \\
Selbst wenn das technische System auch beim Internet Eingang in die Selbstbeschreibung des sozialen Systems findet, so ist diese Eigenschaft eingängiger an dem Instant Messenger ICQ aufzuzeigen. Das Startup Unternehmen wurde 1996 in Israel gegründet und bietet den Nutzern eine Möglichkeit, zeitverschoben oder in Echtzeit miteinander Nachrichten auszutauschen. Die zuvor angesprochene Unterstützung der Kommunikation der sozialen Komponente durch die technische ist hier überflüssig zu erläutern, da der Chat-Anbieter sein Geschäftsmodell auf genau der Ressource ‚Kommunikation‘ aufgebaut hat. \\
In dem Moment, als sich in den Köpfen der Menschen ICQ als Kommunikationsplattform etablierte und anstatt Adressen oder Telefonnummern der ICQ-Kontakt über die so genannte ICQ-Nummer ausgetauscht wurde, identifizierten sich die menschlichen Akteure des Systems mit der Plattform und integrierten ICQ in die Selbstbeschreibung von sich selbst.

%% file: 4_kap4.tex
	\newpage
\section{Der biokybernetische Systemansatz nach Frederic Vester}\label{kapitel4}
Nachdem im vorherigen Kapitel der für die folgende sozioinformatische Analyse relevante Begriff des sozio-technischen Systems definiert wurde, soll hier zunächst Vesters Systembegriff aufgezeigt werden. Eine Vergleich beider Systembegriffe ist nicht beabsichtigt, da es in vorliegender Arbeit vorrangig um die Frage geht, inwieweit sein an der Natur orientiertes Konzept eines komplexen Systems bei der späteren sozioinformatischen Analyse anwendbar bzw. hilfreich ist.\\
In diesem Kapitel wird also die Absicht verfolgt, die für das Verständnis des Vester’schen Sensitivitätsmodells notwendigen theoretischen Grundlagen zu erläutern. Hierfür wird näher auf Vesters biokybernetischen Denkansatz eingegangen, da die Biokybernetik eine zentrale Rolle in seiner Systemtheorie einnimmt. \\
Am Schluss des Kapitels werden diejenigen Aspekte der ‚Fuzzy Logic‘ genauer beleuchtet, die von Relevanz für das Sensitivitätsmodell von Vester sind.\\
An dieser Stelle sei angemerkt, dass inhaltliche Wiederholungen bestmöglich vermieden wird. Insofern beschränken sich die weiteren  Ausführungen in diesem Kapitel auf die Darstellung der Aspekte, welche die theoretische Voraussetzung bzw. Grundlage des sozio-informatischen Analyseansatz (Kapitel \ref{kapitel5}) bilden. 
\newpage
\subsection{Definition des 'komplexen Systems' nach Vester}\label{kapitel4.1}
Statt des eindimensionalen, monokausalen Denkens braucht es nach Vester eine neue „systemische Sichtweise“ \cite[S.103]{Neuland},  bei der  sich nicht, wie üblich, auf das zu lösende Problem innerhalb eines Systems konzentrieren wird, sondern stattdessen versuchen wird, das System quasi von außen zu betrachten. Der Blickwinkel soll sich auf das System selbst und sein gesamtes Wirkungsgefüge richten, sodass der Ausgangspunkt seiner Theorie daher zunächst die genaue Erfassung des komplexen Systems an sich ist. Es empfiehlt sich also zunächst genauer zu klären, was Vester unter einem ‚komplexen System‘  versteht.\\
Der Begriff ‚System‘ kommt aus dem Altgriechischen: ‚\emph{systema}‘ und bedeutet ein ‚aus mehreren Einzelteilen zusammengesetztes Ganzes‘ \cite[S.81]{Macke}.\\
Vester unterscheidet zunächst zwischen System und einem „Nicht-System“, z.B. einem Haufen Sand, bei dem man „Teile davon miteinander vertauschen“ \cite[S.27]{Neuland} kann, ohne dass sich an dem ‚Nicht-System‘ Haufen Sand etwas ändert.\\
Vester folgt wie Kienle und Kunau \cite[vgl. S.83]{Kienle} zunächst dem allgemeinen Systembegriff, wenn er als wesentliche Eigenschaften eines Systems bestimmt, „daß es erstens aus mehreren Teilen bestehen muß, die jedoch, zweitens, verschieden voneinander sind“ \cite[S.27]{Neuland}.  Der letzte Aspekt dieser Definition, dass die Teile „drittens, nicht wahllos nebeneinanderliegen, sondern zu einem bestimmten Aufbau miteinander vernetzt sind“ \cite[S.27]{Neuland} deutet jedoch bereits, wie noch ausgeführt wird, auf die zentrale Eigenschaft seiner Bestimmung komplexer Systeme. \\
Offensichtlich folgt Vester bei der Entwicklung seines Systembegriffs  den  eingangs schon erwähnten Ausführungen zur ‚Allgemeinen Systemlehre‘ des Biologen Ludwig von Bertalanffy. Dieser postulierte statt der Analyse von Einzelphänomenen wie in der Physik üblich für die Biologie die „organisierte Komplexität“ \cite[S.34]{Bertalanffy}, die er nicht durch bloße lineare Kopplungen, sondern durch die bestehenden Wechselwirkungen zwischen den Einzelphänomenen als gegeben sah.\\
Wie auch später Vester unterscheidet er in seiner Systemlehre zwischen offenen und geschlossenen Systemen \cite[vgl. S.36]{Bertalanffy}. Ein geschlossenes System verfügt nach Bertalanffy über keine Wechselwirkungen mit der Umwelt, während offene Systeme dagegen ihre interne Organisation durch Umwelteinflüsse verändern können. Durch diese ‚Selbstorganisation‘ \cite[S.3]{Bertalanffy}, gelingt es offenen Systemen, sich in einem dynamischen Umfeld zu stabilisieren.\\  
Ähnlich bestimmt Vester für offene Systeme als entscheidendes Kriterium die Überlebensfähigkeit:  „Lebensfähige Systeme sind niemals abgeschlossen, sondern immer nach außen offen“ \cite[S.29]{Neuland}. Demnach existieren offene Systeme nie für sich allein, sondern stehen stets im Austausch mit den sie umgebenden anderen Systemen. Wesentlich ist hierbei für Vester das Prinzip der ‚Selbststeuerung‘ lebensfähiger Systeme, auf das noch genauer in \ref{kapitel4.2} eingegangen wird. \\
Der noch zu klärende Begriff ‚komplex‘ leitet sich aus ‚complexus‘,  dem Partizip Perfekt des lateinischen Verbs ‚complecti‘ her, was so viel wie ‚umschlingen, umfassen, zusammenfassen‘ bedeutet.  Die Bedeutung des Gebrauchs im Vester’schen Sinn erschließt sich allerdings erst, wenn man etymologisch weiter auf den altgriechischen Ursprung des Worts zurückgeht. Demnach geht es auf das altgriechische Partizip ‚plektós‘ von  ‚pléko‘ zurück, dem altgriechischen Ausdruck für ‚flechten‘. ‚Komplex‘ bedeutet hier also wörtlich ‚verflochten, verwoben‘, was der Vester’schen Auffassung exakt entspricht.\\
Bei der Bestimmung eines komplexen Systems konkretisiert Vester den dritten Aspekt der Vernetzung in seiner schon erwähnten allgemeinen Systemdefinition. Demnach verfügt ein komplexes System wie jeder Organismus „aus mehreren verschiedenen Teilen (Organen), die in einer bestimmten dynamischen Ordnung zueinander stehen und zu einem Wirkungsgefüge vernetzt sind. In dieses kann man nicht eingreifen, ohne dass sich die Beziehungen aller Teile zueinander und damit der Gesamtcharakter des Systems ändern würde“ \cite[S.25]{KunstVester}.  Insofern muss jeder Problemlösungsprozess zunächst beim System selbst ansetzen, da jeder Eingriff  Auswirkungen auf das Gesamtsystem hat, dessen vernetzte Beziehungen sich zudem scheinbar oft  akausal verhalten, also nicht immer direkt logisch erklärbar sind.  Nach Vester lässt sich daher ein komplexes Systems nicht vollständig aus den Eigenschaften der Komponenten des Systems erklären, „denn komplexe Systeme verhalten sich nun einmal anders als die Summe ihrer Teile” \cite[S.25]{KunstVester}. 
\newpage
\subsection{Herkunft und Bedeutung des Begriffs ‚Biokybernetik‘}\label{kapitel4.2}
Vester folgt in seinem Ansatz der ursprünglichen Bedeutung von Kybernetik, wie sie von ihrem Begründer Norbert Wiener in seinem 1948 erschienenen Buch ‚Cybernetics or Control and Communication in the Animal and the Machine‘ \cite{Wiener} verstanden wird. Wiener bezog sich in seiner kybernetischen Forschung zur Frage nach selbstständiger Steuerung bei sich selbst regulierenden Systemen sowohl auf technische als auch auf lebendige Systeme bzw. Organismen.  Im Laufe der zweiten Hälfte des 20. Jahrhunderts fand der  Begriff ‚Kybernetik‘ seinen Weg in den deutschen Sprachgebrauch und wurde nach Vester als ein „Erkennen, Steuern und selbstständiges Ineinandergreifenden von Regeln und vernetzten Abläufen bei minimalem Energieaufwand“ \cite[S.124]{KunstVester} definiert. Er kritisiert in dem Zusammenhang, dass Kybernetik dabei häufig als „Regeltechnik und Computersteuerung“ missverstanden wurde, obwohl nach seiner Auffassung nichts „unkybernetischer als die Rechenweise eines Computers“ \cite[S.124]{KunstVester} sei. Dieser Ansicht ist jedoch nur bedingt zu folgen. Zwar hat diese Aussage ihre Berechtigung, wenn man einen Computer isoliert nur als Apparat betrachtet. Allerdings wird dann der Umstand nicht berücksichtigt, dass ein ‚Computer‘ ja erst in Verbindung mit dem Menschen funktioniert, der diesen erst einmal einschalten und dann mit entsprechenden Daten ‚füttern‘ muss, damit er zu arbeiten beginnt. Gestützt wird dies auch durch Günter Ropohls  Beschreibung eines Computers als „Teil einer Mensch-Maschine-Einheit“ \cite[S.58]{Ropoh} bei dessen Bestimmung als sozio-technisches System, worauf bereits in  Kapitel drei näher eingegangen wurde. \\\\
Der von Vester gewählte Begriff ‚Biokybernetik‘ ist eine Wortschöpfung, der die beiden altgriechischen Begriffe „\emph{bios}“ für ‚Leben‘ und „\emph{kybernétes}“ für ‚Steuermann‘ kombiniert. Wie im Folgenden ausgeführt wird, verweist Vester durch die Wahl dieses Begriffs bereits auf zwei zentrale  Aspekte seines Konstrukts.\\
Mit dem Begriff ‚Biokybernetik‘ beabsichtigt Vester durch das Präfix ‚Bio‘ den seiner Meinung nach vernachlässigten Aspekt der Kybernetik hinsichtlich ihrer Bedeutung für lebendige Systeme bzw. Organismen zu betonen. \\
Vesters Weltbild geht von der Überzeugung aus, dass Mensch und Natur nicht voneinander getrennt zu betrachtende Systeme sind: „Es gibt nicht hier den Menschen – dort die Natur. Wir selbst sind Natur, (…) sind ein Teil von ihr“ \cite[S.30]{KunstVester}. Die Betrachtung der Natur als übergeordnetes, alles Leben auf diesem Planeten umschließendes „Supersystem“ sieht er durch die Tatsache gestützt ihres erfolgreichen Überdauerns, denn das „Unternehmen Biosphäre hat es seit vier Milliarden Jahren geschafft, sich allen Widernissen zum Trotz auf einem Planeten zu behaupten und sogar noch weiterzuentwickeln. (...) Kein Lebewesen kann für sich allein existieren. Nur die enge Vernetzung zwischen allen Lebewesen macht ein Überleben möglich“ \cite[S.108]{KunstVester}. Dabei hat das gigantisch komplexe System ‚Natur‘ ihr eigenes Überleben keineswegs durch detaillierte Vorprogrammierung oder zentrale Steuerung erreicht, sondern dies lediglich durch Impulsvorgaben zur Selbstregulation geschafft.  Bestand haben in ihr nur lebensfähige Systeme; was nicht funktioniert, wird aussortiert bzw. geht zugrunde. 
Vester zog daraus den Schluss, dass für komplexe Systeme die Überlebensfähigkeit oberste Priorität haben müsse und dies am besten durch eine selbstständige Steuerung erreicht wird, was sich dann, miteinander kombiniert, in seiner  Begriffswahl ‚Biokybernetik‘ widerspiegelt.
\newpage
\subsection{Das Prinzip der Selbststeuerung in kybernetischen Regelkreisen}\label{kapitel4.3}
 \begin{figure}[htb]
 \centering
 \includegraphics[width=0.7\textwidth,angle=0]{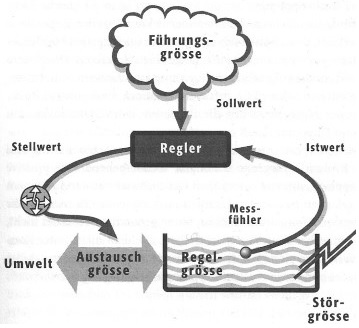}
 \caption{Regelkreis mit den g{\"a}ngigen kybernetischen Bezeichnungen \cite[S.128]{KunstVester}}
\label{fig:vester_rk}
\end{figure}
Eine zentrale Rolle bei der Selbstregulierung spielt dabei nach  Vester der kybernetische Regelkrei
s, der durch seine  „stabilisierende Dynamik“  \cite[S.125]{KunstVester} selbstständig Änderungen an der ‚Regelgröße‘ vornehmen kann und seiner Meinung nach dadurch die Überlebensfähigkeit eines komplexen Systems garantiert. \\\\
Vesters Schaubild eines kybernetischen Regelkreises (Abbildung \ref{fig:vester_rk}) lässt sich recht einfach am Beispiel einer durch einen Thermostaten geregelten Heizung erläutern: \\

\begin{figure}[htb]
 \centering
 \includegraphics[width=0.35\textwidth,angle=0]{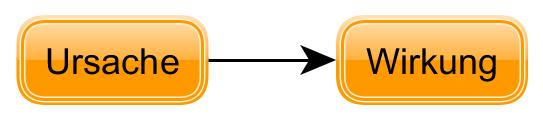}
 \caption[Ursache-Wirkung-Beziehung]{Ursache-Wirkung-Beziehung}
\label{fig:ursache_wirkung}
\end{figure}

Die ‚Führungsgröße‘ gibt dabei den ‚Sollwert‘ an, in dem Fall die gewünschte Raumtemperatur, die mittels eines ‚Reglers‘, hier also eines Thermostats konstant zu halten oder ‚kybernetisch‘ ausgedrückt, zu regeln ist. Informationen zur ‚Regelgröße‘, der aktuellen Raumtemperatur, bekommt der Regler durch den so genannten ‚Messfühler‘ (Thermometer).  Der auf diese Weise ermittelte ‚Istwert‘ wird vom Regler nun mit dem vorgegebenen Sollwert verglichen. Wird eine Abweichung infolge der Einwirkung einer ‚Störgröße‘ festgestellt, z.B. durch ein geöffnetes Fenster, so wird eine entsprechende Anweisung (‚Stellwert‘) an das ‚Stellglied‘, hier an das Zulaufventil des Heizungswassers, übermittelt. Das Ventil korrigiert nun über die ‚Austauschgröße‘, die in dem Fall der zugeführten Menge an Warmwasser entspricht, die Raumtemperatur bzw. die Regelgröße. Diese wird nun wiederum vom Messfühler erfasst, sodass nach Vester „das zu regelnde System mit sich selbst rückgekoppelt ist“ \cite[S.125]{KunstVester}.
Die auf das System einwirkende Führungsgröße kann hier nun auch die Regelgröße eines anderen Regelkreises sein, wodurch sich direkte und indirekte Abhängigkeiten ergeben und ein Regelkreis nicht als isoliert oder abgeschlossen zu betrachten ist.\\\\
\textbf{Rückkopplungen:}\\
Regelkreise funktionieren Vester zufolge auf dem einfachen Prinzip der Rückkopplung, die sich durch eine meist innerhalb eines Systems stattfindende Selbstbeeinflussung durch aufeinander einwirkende Prozesse auszeichnen.\\
Zur Erklärung dieses wesentlichen Prinzips in  Vesters biokybernetischen Ansatz werden hier die  Ursache–Wirkung-Beziehungen (Abbildung \ref{fig:ursache_wirkung}) herangezogen. \\

Die klassische Ursache-Wirkung-Beziehung, bei der  
auf eine Ursache unausweichlich und direkt eine Wirkung folgt, bildet die Grundstruktur des monokausalen Denkens.\\\\

Daraus entwickeln sich nach Derek K. Hitchins  durch steigende Komplexität zunächst  Wirkungsketten \cite[vgl. S.17]{Hitchins}, in denen entsprechend Ursache-Wirkung die Beziehungen verkettet werden, um Reaktionsabfolgen darzustellen und Ursachenforschung zu ermöglichen.  \\\begin{figure}[htb]
 \centering
 \includegraphics[width=\textwidth,angle=0]{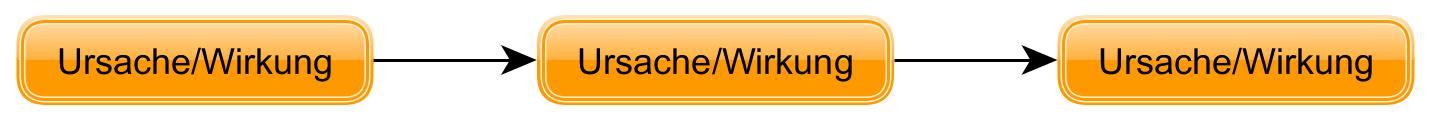}
 \caption[Wirkungs- oder Kausalkette]{Wirkungs- oder Kausalkette}
\label{fig:wk}
\end{figure}

Solche Wirkungsketten bezeichnet Hitchins als „process model“ \cite[S.18]{Hitchins}, Dietrich Dörner spricht hier von „Kausalketten“ \cite[S.54]{Doerner}. \\

 Hat eine solche Wirkungskette Einfluss auf sich selbst, bildet also, wie in \\ Abbildung \ref{fig:wk} gezeigt, einen Wirkungskreis aus Ursache-Wirkung–Beziehungen, so nennt Hitchins dies  „Pipeline/Circe Model“ \cite[S.18]{Hitchins}. \\
 \begin{figure}[htb]
 \centering
 \includegraphics[width=\textwidth,angle=0]{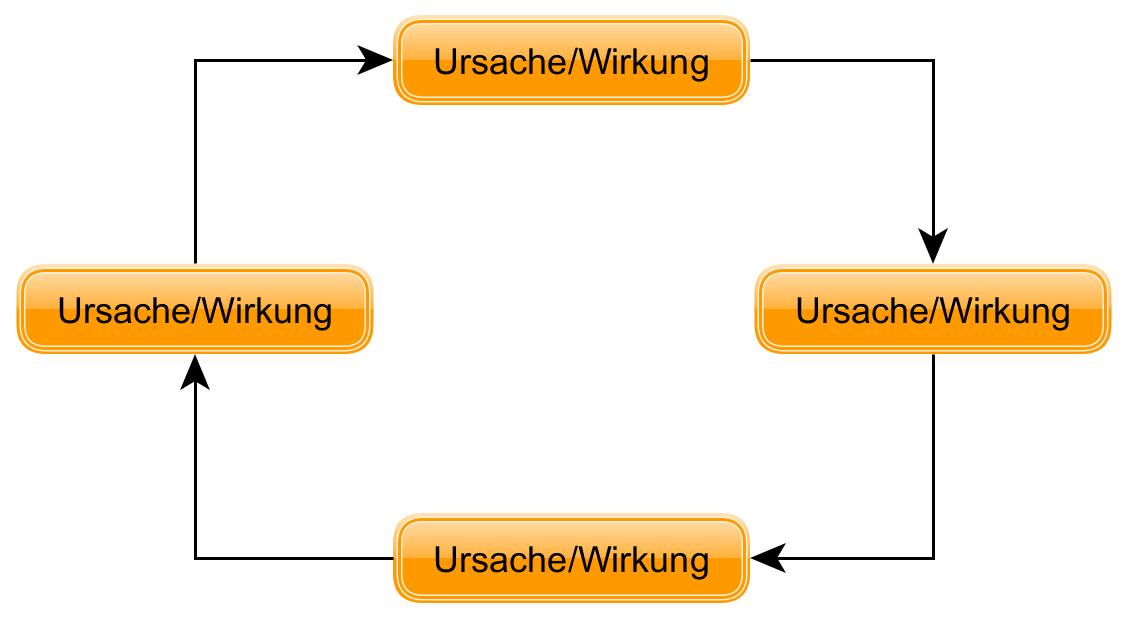}
 \caption[Wirkungskreis]{Wirkungskreis}
\label{fig:kreis}
\end{figure}
 Dieses Modell zeichnet sich durch eine Art Rückkopplung in einem entsprechend komplexen System aus, bei dem eine Ursache auf sich selbst zurückfallen kann \cite[S.18]{Hitchins}. Für Vester ist ein solches System, wie oben aufgezeigt, „mit sich selbst rückgekoppelt“ \cite[S.125]{KunstVester}. Durch den Kreisprozess werden nach Vester Ursache und Wirkung relativiert: „Die Kausalität verliert ihre festgelegte Richtung, weil Ursache und Wirkung verschmelzen, im Kreisprozeß ihre Rollen vertauschen“ \cite[S.129]{KunstVester}.\\\\
 Vester unterscheidet grundsätzlich zwei Arten von Rückkopplung:\\
Ist eine Rückkopplung sich selbst verstärkend, besteht sie also ausschließlich aus gleich orientierten   Auswirkungen, so spricht Vester von einer ‚positiven Rückkopplung‘ \cite[vgl. S.211]{KunstVester}. Eine solche Rückkopplung katalysiert eine einflussnehmende Größe im System oder eine Entscheidung bis ins Extreme, da sich die positiven oder negativen Einflussnahmen immer weiter hochschaukeln, bis sie ein physikalisches Maximum erreichen oder ins Unendliche gehen. Derartige Rückkopplungen sind allerdings für komplexe Systeme gefährlich, selbst wenn gewisse Prozesse eine solche positive Rückkopplung als Katalysator benötigen, um Reaktionsketten in Bewegung  zu setzen \cite[vgl. S.126]{KunstVester}.
Zur Veranschaulichung eines solchen Vorgangs nennt Vester die Metamorphose einer Raupe zu einem Schmetterling, die als Anstoß zunächst eine ‚positive Rückkopplung‘ benötigt, um vom bisherigen Gleichgewichtszustand als Raupe den Veränderungsprozess innerhalb des Kokons zu beginnen und zu durchleben \cite[vgl. S.126]{KunstVester}.
Trotzdem muss sie auch der ihr „übergeordneten negativen Rückkopplung“ \cite[S.126]{KunstVester} gehorchen, um funktionieren zu können bzw. lebensfähig zu sein. Nach Vester bewirken nur jene Regelkreise, die sich durch eine negative Rückkopplung auszeichnen, also auf eine gegenläufige Regulierung hinweisen, eine Selbstregulation im System \cite[vgl. S.212]{KunstVester}. Sie sorgen „für Stabilität gegen Störungen und Grenzüberschreitungen“ \cite[S.128]{KunstVester} und sind langfristig überlebensfähig. \\
Grundsätzlich betont Vester, dass ein System deutlich mehr negative als positive Rückkopplungen ausweisen sollte, um gegenüber Störungen stabil zu bleiben. Auf diesen Sachverhalt wird jedoch an späterer Stelle (Kapitel \ref{kapitel5.8}) noch konkreter eingegangen. 
\newpage
\subsection{Ziel der Biokybernetik}\label{kapitel4.4}
Die Natur als Vorbild vor Augen, sieht Vester, wie schon erwähnt, den entscheidenden Vorteil seines biokybernetischen Modells darin, dass der Steuermann selbst, also der ‚Regler‘, Teil des Systems ist, wobei seine Aufgabe innerhalb des Regelkreises nur in der „Impulsvorgabe zur Selbstregulation“ \cite[S.110]{KunstVester} besteht. Das Ziel sollte dabei nach Vester nicht im Erreichen eines bestimmten Zustands liegen, sondern in der Förderung der Lebensfähigkeit eines Systems \cite[S.110]{KunstVester}, denn nach seiner Ansicht besteht die einzige Zielsetzung eines komplexen Systems, in welcher Form es auch vorliegt, in der Steigerung der Überlebensfähigkeit.\\
Er betrachtet die Natur als übergeordnetes, gigantisches komplexes System	, das uns „die beste Orientierungshilfe“  bieten würde, „um zügig zu neuen energie- und rohstoffsparenden kybernetischen Lösungen zu kommen“ \cite[S.142]{KunstVester}. Durch ihre selbstständige Regelung ineinandergreifender, vernetzter Abläufe sollte sie uns nach Vester in vielen Bereichen als Vorbild dienen. So wälze die Biosphäre Äonen hunderte Milliarden Tonnen an Material und hat ihr Überleben bis heute gesichert. Garantiert wird diese Überlebensfähigkeit gerade durch ihr gänzlich konträres Vorgehen im Vergleich zur wachstumsorientierten heutigen Zeit. Im Gegensatz zu unserer Wachstumsorientiertheit garantierte ihr gerade das Nullwachstum an Biomasse von rund 2.000 Milliarden Tonnen \cite[vgl. S.118]{KunstVester} nicht nur das Fortbestehen des Gesamtsystems, sondern ermögliche in den Jahrmillionen darüber hinaus auch noch eine stetige Weiterentwicklung ihrer Subsysteme bzw. oder Lebewesen zu immer höheren Formen. Dies war nach Vester nur möglich, da  „das Management der Natur eine Handvoll kybernetischer Grundregeln befolgt hat“ \cite[S.113]{KunstVester}.\\\\
Am Vorbild  der Natur entwickelte Vester acht Grundregeln der Biokybernetik \cite[S.127ff]{KunstVester}, 
die ihm zufolge allgemeingültig für alle lebensfähigen Systeme wären, da sie entscheidende Fähigkeiten für ihre Selbstorganisation darstellen. Der Grad der Überlebensfähigkeit oder, wie Vester es auch nennt, der „kybernetischen Reife“ \cite[S.256]{KunstVester} eines Systems ließe sich daran ablesen, inwieweit es sich „die biokybernetischen Grundregeln mehr und mehr zu eigen macht und somit an Robustheit gegenüber äußeren Störungen gewinnt“ \cite[S.256]{KunstVester}. Außerdem würde ihre Umsetzung jedem Projekt helfen, „eine höhere ‚kybernetische Reife‘ zu erlangen“ \cite[S.128]{KunstVester}.\\
Da die Regeln bei jedem komplexen System anwendbar und das ganze System erfassen sollen, würde dies die Bewertung eines komplexen Systems laut Vester enorm vereinfachen. \\\\
Die von Vester postulierte universale  Anwendbarkeit dieser Regeln auf sozio-technische Systeme wird im Anschluss an die soziinformatische Analyse in Kapitel \ref{kapitel6} durchgeführt, in dem die Regeln auch konkret im Einzelnen dargestellt werden. 
\newpage
\subsection{Fuzzy Logic}\label{kapitel4.5}

Die Fuzzy Logic, vom englischen Begriff „fuzzy“ für ‚verwischt‘ oder ‚verschwommen‘, stellt eine Erweiterung der klassischen binären (zweiwertigen) Logik (wahr - falsch, 0-1) dar  und geht bereits auf den griechischen Philosophen Platon zurück, der bereits die Vermutung aufstellte, dass es zwischen Wahr und Falsch etwas dazwischenliegendes geben müsste.\\ 
Begründet wurde dieser Gedanke von J. Lukasiewicz (1878-1956), welcher der klassischen Logik zunächst einen dritten Zustand ‚1 - 2‘ oder ,possible‘ hinzufügte und in seinen weiteren Forschungen darum bemüht war, eine unendlichwertige Logik einzuführen. \\
Dies legte den Grundstein für den amerikanischen Mathematiker und Informatiker Lotfali Askar-Zadeh, der das uns heute bekannte ‚Fuzzy Set‘ \cite{Lotfi} einführte, das als Grundgerüst der Fuzzy Logic gilt. Diese unscharfe Menge (englisch: fuzzy set) unterscheidet sich von der klassischen Mengenlehre, in der ein Element einer Grundmenge in einer definierten Menge entweder enthalten ist oder nicht, in dem Maße, dass ein Element zusätzlich eine Zugehörigkeitsfunktion besitzt, die beschreibt zu welchen Grad ein Element in der aufgestellten Menge enthalten ist. Da diese Funktion  in das reellwertige Intervall von [0,1] abbildet gibt es hier die unendlich viele Möglichkeit die gewünschte Unschärfe bezüglich der Zugehörigkeit eines Elements darzustellen.\\
In Vesters Theorie wird diese Unschärfe genutzt, um die  unexakten Teile der Wirklichkeit zu erfassen. Durch die scharfe Schranken der klassischen Logik lassen sich Unklarheiten, Zweideutigkeiten und Verallgemeinerungen nicht wirklich erkennen. Dies sind typische Phänomene für soziale Systeme sowie Umwelteinflüsse, die deshalb nach Vester dringend in die aussagekräftige Analyseverfahren einbezogen werden sollten. Außerdem bietet die Fuzzy Logic bei Analyseverfahren einen entscheidenden Vorteil, denn „durch die Einbindung qualitativer Daten bei der >>Fuzzy Logic<<-Simulation (…) hat man die Garantie, dass sie zwar mehr oder weniger unexakte, dafür aber niemals falsche Konzepte der Realität anbietet“ \cite[S.227]{KunstVester}.

%% file: 5_kap5.tex
\newpage
\section{Sozioinformatischer Analyse zur Technikfolgenabschätzung am Beispiel des Unternehmens ‚UBER‘}\label{kapitel5}
Der vorliegende sozioinformatische Analyseansatz zur Technikfolgenabschätzung  eines sozio-technischen Systems orientiert sich in der methodischen Vorgehensweise der Analyse am  Sensitivitätsmodell von Frederic Vester.\\
Wie bereits in Kapitel \ref{kapitel4} ausgeführt, postuliert Vester für sein Modell ein „universelles Anwendungsspektrum“ \cite[S.271]{KunstVester}. Daher soll entsprechend der ersten Zielformulierung (vgl. Kapitel \ref{kapitel1.2}) während der sozioinformatischen Analyse das Modell bei der Anwendung auf ein sozio-technisches System hinsichtlich der Frage untersucht werden, inwieweit sich dieses Konzept auch für die Analyse eines sozio-technischen Systems eignet. Dabei sollen sowohl die Anwendbarkeit der Planungsschritte beim methodischen Vorgehen als auch die Eignung des Sensitivitätsmodells für eine Technikfolgenabschätzung bei sozio-technischen Systemen geprüft werden.\\
Aus diesem Grund scheint es angebracht, die Grundzüge des Sensitivitätsmodells zu Beginn des Kapitels \ref{kapitel5.1} im Hinblick auf die von Vester für sein Modell vorgeschlagenen neun Arbeitsschritte kurz darzustellen.\\
Im Anschluss wird in Kapitel \ref{kapitel5.2}  die Durchführung der einzelnen Arbeitsschritte der sozioinformatischen Analyse erläutert, bei der Vesters Sensitivitätsmodell auf das \\Startup Unternehmen ‚UBER‘ angewendet werden soll,  um sich über mögliche Folgen und Reaktionen ein Bild zu verschaffen. 
Das erste Teilkapitel befasst sich mit der Systembeschreibung, wobei das Unternehmen ‚UBER‘ vorgestellt wird. ‚UBER‘ versteht sich als Dienstleister zur Vermittlung von Fahrdiensten und existiert mittlerweile weltweit in verschiedenen, der jeweiligen Gesetzeslage in den Ländern weitgehend angepassten Varianten. Die vorliegende Analyse setzt jedoch den Schwerpunkt auf das UBER’sche  Geschäftsmodell in Deutschland, das während der Systembeschreibung des Unternehmens in \ref{kapitel5.2} vorgestellt wird.\\
Das methodische Vorgehen bei der Erstellung einer systemrelevanten Datensammlung zur Beschreibung des Systems ‚UBER‘ wird in \ref{kapitel5.3} anhand einer chronologischen Darstellung der jeweiligen Arbeitsschritte erläutert.\\
Auf der Grundlage der gesammelten Datensätze kann im nächsten Schritt \ref{kapitel5.3} die für die spätere Technikfolgenabschätzung des Systems relevante Einordnung der innovativen sozio-technischen Komponenten und Ideen in die aktuelle Forschung und das wirtschaftliche Gefüge vorgenommen werden. Angewandt auf das Unternehmen ‚UBER‘ wird in dem Zusammenhang das ‚Share-Economy‘ - Konzept näher erläutert, das der innovativen Geschäftsidee des Unternehmens zugrunde liegt. Basierend auf den bisherigen Ergebnissen kann anschließend das Unternehmen UBER als sozio-technisches System definiert werden. \\
Im Anschluss an die Analyse wird mit den zuvor erstellten Datensätzen entsprechend des Vester’schen Sensitivitätsmodells das Wirkungsgefüge für das sozio-technische System ‚UBER‘ entwickelt, sodass dann mittels einer Regelkreisanalyse die Überlebensfähigkeit des sozio-technischen Systems um UBER beurteilt und mögliche Konfliktpotentiale aufgezeigt werden können.\\
Die biokybernetischen Regeln werden in Kapitel \ref{kapitel6} erläutert und hinsichtlich ihrer Anwendungsmöglichkeit auf sozio-technische Systeme  geprüft.
\newpage
\subsection{Das Sensitivitätsmodell  von F. Vester}\label{kapitel5.1}
Vester versucht mit dem von ihm entwickelten Sensitivitätsmodell konkrete Handlungshilfen für systemgerechtes Planen und Handeln anzubieten. Mittels dieses systemischen  Modells lassen sich einzelne Faktoren eines komplexen Systems auch untereinander in Verbindung setzt, sodass präzise Problemlösungsstrategien gefunden werden können.\\
Mit der Bezeichnung ‚Sensitivität‘ verweist Vester auf die Möglichkeiten seines Modells, „bereits die geringsten Regungen eines komplexen Systems auf innere oder äußere Einflüsse“ \cite[S.158]{KunstVester} zu registrieren.\\
Zum methodischen Vorgehen schlägt er neun Arbeitsschritte vor:\\\\
\textbf{1. Systembeschreibung }\\
 Zunächst gilt es, das System unter biokybernetischen Gesichtspunkten, „im Sinne der übergeordneten Zielsetzung ‚Erhöhung der Lebensfähigkeit“ \cite[S.162]{KunstVester} zu beschreiben.  Für Vester ein Vorgang, in den bereits „alle (…) Betroffenen“ \cite[S.162]{KunstVester} integriert sind, um frühzeitig eine breitgefächerte Übersicht über das System zu gewinnen. Hierbei fließen für Vester „Statistiken, (…) und Finanzgutachten ebenso (…) wie Schilderungen von Mißständen, Wünsche und Meinungen“ \cite[S.162f]{KunstVester} in die Beschreibung mit ein. \\\\
\textbf{2. Erfassung der Einflussgrößen}\\
Es gilt nun eine Menge von „Schlüsseldaten und Einflussfaktoren, die für das Systemverhalten eine Rolle spielen“ \cite[S.163]{KunstVester} aufzustellen. Zu beachten ist, dass Vester Wert auf den variablen, also veränderlichen Charakter dieser Faktoren legt. Eine reine Quantifizierung auf Zahlenwerte ist hier ausdrücklich nicht gefordert,  da für die so genannten „weichen Daten“ \cite[vgl. S.163]{KunstVester} legt, welche sich nur in einer qualitative Bezeichnung widerspiegeln, worauf aber in Kapitel \ref{kapitel5.4} genauer eingegangen wird.\\\\
\textbf{3. Prüfung auf Systemrelevanz }\\
Um in einer gesamtheitlichen Betrachtung nicht vollständig den Überblick zu verlieren, gilt es schon in der Betrachtung und Erfassung der für das System relevanten Einflussgrößen eine Reduktion durchzuführen. Vester schlägt daher den „Prozeß einer systemgerechten Auswahl“ \cite[S.163]{KunstVester} vor, an dessen Ende ein repräsentativer Satz von Einflussgrößen entsteht. „Dabei gilt es, die bis dahin gesammelten Variablen systematisch aus verschiedenen Blickwinkeln abzutasten“ \cite[S.163]{KunstVester}. Reduziert wird auf eine Anzahl von 20-30 Faktoren \cite[vgl. S.164]{KunstVester}, um die folgende Betrachtung durchführbar zu gestalten, dabei aber trotzdem „keiner Frage an das System ausweicht“ \cite[S.164]{KunstVester}.\newpage
\textbf{4. Hinterfragung der Wechselwirkung }\\
Durch Hinterfragung der Auswirkung aller Einflussgrößen aufeinander, entsteht ein Bild der Abhängigkeiten und Einflüsse untereinander. Vester verweist hier auf den von ihm entwickelten „Papiercomputer“ \cite[S.164]{KunstVester} welcher sich als Werkzeug für diese Betrachtung anbiete, da sich als Ergebnis eine so genannte ‚Einflussmatrix‘ entwickeln würde, worauf jedoch in dieser Arbeit nicht genauer eingegangen wird. \\\\
\textbf{5. Bestimmung der Rolle im System}\\
Durch die Betrachtung der in Kapitel \ref{kapitel4} erstellten Kriterienmatrix prognostiziert Vester ein direktes Verständnis über die Position einer jeden Variable im System \cite[vgl. S.164]{KunstVester}. So kann Vester entgegen einer klassischen, starren Betrachtungsweise die innere Dynamik eines solchen Systems abbilden und steuernde Prozesse und Komponenten, so genannte Steuermänner innerhalb des Systems identifizieren. Diese entnehmen die Istwerte und  regeln die Sollwerte des Systems selbst. Durch die Identifizierung solcher Komponenten ist es möglich, sowohl deren Steuermöglichkeit als auch dessen latente Risiken und Chancen zu erkennen. Sollten Variablen sich an diesem Punkt als sehr träge herausstellen, ist ein Auswechseln jederzeit möglich, denn dem Sensitivitätsmodell ist ein ausgeprägtes rekursives Verhalten inne, d.h. ein korrigierendes Wiederaufgreifen eigener Strukturen zur effizienten Problembewältigung, was somit eine Flexibilität gewährleistet, um auf Veränderungen und neue Erkenntnisse reagieren zu können. „Dazu wird die Stärke der Wirkung jeder einzelnen Variable im Fall ihrer Veränderung auf jede andere abgeschätzt“ \cite[S.164]{KunstVester}.\\\\
\textbf{6. Untersuchung der Gesamtvernetzung }\\
Ist das System erfasst, wird es durch eine von der konventionellen Betrachtungsweise abweichende Diagnose untersucht. Die aufgestellten Einflussgrößen  werden durch intensive Betrachtung des Systems in ihren  Wechselwirkungen wahrgenommen und auf grafischer Ebene eingezeichnet. Anhand der so gewonnen Abhängigkeiten untereinander \cite[vgl.S.168f]{KunstVester}, werden schnell Rollen und Aufgaben der einzelnen Variablen ersichtlich, die in einer linearen Ursache-Wirkung-Beziehung nicht auftauchen würden. Vester charakterisiert daher keine lokalen Kausalketten, wie z.B. direkte Abhängigkeiten, sondern trifft Aussagen über das Systemverhalten. \newpage
\textbf{7. Kybernetik einzelner Szenarien} \\
Da sich im Wirkungsgefüge meist Teilbereiche herauskristallisieren, „die man gesondert auf ihre Kybernetik hin untersuchen möchte“ \cite[S.167]{KunstVester}, bietet Vester auch hier ein Programm an, welches mit Hilfe einer tabellenbasierten relationalen Datenbank die Verbindung zum Gesamtsystem herstellt. Je nach Größe des Teilsystems und der veränderbaren Variablen empfiehlt Vester so genannte Policy-Tests und Regelkreisanalysen, um „das Systemverhalten und die Folgen bestimmter Eingriffe zu simulieren“ \cite[S.167]{KunstVester}.\\\\
\textbf{8. Wenn-dann-Prognosen und Policy-Tests}\\
Während des  Aufbaus entsprechender Simulationen in Zusammenarbeit mit den Betroffenen bieten sich Policy-Tests und Wenn-dann-Prognosen an „um Tendenzen, Grenzwerte und Reaktionen des Systems auf bestimmte Eingriffe hin auszumachen und zu testen“ \cite[S.167]{KunstVester}. Konkrete Verhaltensweisen des Systems sowie seiner Teilsysteme können zwar nicht prognostiziert werden, dafür aber ein möglicher Systemzustand, der aus einer Interaktion mit dem Gesamtsystem entstehen könnte. Auch hier wird ein ganzheitliches Vorgehen von Vester gefordert, um die Simulationen und daraus resultierende Lösungsstrategien auf ihre Eignung zu überprüfen.\\\\
\textbf{9. Systembewertung und Strategie}\\
Auf dieser letzten Ebene wird eine ausführliche Beurteilung des Systems bezüglich seiner Überlebensfähigkeit und Selbstregulation auf  der Grundlage der  biokybernetischen Grundregeln durchgeführt, worauf in Kapitel \ref{kapitel6} genauer eingegangen wird. Zuvor gesammelte Erkenntnisse über das System werden an diesem selbst überprüft, bis ein Modell erstellt werden kann \cite[vgl. S.169]{KunstVester}.  	

\newpage
\subsection{Systembeschreibung: Vorstellung des Unternehmens ‚UBER‘ }\label{kapitel5.2}
Im Folgenden wird die Durchführung der sozioinformatischen Analyse anhand der chronologischen Darstellung der einzelnen Analyseschritte dargestellt, wobei im Einzelnen die Anwendbarkeit des methodischen Vorgehens vom Sensivitätsmodell auf die Analyse des sozio-technischen Systems ‚UBER‘ diskutiert wird.\\\\
Als ersten Arbeitsschritt sieht Vester für sein Sensitivitätsmodell eine ausführliche „Systembeschreibung“  \cite[S.162]{KunstVester} vor, was auch dem geplanten Vorgehen des sozioinformatischen Analyseansatz entspricht, da zu Beginn eine grundlegende Sammlung aller Informationen sowie  Rahmendaten auch für sozio-technische Systeme die Grundvoraussetzung für die weitere Betrachtung ist.\\
Insofern steht am Anfang zunächst die genaue Erfassung des Systems, was bei einem komplexen System wie ‚UBER‘  bedeutet, eine Vielzahl von Daten, auch über die Abhängigkeiten von dessen Umwelt, zu erfassen und zu protokollieren.\\
UBER ist ein 2009 in den USA gegründetes Startup Unternehmen, welches sich als Online-Vermittlungsdienst von Fahrgästen sowohl an Mietwagen mit Fahrer (\emph{UberBlack}) als auch an private Fahrer (\emph{UberPop}) zur Personenbeförderung versteht. 2014 ist UBER bereits in 45 Ländern vertreten und bietet in Deutschland neben Berlin seine Dienste noch in vier weiteren Großstädten an. Die folgende Analyse konzentriert sich auf den - nicht nur - in Deutschland umstrittenen Sektor ‚UberPop‘, also die Sparte von UBER, in der private Fahrzeughalter UBER-Kunden ‚mitnehmen‘.
Bei UBER  kann ein Nutzer über eine App in Städten, in denen der Dienst angeboten wird, sein Fahrtziel und den Ausgangsort eingeben und sich den potentiellen „Fahrpreis“ kalkulieren lassen. Wird die Fahrt über die App gebucht, so erhält der UBER-Fahrer über sein Smartphone eine entsprechende Benachrichtigung und kann den Kunden befördern. Von der Begrifflichkeit versteht sich UBER hier nicht als Taxizentrale, sondern als eine Art Mitfahrzentrale, also als innovativen Dienstleister im Sharing Economy-Bereich, worauf später (vgl. Kapitel \ref{kapitel5.6}) noch genauer eingegangen wird. \\
Auch wenn UBER eine Webseite mit ähnlicher Funktionalität wie die APP anbietet, so wird aus Redundanzgründen sowie der unhandlichen Bedienung einer mobilen Version der Webseite in dieser Arbeit auf ihre nähere Betrachtung verzichtet. Es wird also bei der für ein sozio-technisches System wichtigen Betrachtung der technischen Komponente das Hauptaugenmerk auf die APP und die vorhandene Kommunikation mit der Serverinfrastruktur gelegt, die UBER selbst betreibt. \\
Das Startup Unternehmen wurde durch mehrere Finanzierungsrunden zwischen 2009 und 2014 finanziert und erweitert, währenddessen  über 5 Milliarden US\$ eingesammelt wurden und das Unternehmen in seiner Struktur streng den Kalkulationen der so genannten ‚Shareholder‘, zu Deutsch den Anteilseignern, folgt, deren Interessen natürlich eher profitorientiert sind. Dementsprechend reagiert UBER bei der Fahrpreisberechnung auf Angebot und Nachfrage, um so den Verdienst zu optimieren. Darüber hinaus  verbleiben  20\% der Fahrkosten direkt bei UBER und werden nicht an den Fahrer ausgezahlt \cite{Uber}.\\
Die Anpassung der angebotenen Dienste von UBER hinsichtlich der aktuellen Gesetzeslage in Deutschland gab es rechtliche Probleme, was im späteren Verlauf dieses Kapitel  noch genauer erläutert wird. 
Abschließend sei noch angemerkt, dass hier selbstverständlich keine Technikfolgenabschätzung eines neuen sozio-technisches Systems verfolgt wird, da das Unternehmen UBER ja bereits in sein Umfeld integriert wurde und dort seit einigen Jahren interagiert. \\
\newpage
\subsection{Datensammlung über das sozio-technische System zur Erstellung eines Variablensatzes}\label{kapitel5.3}
Vester sieht in seinem Sensitivitätsmodell als nächste Schritte die „Erfassung der Einflussgrößen“ sowie deren „Prüfung auf Systemrelevanz“ \cite[vgl. S.163]{KunstVester} vor, die als Zwischenziele auch in vorliegender Analyse für die Erstellung eines aussagekräftigen Variablensatzes verfolgt werden. 
Die gesammelten Daten werden hier auf ihre Systemrelevanz überprüft, einfache  Rahmendaten über die Größe und Ziele zunächst einmal präzisiert, um das wirkliche Streben im später aufgebauten Vester’schen Wirkungsgefüge des gesamten System  zu erkennen.  Zusätzlich sollten alle involvierten menschlichen Akteure mit ihrer Rolle im System sowie ihren das System beeinflussenden Faktoren in diese Datensammlung aufgenommen werden. \\
Hier wird noch von jeglicher Wertung oder Analyse Abstand genommen, da es sich zu diesem frühen Zeitpunkt zunächst um eine reine Ermittlung der Variablen und Konstanten des Systems handelt. Es gilt ein erstes Grundgerüst vom System zu erstellen und sämtliche gesammelten und erfassten Informationen vorläufig zu strukturieren und aufzulisten, damit in den folgenden Arbeitsschritten darauf zurückgegriffen werden kann. \\\\
Das Ziel dieses Arbeitsschritts ist die Erstellung eines aussagekräftigen Variablensatzes,  sodass zuerst der Begriff einer „Variable“ für diese Arbeit definiert werden muss. \\
Jede Variable wird für die Ermittlung der Einflussgrößen des Systems durch einen Variablennamen betitelt, welcher als Kurzbegriff für eine Systemkomponente fungiert. „Deshalb gehört zu jeder Variable eine Beschreibung der Indikatoren, mit denen sie näher bestimmt wird“ \cite[S.184]{KunstVester}.  Durch eine Betrachtung der Indikatoren ist es möglich, den genaueren Charakter einer Variablen zu verstehen. 
Da für die spätere Betrachtung eine Quantifizierbarkeit wichtig ist, sollten qualitative Variablen, wie zum Beispiel „Interesse von Shareholdern“ zumindest in ihren Indikatoren numerisch quantifizierbar sein. Eine Einflussgröße, welche sich nicht in einem messbaren numerischen Maß ausdrücken lässt, würde abstrakte Erklärungsansätze sowie Verhältnismäßigkeiten nicht anwendbar machen lassen und somit der gesamten Anwendbarkeit des geschaffenen Modells im Wege stehen. Auch wenn nicht numerische Faktoren sehr wohl  einen Einflussbereich besitzen, wird in dieser Arbeit aus den oben genannten Gründen davon Abstand genommen. \\
Manche Einflussfaktoren haben durch ihren Charakter eine qualitative Ausrichtung, welcher sich dann auch im Variablennamen wie zum Beispiel „Druck von Taxiunternehmen gegen UBER“ widerspiegelt. Daher ist darauf hinzuweisen, dass diese in der folgenden Betrachtung variabel sind, sich also auch in der  Bewertung umkehren lassen. Also die ihr zugeschrieben quantifizierbaren Indikatoren eine Auswertung in den positiven sowie negativen Bereich zulässt wodurch in diesem Beispiel auch eine Unterstützung von Taxiunternehmen in der Variable  „Druck von Taxiunternehmen gegen UBER“ abbildbar ist.  \\
Weiterhin ist es für die spätere Betrachtung obligatorisch, dass die Namen, welche beim Aufstellen der Variablen erarbeitet werden, eine Bewegungsrichtung innehaben, damit spätere Aussagen über sie: ‚nimmt zu‘ oder ‚nimmt ab‘ eindeutig sind. Vester bringt hier als Negativbeispiel den Variablennamen „Management“ \cite[S.186]{KunstVester}, welcher keine sinnvolle Aussage über Zu- oder Abnahme zulässt. Sinnvoller wäre es hier auf das System einzugehen und sie z.B. „Effizienz des Managements“ \cite[S.186]{KunstVester} zu nennen. \\
Entgegen Vesters Sensitivitätsmodell ist es möglich, wenn nicht sogar für die Betrachtung sozio-technischer Systeme notwendig, auch quantitative Variablennamen zuzulassen, sollten diese eine elementare Einflussgröße darstellen. Vesters Argument: „Dies würde ja eine Konstante vortäuschen“ \cite[S.186]{KunstVester} wird durch folgendes Beispiel widerlegt. Die später erörterte Variable „Preis pro Kilometer“ (Abbildung \ref{fig:var_1}) baut eine direkte Quantität auf, welche durch die in der Beschreibung genannten Indikatoren „Angebot“ und „Nachfrage“ variabel gehalten wird und damit Bestandteil des Systems ist. \\
Da Vester für den so genannten Variablenraum, also die Menge aller systemrelevanten Variablen, eine Obergrenze von 20 Variablen aufstellt, ist es bei der Betrachtung komplexer sozio-technischer Systeme erforderlich, eine Hierarchisierung der Variablen vorzunehmen, was eine Integration von komplexen Systemen unter einem Variablennamen zulässt. Am Beispiel des oben genannten „Preis pro Kilometer“ ist zu erkennen, dass die Indikatoren „Angebot“ sowie „Nachfrage“ zum einen eigene Systeme darstellen, welche sich auftrennen ließen, andererseits jedoch ihre Kernorientierung und Einflussnahme in der Bestimmung des Kilometerpreises der UBER-APP und dem entsprechenden Algorithmus liegt, der diese zwei Einflussgrößen zur Berechnung heranzieht. Da diese Subsysteme meist jedoch auch mit anderen Subsystemen oder Variablen korrelieren, wird hier zwar versucht eine Dopplung zu vermeiden, was allerdings manchmal nicht zu umgehen ist, wie die Überschneidung des Indikators ‚Angebot‘ mit der Variable ‚Anteil der UBER-Fahrer am Verkehr‘ verdeutlicht. \\
Es empfiehlt sich nicht nur im Hinblick auf die Fragestellung, ob es sich beim Unternehmen ‚UBER‘ um ein sozio-technisches System handelt, die Datensammlung für die sozialen und technischen Aspekte des Systems ‚UBER‘ voneinander  getrennt vorzunehmen. Um den initialen Variablenraum  auszuarbeiten, der für den späteren Aufbau des Vester’schen Wirkungsgefüges von entscheidender Bedeutung ist, werden also zuerst die sozialen Aspekte des Systems um ‚UBER‘ herangezogen (Kapitel \ref{kapitel5.4}), im Anschluss  dann die technischen erarbeitet (Kapitel \ref{kapitel5.5}).
\newpage 
\subsection{Soziale Aspekte des Systems}\label{kapitel5.4}
Grundsätzlich gilt es nach Vesters Methode, in den verschiedenen Arbeitsschritten mit allen beteiligten Personen in direkter Kommunikation alle für das System empfindlichen Anknüpfpunkte zu finden, um eine Erfassung sämtlicher Einflussgrößen bestmöglich zu gewährleisten. Dieser Prozess lässt sich für die Analyse der sozialen Aspekte in verschiedene Schritte unterteilen, welche sich wie folgt darstellen: Zu Beginn wird eine ausführliche Zielbeschreibung des Gesamtsystems in Teilziele und -gebiete aufgestellt und das entstehende Grobkonzept nach Schlüsselfaktoren durchleuchtet. Eine genaue Befassung mit den Zielen des Systems ist für das spätere Modell sehr wichtig, um seine Anwendbarkeit, Verhaltensweise und Reaktionen besser abbilden zu können. Da eine Aufstellung aller Komponenten mit ihren spezifischen Charakteristika den Rahmen dieser Arbeit sprengen würde, werden hier nur exemplarisch einige Komponenten und Akteure im System UBER aufgegriffen und dargelegt.\\
Auch der vorgesehene direkte Austausch mit alle Beteiligten entfällt aus verständlichen Gründen.
Für UBER als durch Shareholder finanziertes Start-up Unternehmen steht an oberster Stelle klar der Profit und die Ausnutzung aller gewinnversprechenden Aspekte der innovativen Geschäftsidee, woraus sich sowohl Wachstum des Unternehmens als auch Bekanntheitsgrad als Teilziele und Messgröße ableiten lassen.  
Um die Ziele weiter zu bestimmen und schlussendlich die beteiligten Einflussgrößen zu erkennen, ist es sinnvoll, sich bereits jetzt mit den beteiligten Akteuren im System zu befassen, da diese maßgeblich den Charakter des zu analysierenden Systems abbilden und formen. Als Akteure treten in sozio-technischen Systemen, wie bereits in Kapitel \ref{kapitel3} beschrieben alle Menschen, Organisationen, Firmen und sonstigen Akteure auf, die direkt oder indirekt das Verhalten des Systems beeinflussen. \\
Als menschliche Akteure gibt unter anderem den aktiven Kunden, der seinen Beförderungswunsch über die App auf seinem Smartphone kommuniziert und somit den Dienst in Anspruch nimmt und zum anderen den UBER-Fahrer, der über die App seine „freien Plätze“ anbietet. Natürlich agiert ‚UBER‘ als Unternehmen in diesem Zusammenhang selbst als Akteur in seinem eigenen System und darf nicht vergessen werden. Auch wenn seine „private“ Zielsetzung mit der des Gesamtsystems übereinstimmt, tritt UBER hier als Akteur mit eigenen Interessen und Impulsen auf.\\
Betrachtet werden nun Zielsetzung und Handlungsorientierung der einzelnen Akteure, um daraus auf einen initialen Einflussgrößensatz zu schließen. \\
Beginnend beim Kunden zeigt sich zunächst der finanzielle Aspekt als ausschlaggebend,  woraus sich zwei Schlüsselfaktoren schließen lassen. Zum einen der Preis für die gewünschte Fahrt, der sich als „Preis pro Kilometer“ abbilden lässt. Auf der anderen Seite steht das Interesse des Kunden, seine geplante Fahrt über UBER zu buchen. Um die oben geforderte Quantifizierbarkeit zu erhalten, wird dieses Interesse anhand der Variable „Marktanteil UBERs am öffentlichen Nahverkehr“ gemessen, da hier jede andere Transportmöglichkeit in direkter und indirekter Konkurrenz mit UBER beachtet wird.\\
Der nächste Aspekt, der den Kunden in seiner Entscheidung, mit UBER zu fahren, beeinflussen könnte, ist die Wartezeit auf eine potentielle Mitfahrgelegenheit zu seinem Wunschziel. Diese lässt sich als  absolute Zahl der Fahrer, welche die Verdienstmöglichkeit über UBER nutzen, abbilden.  Da ein solcher Absolutwert jedoch beim genaueren Betrachten nicht in direkter Beziehung zur Wartezeit steht, wäre es für das spätere System sinnvoller mit einer Relation zu argumentieren, nämlich von UBER-Fahrern zum gesamten Verkehrsaufkommen in der entsprechenden Stadt. Konkret ausgedrückt, wirken sich 1.000 UBER-Fahrer in einer 10.000 Einwohnerstadt anders auf die Wartezeit aus als in einer Stadt mit 1.000.000 Einwohnern.\\ 
Die Interessen und Ziele des Fahrers dagegen bewegen sich auch zunächst im finanziellen Bereich, also ist auch für sie der „Preis pro Kilometer“ von Relevanz, jedoch lassen sich die Wartezeiten der Fahrer auf ein Mitfahrgesuch zum Beispiel über die Anzahl der Benutzer der App in einer entsprechenden Stadt in Verbindung mit dem „Marktanteil UBERs am öffentlichen Nahverkehr“  abbilden. Wichtig bei den App-Nutzern ist wieder eine Relation, um oben genannte Problematik direkt im Kern zu umgehen, weswegen sich dies am besten in „Anteil der Smartphone Nutzer mit UBER APP“ umsetzen lässt. \\ 
UBERs eigenes Streben als Akteur in seinem komplexen System bedarf einer genaueren Aufschlüsselung. Da UBER mit seinem Dienst bei höherem Bekanntheitsgrad\\ tendenziell bessere Umsatzzahlen erwirtschaften kann, ist es auch wichtig, diese Einflussgröße in den Pool der Variablen aufzunehmen, um später auf Veränderungen der Bekanntheit im Modell reagieren zu können und diese abzubilden.
Bei dem „Bekanntheitsgrad von UBER“ handelt es sich um eine qualitative Variablen Bezeichnung, sodass exemplarisch an dieser eine Betrachtung der einzelnen Indikatoren vorgenommen werden soll. Wichtig hierbei ist, dass die Indikatoren keine reine Umformulierung des abstrakten Variablennamen sind, sondern zur Charakterbildung, wie Vester es nennt, beitragen, also ein näheres Verständnis über die Bedeutung der Einflussgröße eröffnen. Einzelne Indikatoren, durch welche sich die Bekanntheit eines Unternehmens abbilden lässt, sind zum einen das  Verhältnis von Menschen, die UBER kennen zu denen, die den Namen noch nie gehört haben. Um aber die Variable näher an das Unternehmen UBER anzupassen, wäre ein zweiter Indikator nützlich, nämlich „Quote der Menschen, die UBER und seinen Dienst kennen“ und als letztes „Anteil der Menschen, die bereits mit UBER gefahren sind“. Aus diesen drei Indikatoren lässt dich der qualitative Wert zur Bekanntheit dem System entsprechend messen, wobei sämtliche Indikatoren numerisch quantifizierbar sind.\\
Weiterhin lässt sich das „Unternehmenswachstum“ anhand der Indikatoren „Umsatzwachstum“ und „Fremdkapitalquote“ bei einem Dienstleistungsunternehmen aussagekräftig abbilden. 
Nachdem ein initialer Satz an Variablen erstellt wurde, sollten diese eigentlich erneut diskutiert und auf ihren Variablen - Charakter bewertet werden, also ob sie für das System  veränderlich sind und diese Veränderung auch Auswirkungen im System anstößt.  Eine solche Diskussion sollte im Plenum der beteiligten Personengruppen geschehen und ist deshalb in dieser Arbeit nicht abbildbar. \\
 \begin{figure}[htb]
 \centering
 \includegraphics[width=\textwidth,angle=0]{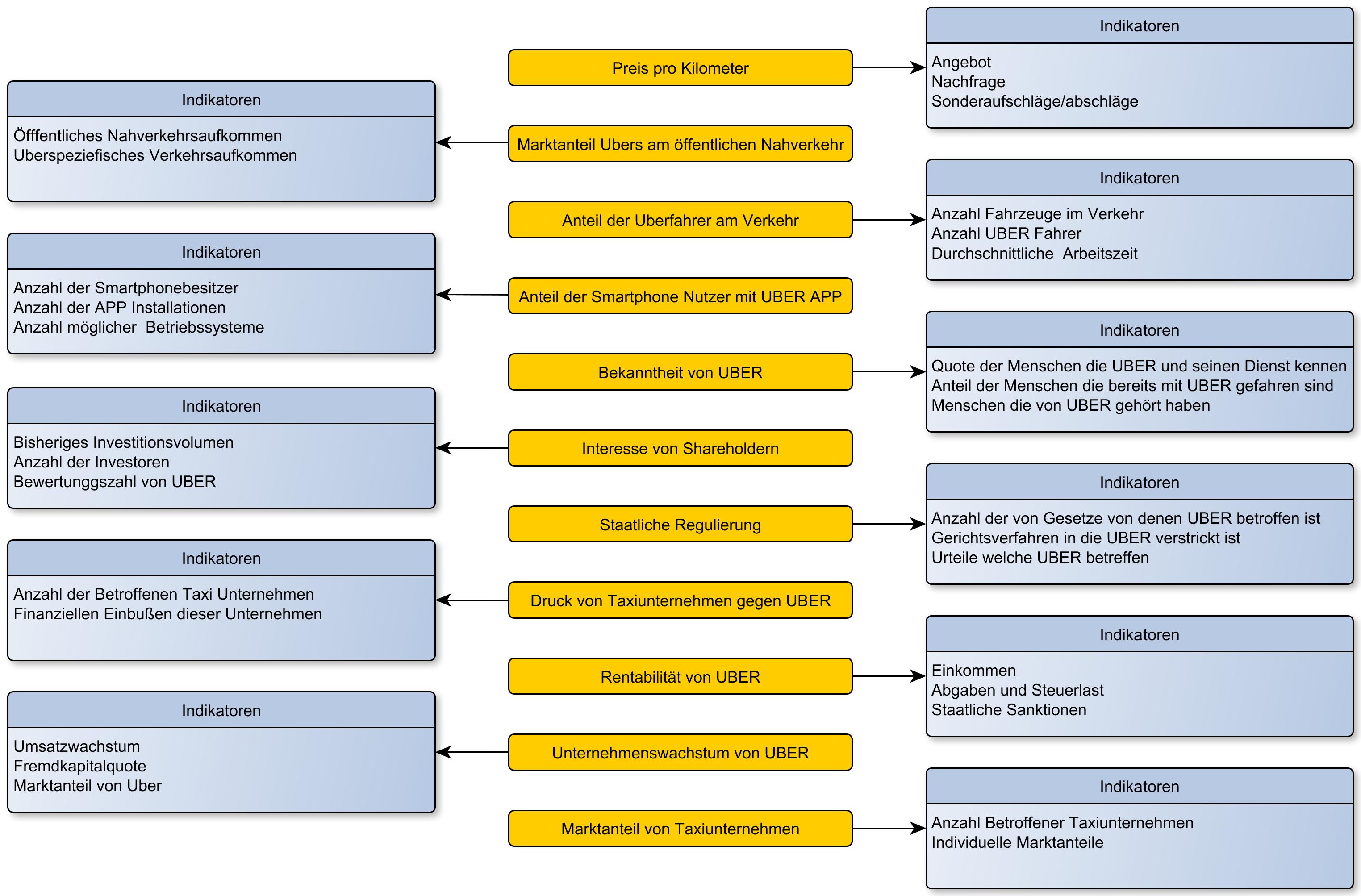}
 \caption[Variablensatz nach Betrachtung der sozialen Komponenten]{Variablensatz nach Betrachtung der sozialen Komponenten}
\label{fig:var_1}
\end{figure}
Als Ergebnis resultiert ein Variablensatz (Abbildung \ref{fig:var_1}), der alle bisher betrachteten Einflussgrößen enthalten sollte und bereits für die spätere Arbeit aufbereitet wurde, also in einem Knotenformat vorliegt, der das Erstellen des Wirkungsgefüge begünstigt. Vester baut in seine Analyse auf einen ständigen Rückgriff auf vorhergehende Arbeitsschritte und Ergebnisse, weshalb Zwischenergebnisse stets abgespeichert werden sollten oder durch eine Versionsverwaltung rekonstruierbar sein müssen.  
\newpage
\subsection{Technische Aspekte des Systems}\label{kapitel5.5}
Zur Vervollständigung der begonnenen Datenerfassung werden im Folgenden die technischen Komponenten des Systems analysiert. An dieser Stelle sollte noch betont werden, dass es gerade im Hinblick auf die geplante Charakterisierung des Systems als potentielles sozio-technisches System wichtig ist, die vorliegenden technischen Aspekte genauestens  zu erfassen und ihre Auswirkungen in und um das System zu erkennen. Da ein System mehrere technische Komponenten enthalten kann, darf auch hier nicht auf eine ausführliche Dokumentation verzichtet werden, um eine Wiederverwertbarkeit der Datensammlung zu gewährleisten.\\
Für  sämtliche technische Komponenten sollten sowohl der funktionale Katalog an Reaktionen, als auch seine Verhaltensweisen sowie die enthaltenen, von Menschen geschaffenen technischen Artefakte aufgezählt werden. Wie bereits ausgeführt, sollte eine technische Komponente zwar in ihrer Verhaltensweise deterministisch agieren, trotzdem empfiehlt es sich,  während der Datenerfassung auf mögliche Unregelmäßigkeiten zu achten. Es besteht immer die Möglichkeit, dass ein Teilsystem eine bestimmte Funktionalität eigentlich erfüllen sollte, aber  durch fehlerhafte Umsetzung oder menschliches Versagen im Planungsprozess eine gewünschte Reaktion ausbleibt oder unerwartete Nebenwirkungen auftreten, die sich potentiell auf das gesamte System auswirken können und es deshalb zu erfassen  gilt. 
Auf der technischen Ebene ist eine Hierarchisierung der einzelnen Komponenten sinnvoll, um intern sowie extern angebotene Funktionalitäten voneinander abzugrenzen und einzelne Komponenten als Gesamtgefüge von Teilen  mit einem speziellen Funktionscharakter zu erkennen und für die kommende Analyse aufzubereiten.\\\\Um die technischen Komponenten im  UBER-Komplex zu analysieren, empfiehlt sich eine Betrachtung der oben skizzierten Kundensituation. UBER bietet eine Applikation (App) für verschiedene Smartphones und Betriebssysteme an, um so eine möglichst breite Masse an Kunden zu erreichen. Kundenfreundliche Bedienungsstrategien gehören hier genauso zum Produktumfang wie die angebotenen Funktionalitäten. Der Kunde kann also direkt über sein Handy eine Fahrt buchen und mit dem gewünschten Fahrer Kontakt aufnehmen. Zusätzlich zu dieser Funktion findet eine Kommunikation mit dem Server UBERs statt, welche eine zweite technische Komponente bildet. Hier werden sowohl die Accounts der Nutzer und der Fahrer mit ihren privaten Daten als auch Algorithmen zur Bestimmung des Fahrpreises gespeichert, um diesen kontinuierlich an die aktuelle Angebot und Nachfrage-Situation in der entsprechenden Stadt effektiv anzupassen. \\
Da auch hier eine ausführliche Beschreibung aller Variablen mit ihren Indikatoren den Umfang dieser Arbeit sprengen würde, wird nur auf spezielle Sonderfälle und Eigenarten des UBER-Systems eingegangen.
Als möglicherweise interessante Variable stellt sich bei den technischen Komponenten im System ‚UBER‘ die „Verbindung der APP zum Rechenzentrum“ dar. Diese hat vor allem durch den Indikator „Konnektivität in den Stadtzentren (Mobilfunknetz usw.)“ einen externen Einflussfaktor, auf den das System ansonsten kaum bis gar keinen Einfluss hat, denn die über einen Drittanbieter aufgebaute Verbindung in das Mobilfunknetz/Internet unterliegt einer Vielzahl von Faktoren, welche in diesem System nicht plan- und modellierbar sind, jedoch massiven Einfluss auf die Kommunikation mit der Serverarchitektur haben. Deshalb ist die Untersuchung der Auswirkungen einer Veränderung dieser Variable von großem Interesse für das gesamte System. \\

\begin{figure}[htb]
 \centering
 \includegraphics[width=\textwidth,angle=0]{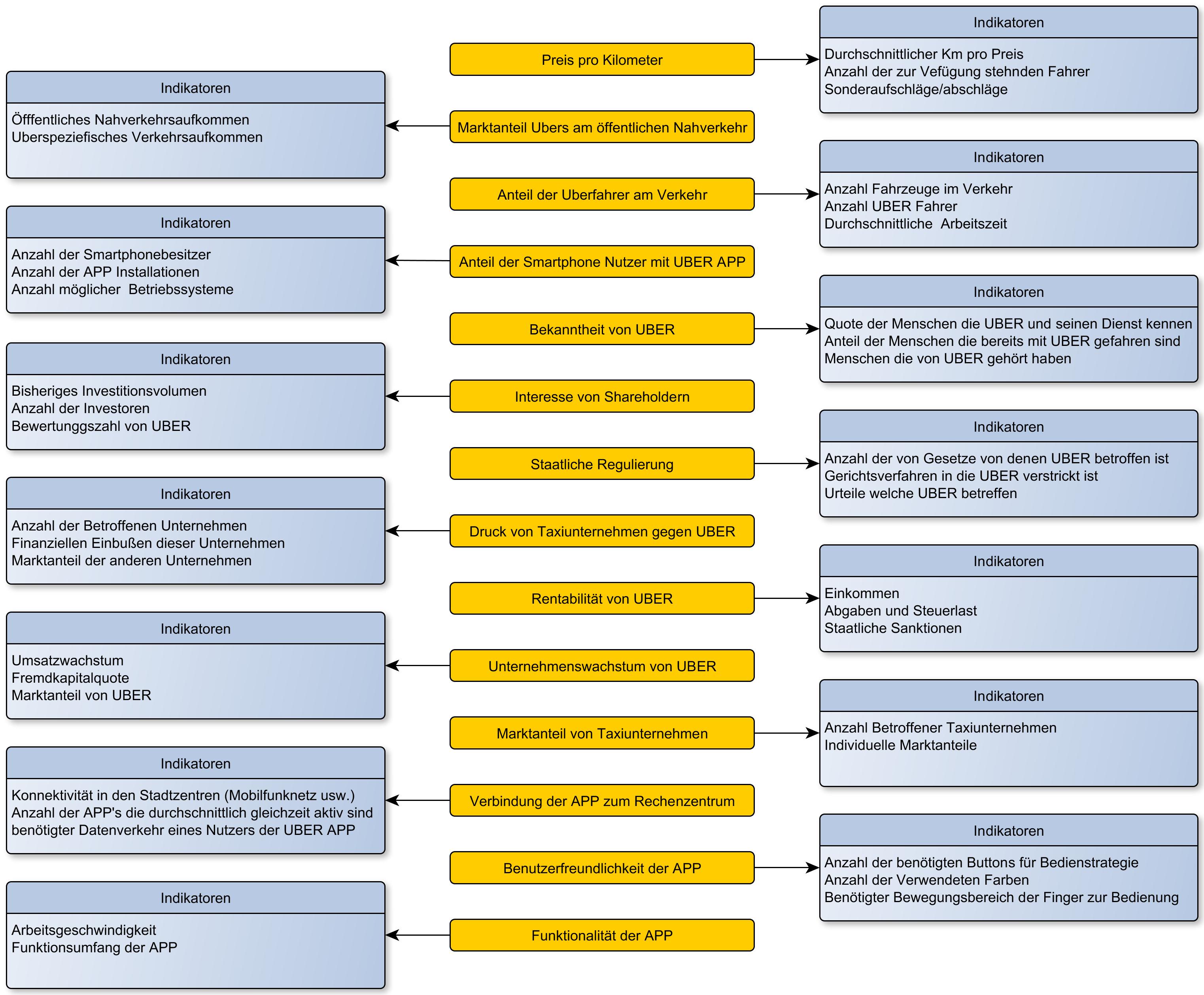}
 \caption[Variablensatz nach Betrachtung der technischen Komponenten]{Variablensatz nach Betrachtung der technischen Komponenten}
\label{fig:var_2}
\end{figure}
Auch zum Abschluss dieser Betrachtung gilt es den erarbeiteten Variablensatz (Abbildung \ref{fig:var_2}) durch eine Diskussion zu validieren. Das Hauptaugenmerk könnte  neben den bereits genannten Kriterien auf den möglichen Interaktionen zwischen den technisch orientierten Variablen und den vorher erarbeiteten liegen, welche sich noch nicht  abbilden lassen. Die Ergänzung von neu gefundenen Variablen ist zu jedem Zeitpunkt ausdrücklich gewünscht, da der Einflussgrößensatz interaktiv erstellt und bearbeitet wird. 

\newpage
\subsection{Analyse möglicher innovativer Ausprägungen des sozio-technischen Systems}\label{kapitel5.6} 
Im Rahmen des ‚Collingridge Dilemma‘ (vgl. Kapitel \ref{kapitel2.3}) schon aufgezeigt, erzeugen innovative technische Neuerungen bezüglich ihrer Folgen sehr häufig eine Vielzahl von bisher unbekannten und potentiell ungewollten Nebenwirkungen. Daher ist es für eine  Technikfolgenabschätzung wichtig, eine genaue Einordnung der innovativen technischen Komponenten und Ideen in den aktuellen Forschungsstand zu erarbeiten. Auch wenn auf den ersten Blick eine solche Innovation in ihren Grundzügen einen großen Nutzen für das Ziel des zu analysierenden Systems bietet, so können nur durch eine genauere Betrachtung der individuellen Auswirkungen auf das Umfeld des Systems und seiner Interaktionen in der späteren Analyse  Aussagen über sein potentielles Verhalten und seinen Charakter getroffen werden. Der objektive Blick auf etwaige technische Neuerungen ist eine nützliche Vorarbeit zur eigentlichen Analyse, die eine Vielzahl von Bezügen zur aktuellen Realität eröffnen kann. \\
Dieser Teilschritt bringt vor allem auch Informationen über Systeme, welche bereits in das lebendige Umfeld integriert wurden, und möglicherweise Aufschluss über daraus entstandene Probleme. Dabei ist es vernachlässigbar, aus diesen potentiell ähnlichen Systemen, Ideen für die Voraussagen über das zu analysierende System zu treffen, sondern es nur zusätzlich unter den erkannten Problemstellungen genauer zu analysieren, um aus den vorhandenen Fehlern und Fehlschlägen zu profitieren. \\
Bei der Anwendung dieses Analyseschritts auf das Unternehmen ‚UBER‘ stößt man unweigerlich auf das Konzept der ‚Share-Economy‘, das der innovativen Geschäftsidee des UBER-Unternehmens zugrunde liegt und daher an dieser Stelle kurz erläutert werden soll.\\
Der Begriff ‚Sharing Economy‘ dreht sich ursprünglich um den Gedanken ‚Ökonomisches Teilen‘ und muss vom Prinzip der Vermietung abgegrenzt werden, bei dem der Kauf eines Produkts durch das Ziel zu verleihen motiviert wird.\\
 Die sich daraus entwickelten Geschäftskonzepte zeichnen sich durch gemeinsame zeitliche begrenzte Nutzung von Ressourcen aus, die nicht dauerhaft benötigt werden. Dieser Gemeinschaftskonsum bedarf einer ausgeklügelten Systematik, die organisiert werden will, um die Auslastung der zeitlich begrenzten Ressource zu optimieren. Auch wird ein ausgeprägtes interdisziplinäres Verständnis benötigt, um das Sharing - Konzept effektiv und vor allem wirtschaftlich zu verbreiten, was die Mitfahrzentralen der verschiedenen deutschen Großstädte aber schon seit Jahren, wenn auch im Einzelnen mit unterschiedlichem Erfolg und Ausmaß praktizieren. \\

Durch die zunehmende Vernetztheit, wodurch der Kreis der erreichbaren Menschen sich deutlich erhöht, ermöglichen Geschäftsmodelle, die auf dem ‚Sharing Economy‘-Gedanken basieren, prinzipiell in unserer Zeit einen effizienteren Umgang mit raren Ressourcen, wobei natürlich auch der finanzielle Anreiz eine große Rolle spielt. Auf den stetig zunehmenden Bevölkerungsteil des so genannten Prekariats, das durch fehlende oder unsichere  Erwerbsfähigkeit von Armut bedroht ist, wirkt das ‚Sharing Economy‘- Konzept besonders attraktiv, nicht nur wegen des ‚Sharings‘, sondern natürlich  auch wegen des damit verbundenen finanziellen Aspekts, was UBER mit seinem Geschäftsmodell geschickt ausnutzt. \\
Der Grundgedanke einer Mitfahrzentrale ist die Vermittlung von Fahrgemeinschaften bzw. Mitfahrgelegenheiten. Ein Fahrer füllt die sonst nicht genutzten Sitze seines Wagens und erhöht somit nicht nur dessen Effizienz, sondern erhält noch eine Beteiligung an den entstandenen Fahrtkosten von den Mitfahrern, die dasselbe Reiseziel oder die selbe (Teil)strecke fahren wie er haben.\\
Da UberPop beförderungswillige Kunden an private Fahrer mit ihren PKW vermittelt, beruft sich das Unternehmen auf das Konzept der Mitfahrzentralen, verschweigt jedoch den Unterschied, dass der UBER-Fahrer ohne seinen Kunden die jeweilige Strecke natürlich nicht fahren würde und somit dieses Geschäftsmodell wenig mit dem ‚Sharing Economy‘- Gedanken gemeinsam hat. Angesichts der massiven Proteste der deutschen Taxifahrer versuchte UBER zwar dann sein Vorgehen dahingehend zu begründen, dass es ja auch durch die Vermittlung der Ressource ‚Auto‘ den beschrieben ‚Sharing Economy- Gedanken‘ aktiv vertrete, dennoch wurde 2014 der Klage von Taxi Deutschland gegen UBER vom Landgericht Frankfurt am Main  schließlich stattgegeben. Auf die rechtlichen Probleme, die sich für UBER in Deutschland ergaben, wird später (Kapitel \ref{kapitel5.8}) noch genauer eingegangen. \\\\
Nachdem man nach eingehender Prüfung der Systemrelevanz aller Variablen einen aussagekräftigen, überschaubaren Variablensatz erstellt hat, schließen sich in Vesters Sensitivitätsmodell zwei Arbeitsschritte an, die sich mit der Beurteilung der Wirkungsbeziehungen befassen: „Hinterfragung der Wechselwirkungen“ \cite[S.192]{KunstVester} sowie „Untersuchung der Rollen im System“ \cite[S.192]{KunstVester}. Da die von Vester zugrunde gelegten Konzepte für das Erstellen einer Einflussmatrix sowie für die entsprechende Methodik zur Auswertung den Rahmen dieser Arbeit sprengen würde, wird auf eine Durchführung verzichtet.
\newpage
\subsection{Das Unternehmen ‚UBER‘ als sozio-technisches System}\label{kapitel5.7}
Nach der  Definition von Kienle/Kunau lässt sich das Unternehmen ‚UBER‘ eindeutig  als sozio-technisches System charakterisieren. Wie bereits ausführlich dargelegt (vgl. Kapitel \ref{kapitel3})  muss ein sozio-technisches System drei Eigenschaften besitzen: \\
\begin{enumerate}
\item Ein soziales System nutzt zur Unterstützung seiner Kommunikationsprozesse ein technisches System.
\item Beide Systeme prägen sich wechselseitig.
\item Das technische System findet Eingang in die Selbstbeschreibung des sozialen Systems.
\end{enumerate}
Wie im Folgenden nachgewiesen wird, erfüllt das System ‚UBER‘ alle drei der an ein sozio-technisches System gestellten Forderungen:\\\\
\textbf{Zu 1:}\\
Anbieter und Nutzer von UBER als die zwei Hauptkomponenten des sozialen Systems gehen  nicht nur „eine besondere Beziehung zu einem technischen System“ \cite[S.96]{Kienle} ein, sondern im Grunde gleich zu zwei Systemen.  Einerseits bildet das technische System ‚Auto‘ den Ausgangspunkt der beiderseitigen Kommunikation nach dem Modell ‚Angebot und Nachfrage‘, andererseits wird die Kommunikation zwischen den sozialen Akteuren durch den Gebrauch eines Smartphones bzw. der App zur Vermittlung gewährleistet. Das soziale System ist also auf die Nutzung dieses technischen Systems für den Kommunikationsprozess angewiesen.\\\\
\textbf{Zu 2:}\\
Die wechselseitige Prägung der beiden Systeme lässt sich an den Konflikten der Geschäftsleitung UBERs als soziales System mit der deutschen Justiz, dem übergeordneten sozialen System, und den daraus jeweils resultierenden Änderungen der UBER-APP, dem technischen System sogar mehrfach nachweisen: Der starke Protest \cite{Focus}  der Taxifahrer gegen UBER beruhte auf einer Vielzahl von Gründen, von denen jedoch zwei wichtige kurz zu erläutern sind. Zum einen brauchten potentielle Fahrer für UBER keinen Führerschein zur Fahrgastbeförderung \cite{Pop} nach dem  Personenbeförderungsgesetz (PBefG) , den so genannten Personenbeförderungsschein, zum anderen waren die Fahrten „durchschnittlich 20 \%“ \cite{Faz} günstiger als bei lizenzierten Taxiunternehmen. Daher klagte Taxi Deutschland, ein Zusammenschluss von Taxizentralen mehrere Großstädte  vor dem Landgericht Frankfurt gegen UBER wegen § 49 PBefG (Personenbeförderung ohne entsprechende Genehmigung), worauf am 18.3.2015 der Klage gegen UBER stattgegeben wurde \cite{Gericht} seinen Dienst, d.h. sowohl das Geschäftsmodell als auch die technischen Aspekte, wie zum Beispiel die APP zu  ändern, was oben aufgeführte Prägung des Unternehmens UBER belegt.\\
Ein weiteres Beispiel für eine gegenseitige Prägung  zeigt sich  im Einfluss der Anzahl der jeweils zur Verfügung stehenden Serverstruktur als technisches System auf die Zahl der potentiellen Nutzer, also das soziale System. Umgekehrt gestaltet dagegen die Nachfrage des sozialen Systems, also die Zahl der Kunden welche Fahrten buchen, den Umfang des technischen Systems, also die Zahl der UBER-Autos. \\\\
\textbf{Zu 3.}\\
Durch die Werbung UBERs sowie die angepasste Verhaltensweise der Menschen, die UBER in ihre Arbeitswege \cite{Newsroom} oder Abendplanungen integrieren, findet zudem das technische System Eingang in die Selbstbeschreibung des sozialen Systems, sodass eine sozio-technische Selbstbeschreibung entsteht und somit die letzte geforderte Eigenschaft nach der Definition von Kienle/Kunau erfüllt ist.

\newpage
\subsection{Wirkungsgefüge}\label{kapitel5.8}
Ein „Wirkungsgefüge“ ist nach Vester Ziel und zentrales Element der Systemanalyse zur „Visualisierung der Systemvernetzung (…), die dem Verständnis der speziellen Kybernetik des Systems, das heißt der Interpretation seiner Wirkungsketten und Rückkopplungen dient“ \cite[S.240]{KunstVester}. \\ 
Zunächst gilt es die Struktur zu verstehen, nach welcher Vester seine Wirkungsgefüge aufbaut und darstellt.  Ihm geht es  hier weniger darum, auf das exakte Ausmaß der Auswirkungen einer bestimmten Veränderung einzugehen, als vielmehr die Abhängigkeiten der einzelnen Variablen in ihrer Grundausrichtung und Interaktion miteinander zu visualisieren.\\\\
\begin{figure}[htb]
 \centering
 \includegraphics[width=0.9\textwidth,angle=0]{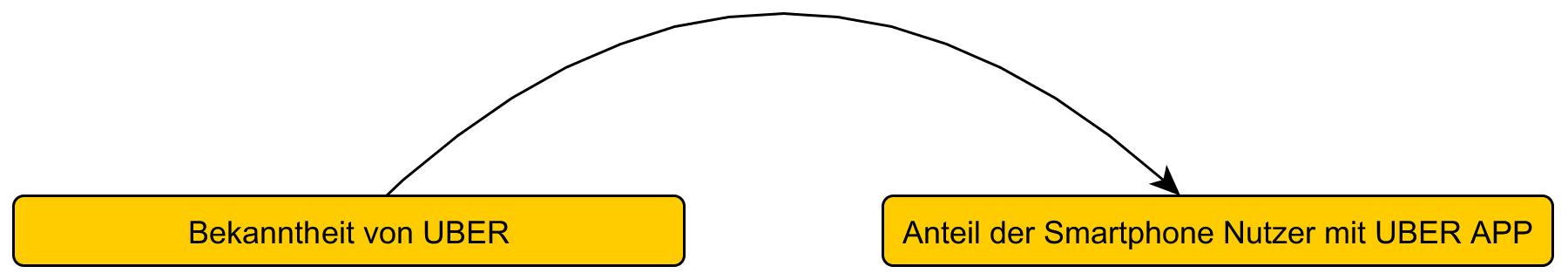}
 \caption[Beispiel gleichgerichte Beziehung]{Beispiel gleichgerichte Beziehung}
\label{fig:Gleichgerichtet}
\end{figure}
Einflüsse von Variablen, repräsentiert durch Knoten untereinander, werden durch Kanten zwischen ihnen dargestellt. Von welcher Natur dieser Einfluss ist, wird durch die Visualisierung der Kante ausgedrückt. Ein durchgezogener Pfeil (Abbildung \ref{fig:Gleichgerichtet})\\
 von der Variable „Bekanntheit von UBER“ zu „Anteil der Smartphone Nutzer mit UBER APP“ steht für eine „gleichgerichtete Beziehung“ \cite[S.211]{KunstVester}. Je größer der Anteil der Smartphone Nutzer mit UBER APP wird, desto größer wird der Bekanntheitsgrad des Unternehmens UBER. \\
 Ebenso hat ein negativer Trend der Variable, aus welcher der Pfeil entspringt, eine direkte negative Auswirkung auf die entsprechend andere Variable mit der eingehenden Kante. \\
 \begin{figure}[htb]
 \centering
 \includegraphics[width=0.9\textwidth,angle=0]{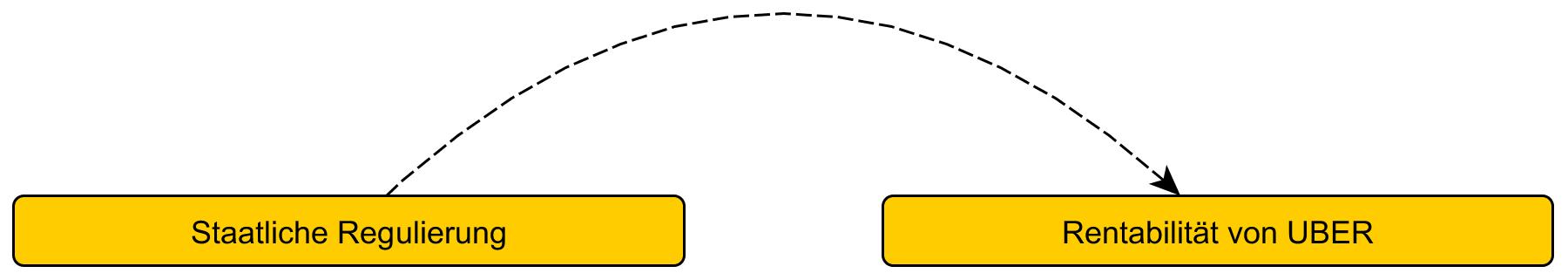}
 \caption[Beispiel gegengerichtete Beziehung]{Beispiel gegengerichtete Beziehung}
\label{fig:Gegengerichtet}
\end{figure}
Umgekehrt verhalten sich Beziehungen, die durch einen unterbrochenen Pfeil (Abbildung \ref{fig:Gegengerichtet}) gekennzeichnet sind. Hier verhält sich die Variable „Rentabilität von UBER“ antiproportional zur Entwicklung von „Staatliche Regulierung“, denn je größer die staatliche Kontrolle über UBER ausgeprägt ist, desto höher sind die Indikatoren „Abgaben und Steuerlast“ sowie „Staatliche Sanktionen“ und entsprechend sinkt die Variable „Rentabilität von UBER“. Eine solche Beziehung bezeichnet Vester als gegenläufig.\\
Wichtig ist, dass bei Vesters grafischen Darstellungen Pfeile nicht bidirektional sind, also eine eindeutige Wirkungsrichtung haben, nach denen sie interpretiert werden müssen. Da für die Rückrichtung andere Bedingungen und Verhältnisse auftreten können, sind hier ein paar Spezialfälle zu betrachten. \\

Die im letzten Kapitel beschriebenen Rückkopplungen lassen sich in diesen Wirkungsgefügen sehr einfach durch Zyklen erkennen. Ein Zyklus oder Kreis ist in der Graphentheorie ein Weg von Kanten in einem Graphen, bei dem der Endknoten mit dem Startknoten übereinstimmt.\\
Finden sich in einem solchen Zyklus nur  Beziehungen vom gleichen Typ, so spricht Vester von einem positivem Rückkopplungszyklus \cite[vgl. 211f]{KunstVester}. Dieser kann in zwei Ausprägungen vorliegen. Zum einen wenn alle Beziehungen gleichgerichtet sind, wie in diesem Beispiel
\begin{figure}[htb]
 \centering
 \includegraphics[width=0.9\textwidth,angle=0]{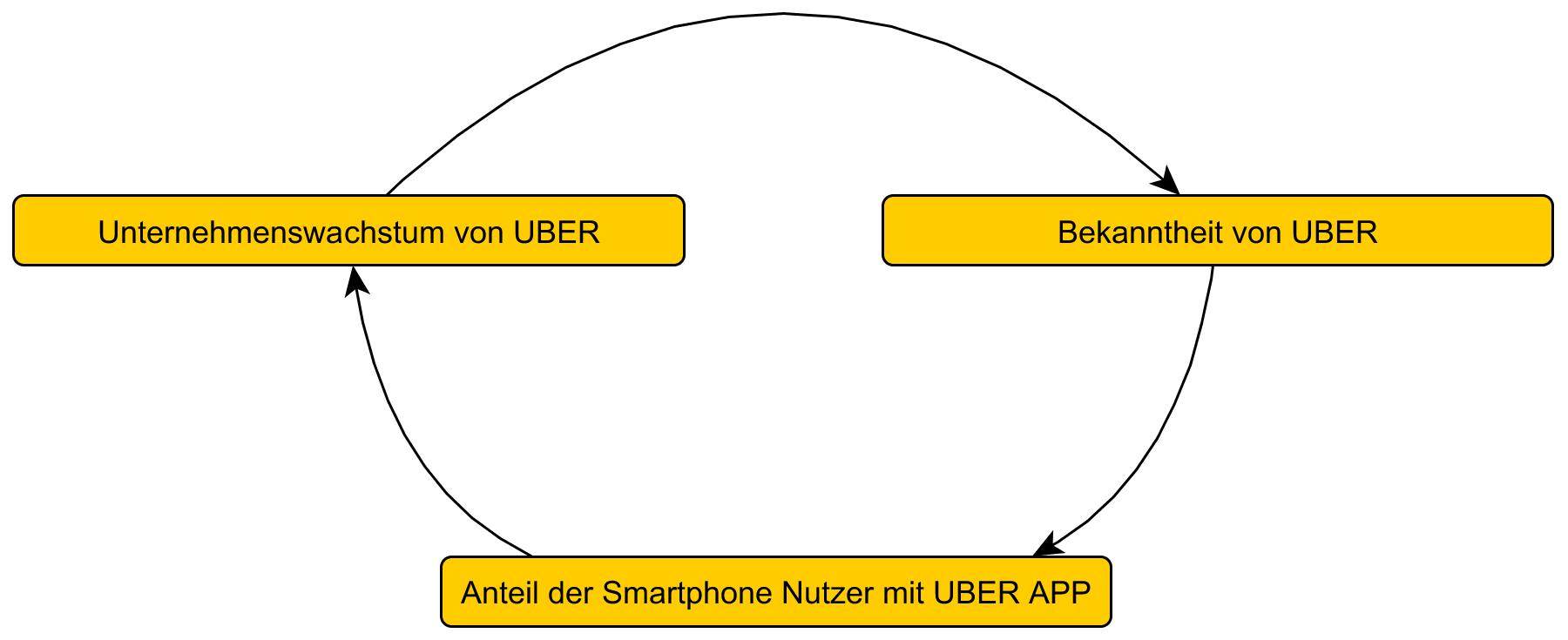}
 \caption[Beispiel positive Rückkopplung aus gleichgerichteten Beziehungen]{Beispiel positive Rückkopplung aus gleichgerichteten Beziehungen}
\label{fig:Pos_1_rk}
\end{figure}

 (Abbildung \ref{fig:Pos_1_rk}), dann steigert eine  zunehmende  „Bekanntheit von UBER“ direkt den „Anteil der Smartphone Nutzer mit UBER APP“, woraus sich ein erhöhtes „Unternehmenswachstum von UBER“ ergibt. Da der Dienst nun offensichtlich mehr genutzt wird, was durch eine weitere gleichgerichtete Beziehung gekennzeichnet ist. Steigt nun eben diese, erhöht sich auch die „Bekanntheit von UBER“. Somit schaukeln sich diese drei Variablen, ausgehend von einem Startimpuls entweder nach oben oder nach unten, völlig unkontrolliert auf, bis ein natürlicher Grenzwert erreicht wird, das System kollabiert oder durch eine darüber gelagerte Rückkopplung fängt diese Spirale auf, wie das Verbot von UberPOP durch das Urteil vom 18.3.2015 vom LG Frankfurt \cite{Gericht}. \\\\
 Ähnliche Probleme entstehen bei einem Zyklus von gegenläufigen Beziehungen \cite[vgl. 211f]{KunstVester}. Wie in diesem Beispiel 
\begin{figure}[htb]
 \centering
 \includegraphics[width=0.9\textwidth,angle=0]{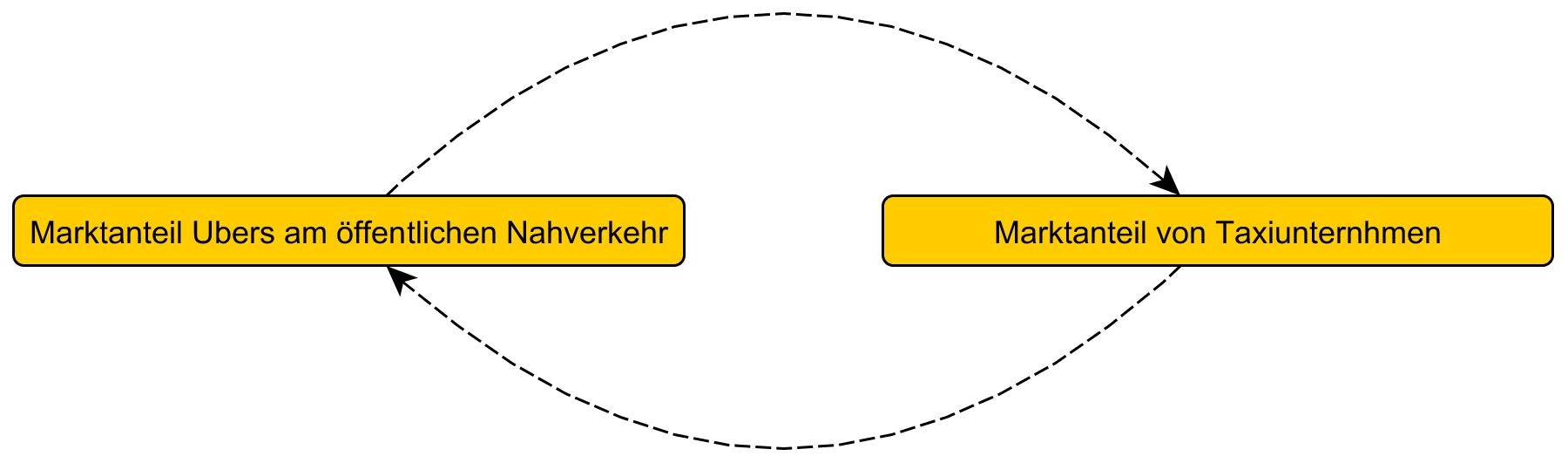}
 \caption[Beispiel positive Rückkopplung aus gegengerichteten Beziehungen]{Beispiel positive Rückkopplung aus gegengerichteten Beziehungen}
\label{fig:Pos_2_rk}
\end{figure}
(Abbildung \ref{fig:Pos_2_rk}) zu erkennen ist, kann sich ausgehend von einem Startimpuls eine dieser zwei Variablen zugunsten der anderen aufwerten. So nimmt zum Beispiel mit steigender Größe von UBER der Marktanteil der Taxiunternehmen ab, wodurch wiederum UBER Chancen und Möglichkeiten eingeräumt werden, noch mehr zu expandieren. Auch dieses Wachstum ist tendenziell nur durch einen externen Grenzwert gedeckelt und somit für das Wirkungsgefüge gefährlich. \\\\
Eine regulierende Wirkung haben dagegen, wie bereits ausgeführt, die so genannten negativen Rückkopplungszyklen, die durch einen Zyklus gekennzeichnet sind, in dem sowohl gegenläufige als auch gleichgerichtete Kanten zu finden sind \cite[vgl. 212]{KunstVester}. \\
Diese wechselseitige Einflussnahme ist am folgenden Beispiel \begin{figure}[htb]
 \centering
 \includegraphics[width=0.9\textwidth,angle=0]{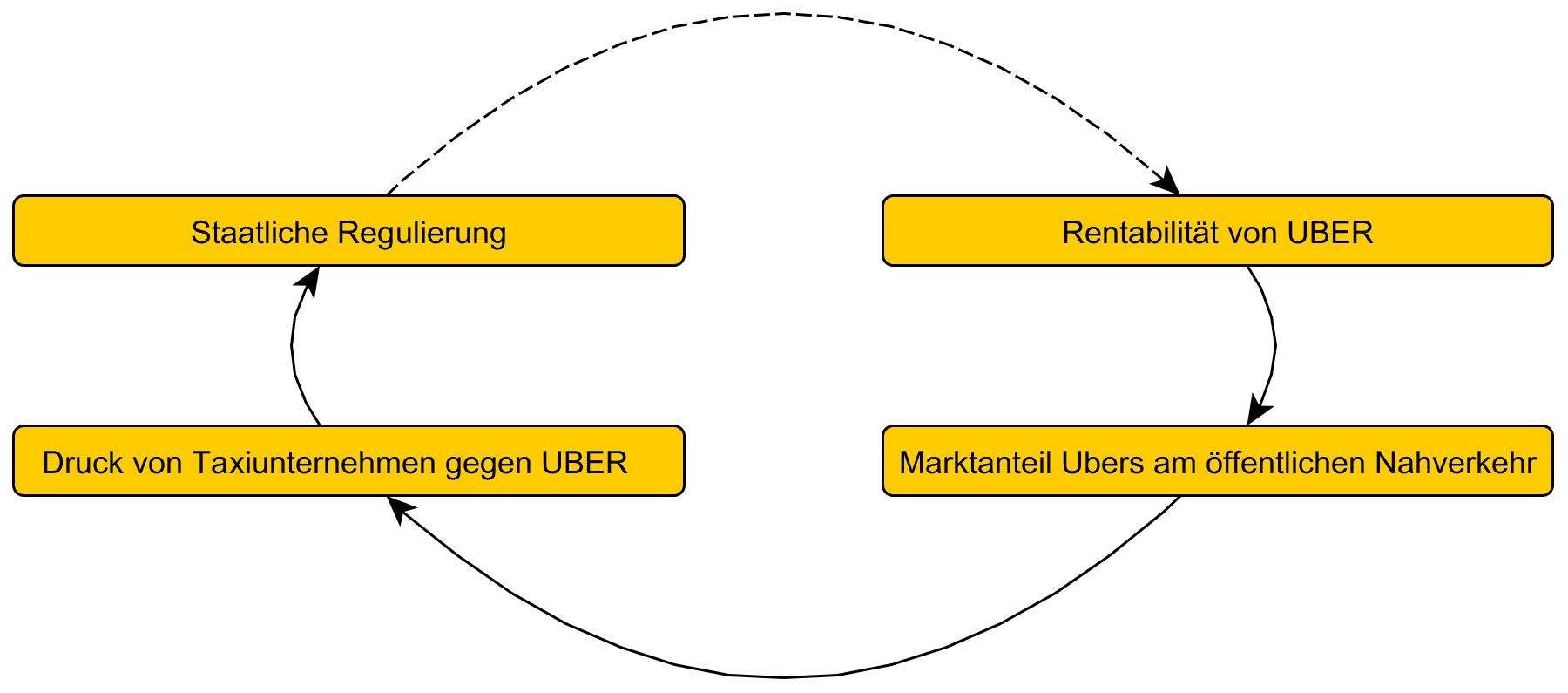}
 \caption[Beispiel negativ Rückkopplung]{Beispiel negative Rückkopplung}
\label{fig:Neg_rk}
\end{figure}
 (Abbildung \ref{fig:Neg_rk}) einfach zu veranschaulichen, denn mit steigendem Marktanteil von UBER am öffentlichen Nahverkehr steigt natürlich der Druck von Taxiunternehmen gegen UBER, welche wiederum an einer erhöhten Regulierung von staatlicher Seite interessiert sind, vor allem an einer angepassten Versteuerung der Einnahmen, die aktuell noch zu einem Großteil ins Ausland laufen \cite{Steuer}, was dann die Rentabilität sinken lässt. \\
Reziprok verhält sich dann die Auswirkung einer solchen zunehmenden Regulierung von staatlicher Seite für das Unternehmen UBER und seine Rentabilität. Denn mit eingeschränkter Handlungsfreiheit und/oder einer verschärften staatlichen Kontrolle sinken die Expansionsmöglichkeiten und das Unternehmen  vermutlich in seinem Einflussbereich und Bekanntheitsgrad. Belegen lassen sich diese Behauptungen durch die Reaktionen UBERs auf das oben schon erwähnte Urteil vom 18.3.2015 \cite{Gericht}. Anfänglich ignorierte UberPop das Verbot durch das LG Frankfurt. Als mit drastisch hohen Strafen gedroht wurde, reagierte UBER mit einer drastischen Fahrpreissenkung auf 30 Cent/km/h \cite{Heise}.\\

So klagte „Taxi Deutschland“ vor dem Landgericht Frankfurt am Main, dass ‚UBER‘ gegen das Personenbeförderungsgesetz verstoße, denn die Fahrten, welche über das UBER System vermittelt würden, seien gewerbliche Personenbeförderungen, für die ein Entgelt entrichtet werden muss, das über den reinen Betriebskosten des Fahrzeugs liegt und das allein mit der 20\% Gewinnmarge von UBER selbst überstiegen wird.\\
 Zusätzlich  besitzen die selbstständig gemeldeten Fahrer nicht zwingend einen so genannten ‚Personenbeförderungsschein‘, wenn sie den Vermittlungsdienst und somit den Kundenzustrom nutzen. Dieser wird jedoch für eine gewerbliche Personenbeförderung in Deutschland benötigt .
 Nach dem das Urteil \cite{Gericht} gegen UBER rechtskräftig wurde und UberPop aufgrund fehlender Zulassungen der Fahrer verboten wurde veränderte UBER in Deutschland ihr Angebot. Der neu entstandene Dienst UberX passte sich den gesetzlichen Regeln in Deutschland an. So versicherte der Sprecher von Uber Deutschland, „alle Partner der neuen Vermittlungsplattform besäßen entsprechende Genehmigungen nach dem Personenbeförderungsgesetz“ \cite{go} Auch wurden bei  Uberx, laut Uber, nun alle Fahrzeuge offiziell als Mietwagen zugelassen und entsprechend versichert.  \\\\ 
Hier soll noch ein von Vester nicht bedachtes Phänomen angesprochen werden, das seiner Aussage über negative Rückkopplungszyklen widerspricht. \\
\begin{figure}[htb]
 \centering
 \includegraphics[width=0.9\textwidth,angle=0]{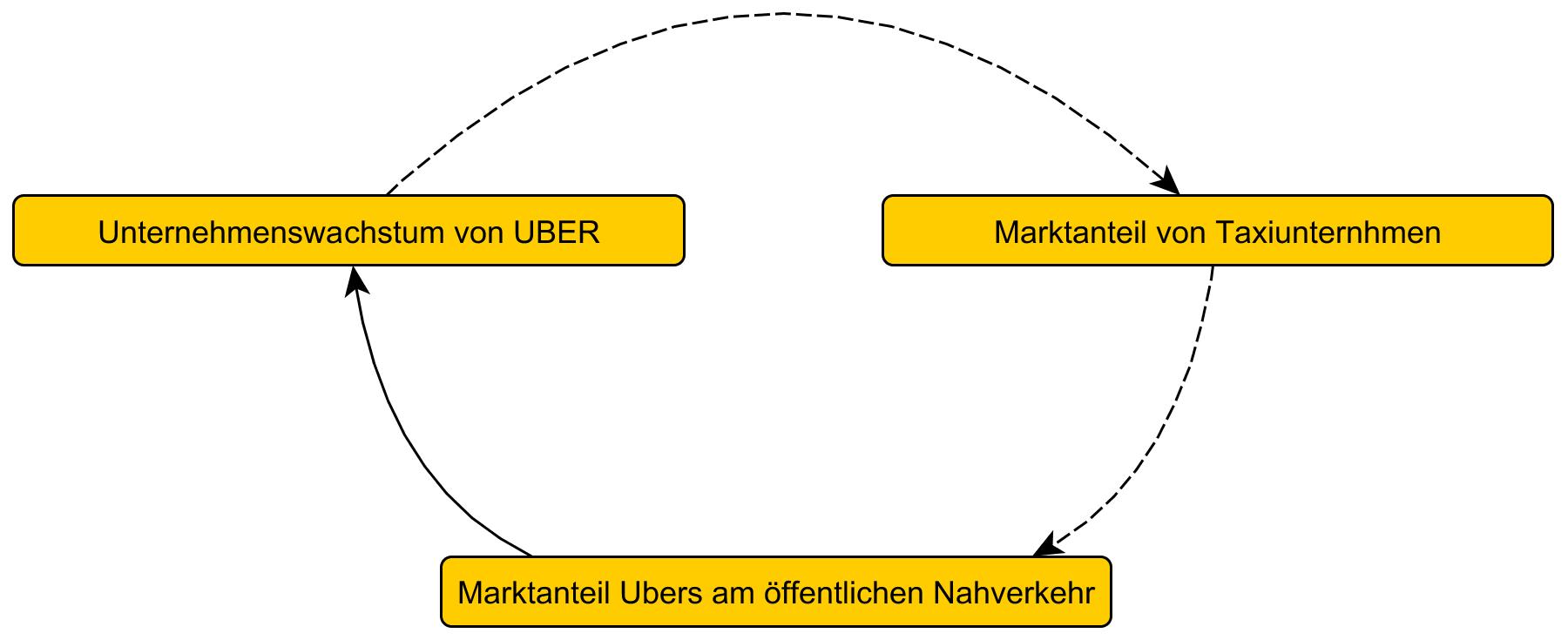}
 \caption[Beispiel negativ Rückkopplung mit Konflikt]{Beispiel negative Rückkopplung mit Konflikt}
\label{fig:konflikt_1}
\end{figure}

\begin{figure}[!h]
 \centering
 \includegraphics[width=1.6\textwidth,angle=270]{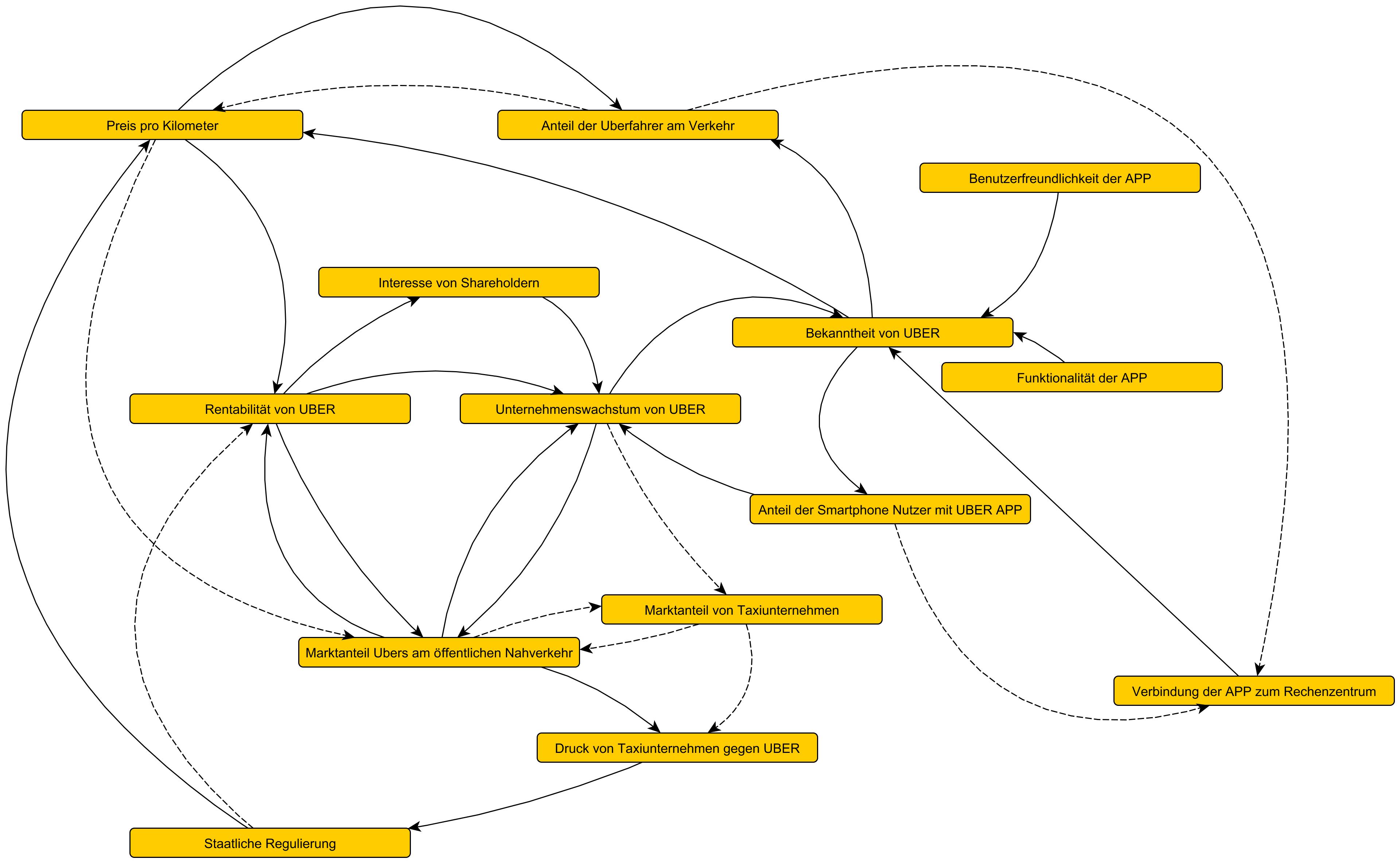}
 \caption[Wirkungsgefüge]{Wirkungsgefüge}
\label{fig:Wirkungsgefüge}
\end{figure}

Wenn in einem Zyklus aus gleichgerichteten und gegenläufigen Beziehungen,  wie in folgendem Beispiel    dargestellt,  eine gerade Anzahl von gegenläufigen Beziehungen auftauchen sollte, so ‚heben‘ diese die von Vester prognostizierte Regulierungswirkung des negativen Rückkopplungszyklus auf und befördern ihn, ausgelöst durch einen Startimpuls, ähnlich einem positiven Rückkopplungszyklus, in einen instabilen Zustand, in dem sich die Variablen ungebremst auf- bzw. abschaukeln. \\
Am System von UBER fällt das an an folgender Stelle (Abbildung \ref{fig:konflikt_1}) des Wirkungsgefüges auf, da hier z.B. bei steigendem „Unternehmenswachstum von UBER“ natürlich der „Marktanteil von Taxiunternehmen“ sinkt. Dadurch steigt entsprechend der eingezeichneten gegenläufigen Beziehung der „Marktanteil UBERs am öffentlichen Nahverkehr“ und somit offensichtlich auch das „Unternehmenswachstum von UBER“. Diese Reaktion kommt zum Tragen, da eine Hintereinander-Reihung einer geraden Anzahl von gegenläufigen Beziehungen einen positiven Input-Impuls auch als positiven Impuls aus der Kette als Output weitergibt, es sich also wie eine gleichgerichtete Beziehung auswirkt.\\\\
Ein weiterer von Vester nicht bedachter Konflikt kann bei einem positivem Rückkopplungszyklus zum Tragen kommen, wenn dieser aus einer ungeraden Anzahl von gegenläufigen Beziehungen besteht, wie folgendes Beispiel
\begin{figure}[htb]
 \centering
 \includegraphics[width=0.9\textwidth,angle=0]{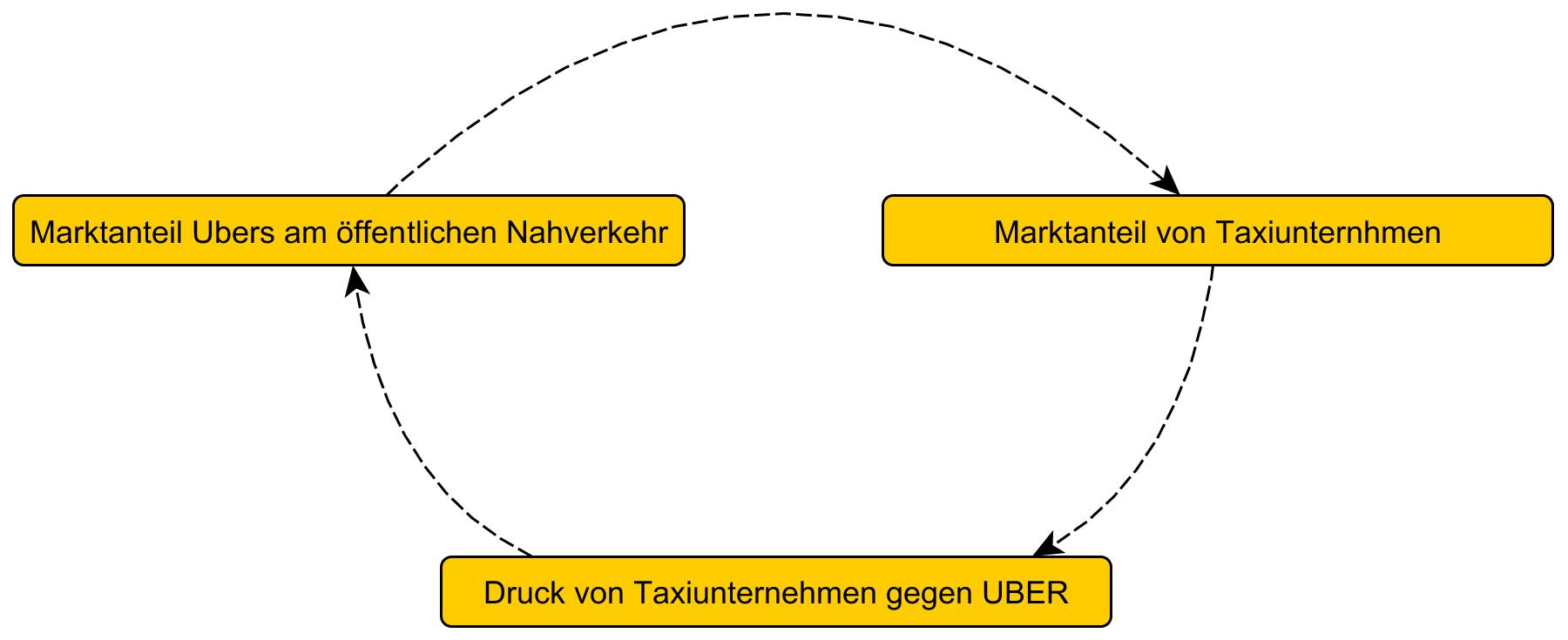}
 \caption[Beispiel positive Rückkopplung mit Konflikt]{Beispiel positive Rückkopplung mit Konflikt}
\label{fig:konflikt_2}
\end{figure}
(Abbildung \ref{fig:konflikt_2}) verdeutlicht. Ausgehend von einer gleichmäßigen Auswirkung von Variablen aufeinander, so ist dies ein positiver Einfluss auf zum Beispiel den „Marktanteil von Taxiunternehmen“, der sich in diesem Rückkopplungszyklus entsprechend als dämpfende Wirkung auf den „Druck von Taxiunternehmen gegen UBER“ auswirkt. Der „Marktanteil UBERs am Nahverkehr“ steigt daraufhin, was wiederum einen negativen Einfluss auf den „Marktanteil von Taxiunternehmen“ hat. Der Einfluss reguliert sich an dieser Stelle selbst, da der Impuls, der in den Regelungskreis eingedrungen ist, abgefedert wird, sodass die von Vester für jeden positiven Rückkopplungszyklus  geltende Wirkung das System, wie oben beschrieben, in eine unausgeglichene Position zu bringen, widerlegt wird. Dieser Effekt muss für spätere Analyseansätze beachtet werden, da ansonsten fälschlicherweise Annahmen oder Aussagen über das entstehende Wirkungsgefüge getroffen werden, die sich jeder Beleggrundlage entziehen.\\\\
Die Erstellung des gesamten Wirkungsgefüges (Abbildung \ref{fig:Wirkungsgefüge})
 von und um das Unternehmen UBER entsteht eigentlich im Dialog mit den beteiligten Personenkreisen sowie deren Vertreten. Die gesamte Betrachtung und Diskussion lässt sich im Rahmen dieser Arbeit jedoch nicht abbilden, weshalb im Folgenden nur auf einzelne Punkte eingegangen wird, die sich im Laufe des Erarbeitungsprozess ergeben haben. \\
Da UBER bereits seit einigen Jahren am Markt agiert, konnten wichtige Ansatzpunkte für die entsprechenden Wirkbeziehungen aus dem Vergangenen verstanden werden. Trotzdem ist hierbei zu vermeiden von der Vergangenheit auf die Zukunft zu schließen, sondern lediglich bisher vielleicht nicht betrachtete Zusammenhänge in die weitere Betrachtung mit einzubeziehen. 

\newpage 
\subsection{Regelkreisanalyse}\label{kapitel5.9}
Die sich an die Aufstellung des Wirkungsgefüge anschließende Regelkreisanalyse erfolgt nach  Vester rein numerisch \cite [vgl. S.216]{KunstVester} und beginnt mit einer Zählung sämtlicher positiver und negativer Rückkopplungszyklen. Dies entspräche hier am Beispiel UBER einer Aufzählung von über 50 verschiedenen Zyklen, die eine deutliche Dominanz von positiven Rückkopplungszyklen aufweisen, was Vester zufolge ein Zeichen für eine nur geringe Überlebensfähigkeit des Unternehmens UBER zu deuten wäre. \\
Dagegen spricht zwar zunächst, dass sich das Unternehmen relativ schnell weltweit etablieren konnte, andererseits  verwickelte  sich UBER durch das rigorose Profitstreben ebenso schnell in zahlreiche Konflikte mit der Justiz der verschiedenen Länder. Für Deutschland trifft das Ergebnis der Analyse insofern zu, als UBER aufgrund des Verbots, Fahrdienste in Deutschland zu vermitteln, solange die Fahrer keine behördliche Genehmigung vorweisen können, gezwungen wurde, das Unternehmen umzustrukturieren. Der Dienst von UberPop entpuppte sich tatsächlich als nicht lebensfähiges System und wurde  eingestellt. Wie bereits ausgeführt, wurde dieser aber durch UberX ersetzt, das nun nach UBERs eigener Aussage konform mit dem deutschen Personenbeförderungsgesetz organisiert ist. Die Fahrdienste sollen jetzt \enquote{durch lizensierte Mietwagenunternehmen durchgeführt werden. Alle Fahrer verfügen über einen Personenbeförderungsschein und sind selbstverständlich voll versichert} \cite{Newsroom2}. \\ 
In Anbetracht der Tatsache, dass Vester zeitlich begrenzte Prognosen  für sein Modell ausschließt, bleibt also abzuwarten, ob sich das Ergebnis der Analyse nicht doch irgendwann als zutreffend herausstellt und  sich UBER in Deutschland auch in seiner aktuellen Ausprägung als UberX als nicht lebensfähig herausstellt.\\
Anschließend sollen nach Vester die Länge der entsprechenden Rückkopplungen betrachtet werden \cite[vgl.S.214]{KunstVester}, um die Reaktionszeit der entsprechenden Kreisläufe abzuschätzen und zu deuten.  Dabei zeigt sich, dass das von den Shareholdern geforderte Wachstum zumeist zu ungebremsten kurzen positiven Rückkopplungszyklen bei den für die Investoren wichtigen Variablen „Rentabilität von UBER“ und das „Unternehmenswachstum von UBER“  führen, um kurzfristige Gewinne für die Investoren zu erwirtschaften. Dies jedoch versetzt das System nach kybernetischen Betrachtungspunkten in keinen stabilen Zustand.\\
Eine abschließende Analyse über die in den Regelkreisläufen häufig vorkommenden Variablen soll nun Auskunft über deren Charakteristik als Hebel oder Stütze des Systems geben, indem aus dem aufgestellten Graph nacheinander jeweils nur eine Variable entfernt und die Rückkopplungszyklen neu ausgezählt werden. Eine Variable, bei der durch ein solches Entfernen viele bis alle negativen Rückkopplungszyklen verschwinden, muss dann für das System eine zentrale Rolle spielen, was die Stabilität gegenüber auf das System einwirkender Impulse darstellt.\\
Bei dem Wirkungsgefüge um UBER würden eine solche zentrale Rolle zum einen der „Marktanteil von Taxiunternehmen“ spielen sowie der durch Angebot und Nachfrage regulierte „Preis pro Kilometer“ darstellen.  Es zeichnen sich also zwei für das System wichtige Stützen ab, die das System vor einem unkontrollierten Wachstum zügeln. Da diese die, zwar wenigen jedoch wichtigen negativen Rückkopplungen, also stabilisierende Wechselwirkungen aufrechterhalten. 

\subsection{2. Gedanken}
Es ist nur konsequent, wenn die vorliegende Analyse, die sich immer wieder auf neuem, Terrain bewegt hat, - angefangen bei der Sozio-Informatik  als noch relativ neuen Zweig der Informatik, über die oft noch nicht einheitlich etablierte Terminologie, bis hin zur Erprobung eines neuen Analyseansatzes, - nicht mit dem obligatorischen Schlusswort endet. Darüber hinaus ist das  Denken ‚zweiter Gedanken‘ nicht nur kreativer, sondern auch produktiver: Es könnten sich schließlich beim Durchspielen der verschiedenen Szenarien, die aus den Ergebnissen der Analyse entstehen könnten, möglicherweise neue Ausblicke eröffnen. 
Besonders interessant wäre sicher, sofern man vorher im Team gearbeitet hat, sich    abschließend gemeinsam noch einmal die gesammelten Informationen zu vergegenwärtigen und dann zusammen mit allen Beteiligten ‚ zweite Gedanken‘ zu entwickeln.\\
Der zweite Gedanke sollte sich an einer Situation oder Aktion orientieren, die sich im Rahmen der für das betrachtete sozio-technische System  beschriebenen Ergebnisse ereignen könnte. Davon ausgehend, sollte eine mögliche Auswirkung oder  Konsequenz aufgezeigt werden.\\
Der Konjunktiv an dieser Stelle ist bewusst gewählt, da zunächst keine Idee verworfen werden sollte, sofern man die Szenarien im Plenum bespricht  und ihre Folgen erörtern will. \\\\
Am Beispiel von UBER birgt zum Beispiel die zeitgenaue Ortung aller Fahrer ein  Problem. Im Normalfall ist eine exakte Ortung der Fahrer für das System vorteilhaft für beide Parteien, Kunde und Fahrer.  Die exakte Ortung des Fahrers aber, verknüpft mit der Möglichkeit, dass der Kunde seine Lieblingsfahrer in einer Liste samt Vor- und Nachnamen führen und bei Interesse Verfügbarkeit und Standort überprüfen kann, wären  Idealbedingungen für kriminelle Energien, diese Situation schnellstmöglich auszunutzen, um z.B. Wohnung/Haus des UBER-Fahrers auszurauben, wenn er gerade eine längere Tour hat. Es gäbe die verschiedensten Varianten, fest steht aber in jedem Fall, dass der UBER-Fahrer durch seine öffentlich zugänglichen Daten ein nicht zu kalkulierendes Risiko eingeht. \\
Damit diese Arbeit mit  positiv geprägten ‚zweiten Gedanken schließt, sei folgendes Szenario beschrieben:
Da die Einstellungskriterien für einen UBER-Fahrer bislang keine Vorkenntnisse oder Berufserfahrung fordern, ist es für Personen mit einem Migrationshintergrund, welche in ihrer alten Heimat keine Ausbildung abschließen konnten, nach  Erwerb eines Führerscheins relativ schnell möglich, einer Tätigkeit nachzugehen. Zusätzlich durch das UBER-Sterne-Bewertungssystem, bei dem sich  Fahrer und Mitfahrer nach einer Fahrt unabhängig voneinander auf einer Skala von 1-5 Sternen bewerten, ergibt sich für den Fahrer über die letzten 500 Fahrten ein Bewertungsprofil \cite{Sterne}, das er pflegen kann und für zukünftige Bewerbungen als potentielle erste Referenz im Umgang mit Kunden sowie als Nachweis für die Qualität seiner Dienstleistung anführen kann. \\
Dass ‚zweite Gedanken‘ natürlich nie „zu Ende“ gedacht werden können, da sie immer weiter aufeinander aufbauen können, zeigt der folgende Anschlussgedanke. \\
Sollte  eine solche Bewertung einen derart hohen Einfluss auf die Fahrer selbst haben, stellt sich die Frage, welche Probleme sie bekommen könnten, wenn Fahrgäste für sie eine schlechte Bewertung abgeben. Da UBER bereits seit einigen Jahren am Markt ist, hat Christian Stöcker in einem Essay für den Spiegel diese Problematik bereits ausgeführt und ist zu der Erkenntnis gekommen, dass die Öffentlichkeit in den nächsten Jahren abwägen muss, ob  „crowdbasiert Qualitätsstandards“ \cite{xxa} geschaffen werden oder ob die Motivation fehlt „völlig normale Alltagstransaktionen ständig bewerten zu müssen“.  Abschließend führt er aus, dass sich bereits durch die Onlinehandelsplattform ‚ebay‘  daran gewöhnt wurde, Fremden im Netz - bis zu einem gewissen Grad - zu trauen. Ohne Bewertungen würde das nicht funktionieren.

%% file: 6_kap6.tex
\newpage
\section{Überprüfung der Gültigkeit der Vester’schen  biokybernetischen Regeln Biokybernetik für sozio-technische Systeme}\label{kapitel6}

Wie in Kapitel 4 bereits ausgeführt, entwickelte Vester acht Grundregeln der Biokybernetik \cite[S.127ff]{KunstVester}, denen das für Vester zentrale „Kriterium der Lebensfähigkeit“ \cite[S.127]{KunstVester} zugrunde liegt, denn sie seien „der Natur abgeschaut“ \cite[S.128]{KunstVester}.\\
Weiterhin postuliert er die allgemeine Gültigkeit seiner Regeln, da sie „sowohl für die Ökosphäre als auch für die Technosphäre und auch für von Menschen geschaffene Systeme wie Unternehmen, Kommunen, Verkehrs- und Energiesysteme“ \cite[S.142]{KunstVester} gelten würden. \\
Es stellt sich daher die Frage, ob diese acht Regeln auch zur Bewertung und Analyse der Überlebensfähigkeit eines sozio-technischen Systems geeignet sind, denn dann wären sie ein wichtiges Handwerkszeug für kommende Analyseansätze. \\ 
Im Folgenden werden die acht Regeln im Einzelnen dargestellt und unter Einbeziehung der bei der sozioinformatischen Analyse gewonnenen Ergebnisse auf ihre Anwendbarkeit bei sozio-technischen Systeme analysiert.     
     
\subsection{Regel 1: Negative Rückkopplung muss über positive Rückkopplung dominieren!}\label{kapitel6.1}
Vester stellt als erste Regel für die Lebensfähigkeit von System die Dominanz von negativen über positiven Rückkopplungszyklen auf, denn eine „negative Rückkopplung bedeutet Selbstregulation durch Kreisprozesse“  und würde  im System Stabilität gegen äußere Störungen und Grenzüberschreitungen bewirken. \\
\begin{figure}[H]
 \centering
 \includegraphics[scale=0.3]{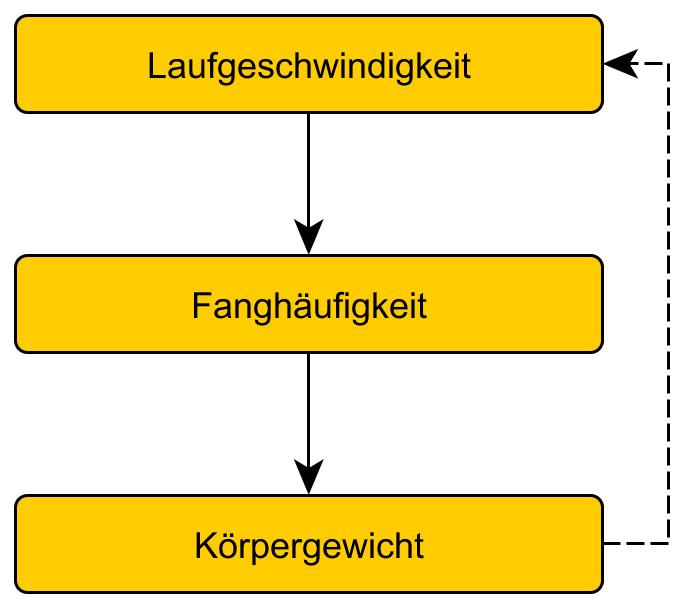}
 \caption[Regelkreis Jäger/Beutetier]{Regelkreis Jäger/Beutetier}
\label{fig:Regelkreis_Beutetier}
\end{figure}
Vesters Verbundenheit mit der Biologie zeigt sich in seiner Tendenz, häufig \\ Beispiele aus diesem Fachbereich heranzuziehen. So wählt Vester auch hier als Beleg seiner These zum  negativen Rückkopplungszyklus ein Beispiel aus der Natur, indem er aus der Beziehung   Jäger (Wolf) und seinem Beutetier (Hase) über das Körpergewicht des Jägers und seine Laufgeschwindigkeit sowie der daraus resultierenden „Fanghäufigkeit“ einen durch negative Rückkopplung bestimmten Regelkreis entwirft \\(Abbildung \ref{fig:Regelkreis_Beutetier}). \\
„Je schneller ein Wolf läuft, desto mehr Hasen kann er fangen. Je mehr Hasen er fängt, desto dicker wird er und umso langsamer kann er laufen“ \cite[S.128f]{KunstVester}. Dieser Regelkreis wird durch negative Rückkopplungen dominiert, was den Fortbestand in diesem kleinen Modell sicherstellt.\\
Demgegenüber betrachtet Vester positive Rückkopplungen als Katalysator und eine Art Starter, welche zwar Prozesse durch Selbstverstärkung in Gang bringen, ohne eine entsprechende Dämpfung würde sich jedoch, nach seinem Verständnis, das System durch unkontrolliertes Wachstum selbst zugrunde richten. Mit dem Beispiel einer Raupe, die für ihre Metamorphose zum Schmetterling den Impuls einer positiven Rückkopplung benötigt, verdeutlicht Vester seine Auffassung positiver Rückkopplungen, die er als eine Art Motor versteht, „um Dinge zum Laufen zu bringen“ \cite[S.126]{KunstVester}.  Allerdings muss Vester zufolge eine solche positive Rückkopplung alsbald von einer negativen abgelöst werden,  ansonsten würde sich das System in eine Richtung immer weiter aufschaukeln und irgendwann kollabieren, was den eingangs erwähnten Bedarf eines übergeordneten negativen Rückkopplungskreises für ihn begründet. \\
Betrachten wir nun ein sozio-technisches System und vor allem seine speziellen Charakteristika, so stellt sich die Frage, ob diese Regel auch hier ihre absolute Gültigkeit behält. Gerade die Einbeziehung von technischen sowie menschlichen Komponenten stellt für eine Vielzahl von Reaktionswegen eine natürliche obere Grenze dar, welche nicht durch eine negative Rückkopplung künstlich herbeigeführt werden muss. 
Am Beispiel des in Kapitel \ref{kapitel3} als sozio-technischen Systems charakterisiertem um und mit dem Unternehmen ICQ zeichnet sich ebenfalls ein völlig konträres Bild bezüglich Vesters These, denn mit zunehmender Größe und Ausbau der individuellen Mitgliederzahlen steigt seine eigene Überlebensfähigkeit, solange sie nicht durch einen künstlichen negativen Rückkopplungszyklus gebremst wird. Um eine natürliche Barriere gegen unkontrolliertes Wachstum muss man sich nicht sorgen, denn die Anzahl der Menschen sowie die benötigten Ressourcen bilden eine natürliche Obergrenze, der sich im allgemeinen zunächst gefahrlos genähert werden kann. 

\newpage
\subsection{Regel 2: Die Systemfunktion muss vom quantitativen Wachstum unabhängig sein!}\label{kapitel6.2}
Als nächste wichtige Voraussetzung für ein kybernetisches System postuliert Vester die Unabhängigkeit der Systemfunktionen vom quantitativen Wachstum. Reines Wachstum einzelner Teilsysteme und deren Komponenten funktioniert nicht, wenn ausgleichende Anpassungen an zwangsläufig neu entstehende Aufgabenstellungen ausbleiben. Das System würde sich nach Vester zwar rein subjektiv vergrößern, dann jedoch der benötigte Vernetzungsgrad, der für Stabilität und Überlebensfähigkeit eines Systems ausschlaggebend sei, fehlen. Als Beleg führt Vester hier wieder den Begriff der „Metamorphose“ an,  denn eine „Raupe wäre ab einer bestimmten Größe nicht mehr lebensfähig“ \cite[S.130f]{KunstVester}.  Sie muss rechtzeitig auf Nullwachstum umschalten, um sich zu verpuppen und zum Schmetterling zu wandeln, sich also intern an die neue Aufgabenstellung anpassen und effizient umstrukturieren.  \\
Hohes Wachstum und immer höherer Vernetzungsgrad können nach Vester dagegen zu Chaos führen, da Regulierungen nicht mehr abgestimmt sind. Aus diesem Grund bilden sich in der Natur oft „Cluster“,  in denen zwar eine hohe Vernetzung herrscht, unter denen allerdings nur wenige, ausgewählte Relationen entstehen.  Zur Veranschaulichung führt Vester das menschliche Gehirn an, welches aus einzelnen Neuronen aufgebaut ist, die schon bald nach der Geburt ausgewachsen sind und nur noch in ihrer Funktionsweise und „Verdrahtung“ angepasst werden \cite[vgl S.131]{KunstVester}. Also ist ein fast Nullwachstum gegeben bei starker Steigerung der kognitiven Prozesse und Fähigkeiten durch Vernetzung während des Aufwachsens des Kindes. Abschließend wird von Vester noch betont, dass sich diese Regel nicht konsequent gegen Wachstum allgemein richte, aber sie warne vor dessen Abhängigkeit \cite[vgl S.132]{KunstVester}. Ungebremstes Wachstum hebe jedoch nach Vester sämtliche Selbstregulatorien der ersten Regel auf und richte ein kybernetisches System zu Grunde, da es sich dann in den Regelkreis des darüber liegenden Systems verlagern würde, was dort irgendwann durch starke und instabile positive Reglungskreise zu katastrophalen Folgen führen könne.  Vester verweist hier zwar nur allgemein auf Beispiele, die man „jeden Tag in der Wirtschaftspresse nachlesen“ \cite[S.132]{KunstVester} kann, jedoch gibt er zuvor bei seiner Erklärung zum Mechanismus des ‚Dichtestress‘ \cite[vgl S.72ff]{KunstVester} ein anschauliches Beispiel für die Gültigkeit dieser Regel im Bereich der Natur, wodurch sich eine zu rasch wachsende Population ab einem gewissen Grenzwert von selbst wieder auf eine überlebensfähige Dichte reduziert.\\
Betrachtet man sozio-technische Systeme, so gilt diese Regel auch hier, auch wenn in einem sozio-technischen System, welches über menschliche Akteure verfügt, soziale Aspekte spezielle Anforderungen an die Wachstumsbedingungen der technischen Komponenten und dessen Funktionskatalog bilden. Durch die Vielzahl unterschiedlicher sozialer Normen, die je nach Art und Größe der jeweiligen Gruppe von Akteuren variieren, muss ein überlebensfähiges sozio-technisches System flexibel auf solche Veränderungen, die sich beim  Wachstum von Personengruppen und deren Kommunikation ergeben, reagieren können.\\
Ein solches Reagieren lässt den Schluss zu, dass diese Fähigkeit zum Überleben für das sozio-technisches System  in seiner Unabhängigkeit vom quantitativen Wachstum gemessen werden kann. Die aufgestellte Gültigkeit dieser Regel wird somit durch den Anteil menschlicher Akteure des Systems erhärtet und bietet in den späteren Analyseansätzen eine Möglichkeit, etwas über die Überlebensfähigkeit des sozi-technischen Systems auszusagen.\\
Am Beispiel des in Kapitel \ref{kapitel5} ausführlich behandelten sozio-technischen System um ‚UBER' zeigt die Variable ‚Bekanntheit von UBER‘ die Reaktion des Systems auf potentielle Entwicklungen um UBER während einer Expansion über eine Ländergrenze hinaus. Hier wurde die Systemfunktion unabhängig von quantitativem Wachstum gestaltet, um so  die eigene Zielsetzung ‚Wachstums- und Expansionspläne‘ umzusetzen und trotzdem profitabel zu bleiben. Kurz gesagt, passt sich UBER in seiner genauen Ausprägung bezüglich der Länder an, in die es expandiert, um maximalen Profit zu erwirtschaften. 
 \newpage 
\subsection{Regel 3: Ein System muss funktionsorientiert und nicht produktorientiert sein!}\label{kapitel6.3}
Die dritte biokybernetische Grundregel betrifft die Ausrichtung des Gesamtsystems. Nach Vester muss ein System funktions- und nicht produktorientiert sein, denn Produkte ändern sich rasch, Funktionen dagegen nicht. In einem komplexen Wirkungsgefüge sollte ein System nie stur nach einem Produkt streben, sondern seinen Fokus auf dessen Funktion und Funktionalität richten, um auf veränderte Umwelteinflüsse flexibel reagieren zu können.\\  
Am Beispiel des durchschlagenden Erfolgs, den VW mit der Marke Käfer weltweit erzielt hat \cite[vgl S.133]{KunstVester} verdeutlicht Vester diese Regel, da sich der Erfolg nicht aus dem Produkt selbst erklären lässt, sondern ihm zufolge durch „den einmaligen flächendeckenden Service erreicht“ \cite[vgl S.133]{KunstVester} wurde. In unserer aktuellen schnelllebigen Zeit kann es sich kein Unternehmen leisten, an einem Produkt festzuhalten und nicht auf Veränderungen im Markt zu reagieren.  Ziel des Managements müsse es Vester zufolge vielmehr sein, sich über die Funktion des Produkts im Wirkungsgefüge klar zu werden. Sobald es erkennt, dass die Funktion in der Erfüllung eines Bedürfnisses liegt, könne man sich an dieser Funktion orientieren und wisse, dass das Produkt  langfristig profitabel ist \cite[S.134]{KunstVester}.\\
Diese Regel zeichnet sich auch bei sozio-technischen Systemen als wichtiges Kriterium für die Überlebensfähigkeit ab. Gerade durch die Integration von sozialen Akteuren im System sind Veränderungen an Anforderungen und Ergebnisse meist an eine Funktion gebunden, sodass ein starres Festhalten an entsprechenden Produkten das System vor Probleme stellt, wie am Niedergang des Instantmessengers ICQ zu erkennen ist. Das 1996 gegründete Unternehmen stellte über seine Software eine damals innovative, einfache Möglichkeit zur Verfügung, über das Internet miteinander zu kommunizieren. Die Unternehmensidee hatte einen durchschlagenden Erfolg: Während der Höchstzeit des ICQ 2001 zählte das Unternehmen über 100 Millionen User \cite {Icq} und galt damals als unangefochtener Standard zur einfachen zwischenmenschlicher Kommunikation über das Internet. Da sich das sozio-technische System um das Unternehmen ICQ jedoch nicht weiter entwickelte bzw. die Funktionalität erst sehr spät in andere Bereiche portierte bzw. in anderen Bereichen überhaupt aktiv wurde und stattdessen am Produkt des Instantmessangers festhielt, etablierten sich mit mobilen Anwendungen, wie z.B. „WhatsApp“ oder dem „Facebook Messenger“ schnell Konkurrenten, welche die gewünschte Funktion Kommunikation auf ein neues ,mobiles‘ Level anhoben. 

\newpage
\subsection{Regel 4: Nutze Fremdenergie zum Arbeiten und deine eigene nur zum Steuern!  }\label{kapitel6.4}
Vester folgt in dieser Regel dem Jiu-Jitsu-Prinzip: Es sollen demnach keine eigenen Energien verschwendet, sondern stattdessen vorhandene (fremde) Konstellationen und Kräfte für die systemeigenen Zwecke genutzt werden,  indem diese durch geringe (eigene) Steuerenergie in die gewünschte Richtung umgelenkt werden. Ein System, welches die vorliegenden Konstellationen nutzen würde, anstatt gegen sie anzukämpfen, profitiere von diesem Zuwachs an Möglichkeiten und fördere so die Selbstregulation, sagt Vester. Da nicht gegen eine existierende Kraft gekämpft, sondern deren Energie für eigene Zwecke ausgenutzt würde, spare man zudem beträchtliche Eigenenergie. Sein Vorschlag für die Anwendbarkeit dieser Regel, statt des Baus teurer Kläranlagen die Selbstreinigungsfunktion vorhandener Auwälder aufgrund der Schwammfunktion nicht versiegelter Böden zu nutzen, wird mittlerweile schon umgesetzt, wie z.B. bei der Kläranlage in Kaiserslautern, was aus der Studie \enquote{Energiesituation der kommunalen Kläranlagen in Rheinland-Pfalz} \cite{Klar} hervorgeht.\\ 
Bevor die Regel auf ihre Gültigkeit bei sozio-technischen Systeme überprüft wird, sei, um Missverständnisse zu vermeiden wird, darauf hingewiesen, dass sich der Begriff ‚Fremdenergie‘ nicht auf eine externe Stromquelle, sondern das Einwirken von externer Seite auf ein System bezieht. \\
Weiterhin  sei angemerkt, dass die dem sozioinformatischen zugrunde gelegte Definition eines sozio-technischen Systems nach Kienle/Kunau in Bezug auf die erforderlichen sozio-technischen Selbstbeschreibungen zur Festlegung der Systemgrenzen geringfügig modifiziert wurde. Da normalerweise die Grenzen durch weitere Selbstbeschreibungen variieren können, wird hier, wie in Kapitel drei bereits ausgeführt, von  festgelegten Grenzen ausgegangen, wodurch sich externe und interne Einflüsse kategorisieren lassen.\\

Formal auf die Bedürfnisse einer Analyse eines sozio-technischen Systems zugeschnitten,  würde diese Regel also dem System eine erhöhte Überlebensfähigkeit zusprechen, wenn es durch effiziente Steuer- und Kontrollorgane innerhalb des Systems von außen einwirkende Impulse verstärken und umlenken kann, um das eigene Systemziel zu erreichen. \\
Einer Pauschalisierung dieser Regel gilt es jedoch im Hinblick auf die technischen Komponenten eines sozio-technischen Systems zu widersprechen,  denn es gibt potentiell komplexe Systeme, in welchem gerade durch Abschottung nach außen die Überlebensfähigkeit des Gesamtsystems erhöht wird.  Am Beispiel einer Serverstruktur des Internets, das bereits in Kapitel \ref{kapitel3}, auf Seite \pageref{ref_internet} als sozio-technisches System definiert wurde, lässt sich zeigen, dass es im Gegenteil völlig kontraproduktiv wäre, würde ein Server nach Regel 4 agieren. Bei einem so genannten DDOS-Angriff  wird ein Server mit einer Vielzahl kurzer Anfragen überschwemmt, die einzig das Ziel verfolgen, ihn dadurch zu blockieren. Mit den zur Verfügung stehenden Mitteln ist es dem Server nicht möglich, die in das System fließende Fremdenergie in Form von Anfragen für das eigene Funktionieren nutzbar zu machen. Eine solche Reaktion würde auch für das System keinen Sinn ergeben, da diese Energien in Ausführung und  Masse rein destruktiver Natur sind. Zur Aufrechterhaltung seiner Lebensfähigkeit muss es sich abschotten und  derartige Anfragen verwerfen, was natürlich in direktem Konflikt zur  vierten Vester’schen Grundregel steht. Insofern muss konstatiert werden, dass die vierte biokybernetische Regel für die Analyse sozio-technischer Systeme keine allgemeingültige Aussage über die Überlebensfähigkeit des Systems zulässt.
\newpage
\subsection{Regel 5: Mehrfachnutzung von Produkten, Funktionen und Organisationsstrukturen}\label{kapitel6.5}
Für ein kybernetisches System ist es nach Vester von hoher Priorität, Produkte, Funktionen und  Organisationsstrukturen mehrfach zu nutzen. Prozesse, die ausschließlich in singulären Bereichen arbeiten, funktionieren seiner Ansicht nach genau entgegen der für die Überlebensfähigkeit eines Systems wichtigen Vernetztheit. Vester zufolge sollten alle in einem System eingesetzten Produkte oder Verfahren ein „multifunktionales System“ \cite[S.136]{KunstVester} bilden, auch wenn die Effizienz einer Komponente durch mehrfache Nutzung für einen bestimmten Prozess vielleicht abnimmt, so würde nach Vester das gesamte System dennoch effektiver arbeiten. Ein solcher Optimierungsvorgang verlange natürlich, so Vester, \enquote{fachüberschreitendes Denken von der Forschung über die Entwicklung bis zum Endprodukt.} \cite[S.136]{KunstVester} 
Anhand des Beispiels eines Fahrzeugantriebs im Verbund mit neuer Haustechnik zeigt er eine interessante, von ihm entwickelte „kybernetische Lösung“ \cite[S.136]{KunstVester} für die Frage, warum ein Stadtauto zur Bewältigung zumeist kurzer Strecken „gewissermaßen ein eigenes Kraftwerk mitschleppen soll, wodurch das Fahrzeug schwer und damit energiewirtschaftlich ineffektiv wird“ \cite[S.136]{KunstVester}.\\
Sein Vorschlag: „Motor in den Keller, Abwärme nutzen und mit dem erzeugten Strom gleichzeitig Batterien aufladen.“ \cite[S.136]{KunstVester}  Auch wenn die für seine ideale Vorstellung benötigten technischen Innovationen zu seiner Zeit - und in vollständig ausgereifter Form bis heute -  noch nicht zur Verfügung standen bzw. stehen,  so zeigt diese Idee doch den Nutzen der oben genannten Regel für ein komplexes System.  \\
Interessant wird dieser Ansatz, wenn man versucht, diesen auch auf sozio-technische Systeme anzuwenden.  Hier zeigt sich, dass gerade durch die technischen Komponenten, welche nach Bedarf genutzt werden können, den sozialen Akteuren eine einfache Möglichkeit verschafft, im System Funktionalitäten und Produkte effizient wiederzuverwenden oder zeitlich reguliert zu teilen. Der Gedanke der ‚Sharing Economy‘ wurde bereits in Kapitel \ref{kapitel5.4} ausführlich erläutert, sodass hier lediglich auf den Umstand verwiesen sei, dass ‚UBER‘ als Unternehmen diese von Vester postulierte Regel direkt in ein Geschäftsmodell umgewandelt hat und somit für die Anwendbarkeit dieser Regel das adäquate Beispiel darstellt. 
\newpage
\subsection{Regel 6: Nutzung von Kreisprozessen: Recycling}\label{kapitel6.6}
    Diese Regel thematisiert  das Thema Recycling, das Vester von der  Natur perfekt umgesetzt sieht, da sie „überhaupt keine ‚Abfälle‘ als solche kennt“ \cite[S.137]{KunstVester},  denn sie habe aus gutem Grund „im Laufe der Äonen einen geschlossenen Materialkreislauf etabliert“ \cite[S.137]{KunstVester}.  Er betont das ressourcensparende Prinzip der Natur durch das „nutzbringende Wiedereingliedern von Abfallprodukten in den lebendigen Kreislauf aller beteiligten Systeme“ \cite[S.137]{KunstVester} und verweist auf die notwendige  Umsetzung durch den  „immer bedrohlicheren Abfallstau“ (Vester, S. 138). \\
In dem Zusammenhang fordert er eine Abkehr „vom eindimensionalen Denken“ \cite[S.138]{KunstVester} und stattdessen „ein Denken in kybernetischen Kreisprozessen“ \cite[S.138]{KunstVester}.\\\\
 Hier fällt auf, dass die Nutzung von ‚kybernetischen Kreisprozessen‘ \cite[S.138]{KunstVester} zwar einen elementaren Baustein für ein funktionierendes Vester‘sches Wirkungsgefüge bildet, diese Regel jedoch nur recht kurz, ohne die üblichen veranschaulichenden Beispiele ausgeführt wird. Auch seine Empfehlung des Recyclings für „mittelständische Betriebe und das Handwerk“ \cite[S.137]{KunstVester} durch „hochinteressante Anwendungsbereiche“ \cite[S.137]{KunstVester} wird nicht weiter konkretisiert, ebenso wenig die profitablen Recyclingmöglichkeiten innerhalb von Industrieunternehmen.\\\\
Bei der Überprüfung dieser Regel auf  ihre Anwendbarkeit bei sozio-technischen Systemen empfiehlt sich eine Unterscheidung folgender zwei Bereiche, die voneinander getrennt untersucht werden sollten.  Zum einen spricht Vester von der Nutzung kybernetischer Kreisprozesse, welche in sozio-technischen Systemen eine große Rolle spielen, zum anderen beruft er sich auf den Recycling – Begriff, welcher aufgrund des meist ausbleibenden materiellen Outputs bei technischen Systemen als Messkriterium für die Überlebensfähigkeit eines sozio-technischen Systems weitestgehend irrelevant ist. \\
Betrachtet man nun zunächst die Rolle von Kreisprozessen, so erkennt man, dass gerade durch technische Komponenten mit einem meist deterministischen Funktionskatalog die Möglichkeit gegeben wird, redundante Aufgabenfelder, deren Ausübung menschlichen Akteuren aufgrund der Monotonie eher schwerfällt, zu bewerkstelligen. So lagert beispielsweise das sozio-technische System um UBER die sich stets wiederholende Fahrpreisberechnung an einen Server, also an eine technische Komponente, aus,  wodurch dieser für den Menschen langweilige und zudem zeitraubende Aufgabenbereich in diesem Kreisprozess entfällt.  \\
Insofern ließe sich die von Vester aufgestellte Allgemeingültigkeit in Bezug auf kybernetische Kreisprozesse für sozio-technische Systeme bestätigen. Die Nutzung solcher Kreisprozesse ermöglicht darüber hinaus eine gewisse Unabhängigkeit gegenüber Fremdeinflüssen, was vor allem in Hinblick auf komplexe Systeme mit integrierter technischer sowie sozialer Komponente einen klaren Vorteil bietet. Die Überlebensfähigkeit wird hier auch dahingehend begünstigt, dass es die Einflussfaktoren von außerhalb des Systems drastisch reduziert, sodass das System stabiler und ,- um Vesters Wortlaut zu entsprechen,-‚kybernetisch reifer‘ reagieren kann.  \\
Da in Kapitel \ref{kapitel5} für das als sozio-technisches System charakterisierte Unternehmen um UBER bereits ausführlich aufgezeigt wurde, wie stark die bereits beschriebenen Kreisprozesse und Rückkopplungen das System stabilisieren und zur Überlebensfähigkeit beitragen, wird an dieser Stelle nicht ergänzend darauf eingegangen. 
Untersucht  man dagegen die Bedeutung des von Vester postulierten Recyclings für sozio-technische Systeme und betrachtet die speziellen, vom technischen Komponenten geschaffenen Outputströme, so muss man feststellen, dass gerade ein technisches System wenig bis gar keinen recycelbaren Output generiert, der wieder genutzt werden könnte, um im übergeordneten System einen in sich geschlossenen Materialkreislauf zu etablieren. 
 Komprimiert lässt sich  also Vesters Regel hier konsequent nur auf nicht materielle Verflechtungen anwenden, da materielle für technische Anwendungen potentiell nicht von großem Belang sind und die  Bewertung der Überlebensfähigkeit des Systems an dieser Stelle schwer fallen könnte.
\newpage
\subsection{Regel 7: Symbiose}\label{kapitel6.7}

Der Begriff ‚Symbiose‘ bezeichnet in der Biologie das „enge Zusammenleben unterschiedlicher Arten zum gegenseitigen Nutzen“ \cite[S.138]{KunstVester} und bildet die Grundlage aller lebendigen Systeme. Die „gegenseitige Nutzung von Verschiedenartigkeit durch Kopplung und Austausch“ \cite[S.138]{KunstVester}, wie Vester die Symbiose umschreibt, transferiert seine biologische Betrachtungsweise auf komplexe Systeme. 
Das Streben nach Symbiose hat nach Vester auf struktureller Ebene viele Vorteile. Gleichartige Monostrukturen würden vermieden, da diese nicht viel Austausch untereinander ermöglichen, was wiederum eine lokale Ressourcenknappheit verhindert. Die benötigte Vielfalt auf sehr kleinem Raum schaffe außerdem, so Vester, eine steigende Funktionalität der einzelnen Komponenten.  Auch würde strukturellen Schwachpunkten, wie einer zentralen Denkeinheit, entgegengewirkt, da effektive Symbiosen dezentrale Strukturen bevorzugen, sodass eine Verteilung der Steuerungsfunktionalität sinnvoll sei. \\
Die Systeme kommen nach Vester durch symbiotische Beziehungen untereinander zu „meistens  hochinteressanten Lösungen, nach denen wir auch immer wieder suchen sollten, da sie ‚kurzfristige Ausbeutung‘ durch ‚stabile Kooperation‘ ersetzen“  \cite[S.139]{KunstVester}.
So führt Vester als Beispiel für die Bedeutung von Symbiosen für Systeme den Menschen an, welcher innerhalb seines Körpers viele Symbiosen aus Zellen, Bakterien und Einzellern eingeht, um das System ‚Mensch‘ am Leben zu erhalten. \\
Vester suggeriert jedoch anhand des gewählten biologischen Beispiels, dass eine solche Symbiose hauptsachlich nach innen gerichtet sei,  also innerhalb eines  komplexen System stattfindet und dort eine Bewertungsgrundlage für seine Regel schafft. Demnach gilt, je höher die symbiotischen Beziehungen innerhalb des Systems oder in seinen Worten die „interne Dependenz“ \cite[S.138]{KunstVester}, umso überlebensfähiger ist das gesamte System. \\\\

Bei der Analyse eines sozio-technischen Systems zeigt sich zwar ebenfalls, dass ein symbiotischer Charakter eines Systems eine Vielzahl von Vorteilen generiert, wobei sich jedoch im Gegensatz zur Vester’schen Sichtweise vor allem symbiotische Beziehungen nach außen positiv auswirken. Gerade die von den Systemgrenzen als außerhalb definierten Kräfte und Verhältnisse schaffen hier einen wichtigen Mehrwert, der sich für die kybernetische Bewertung des sozio-technischen Systems heranziehen lässt.\\
Deutlich wurde dies in der in Kapitel \ref{kapitel5} ausgeführten Analyse des sozio-technischen Systems um UBER, welches sich in keiner Art und Weise symbiotisch in seine Umwelt integriert hat, sondern,- um in Vesters Sprachraum zu bleiben,- parasitär versucht, in den verschiedenen Grauzonen der einzelnen Ländern und ihrer Gesetze maximalen Profit zu erwirtschaften.
UBER agiert mit seinem beschriebenen Geschäftsmodell zwar unter dem ‚Sharing Economy‘- Gedanken, vermeidet real jedoch einen direkten Kontakt zur Regierung oder zu Mitbewerbern im entsprechenden Nahverkehr, was das Unternehmen vielleicht zunächst weniger profitabel gemacht, dafür aber in Synergie mit seiner Umwelt integriert hätte. Die Konsequenzen der  ausbleibenden Kommunikation wurde für Deutschland bereits mit den beschriebenen Unterlassungsurteilen ausgeführt und zeigt das Widerstandsverhalten der Umwelt gegen die neu etablierten Beförderungsmethoden UBERs sowie dessen rigorosen und rein profitorientierten Vorgehens. \\
Eine Abwertung des Unternehmens wird hier nicht beabsichtigt, vielmehr soll das Beispiel  aufzeigen, was mit einem sozio-technischen System passieren kann, wenn es seiner Umwelt gegenüber nicht symbiotisch  eingestellt ist. Sowohl Gerichtsurteile als auch geltend gemachte Schadensersatzansprüche sind ein deutlicher Einschnitt in die Überlebensfähigkeit des Systems UBER und bestätigen insofern die Gültigkeit der von Vester postulierten Regel auch für sozio-technische Systeme.
\newpage
\subsection{Regel 8: Biologisches Design durch Feedback-Planung}\label{kapitel6.8}
Vesters letzter Regel zum biologischen Design lässt sich in mehrerer Hinsicht nicht folgen. 
Danach sollte jedes Produkt, Verfahren und jede Organisationsform zum „Überleben unserer Spezies“ beitragen  und „mit der Biologie des Menschen und der Natur vereinbar sein, also der Struktur überlebensfähiger Systeme entsprechen“. \cite[S.140]{KunstVester} Dies sei nicht nur eine ökologische, sondern Vester zufolge auch eine psychologische und ökonomische Forderung. Planung und Gestaltung der Projekte müssten daher stets „im Feedback mit der lokalen, lebendigen Umwelt“ \cite[S.141]{KunstVester} erfolgen. \\
Auch wenn man die von Vester vorgenommene Einordnung des Menschen als ‚überlebensfähiges System‘ als Zeugnis ungebrochenen Optimismus begrüßen könnte, bleiben dennoch vielerlei Einwände und Fragen offen. \\
Im Gegensatz zu den vorhergehenden Regeln, die insgesamt begründet und zumeist durch entsprechende Beispiele belegt sind, besteht die achte Regel aus einer Aneinanderreihung unbewiesener Behauptungen, basierend auf nicht näher definierten Schlagwörtern. So vermisst man eine Konkretisierung, was unter ‚biologischem Design‘  (s.u.) genau zu verstehen ist, ebenso vage bleibt die ‚Feedback-Planung‘, auch für die plakative Behauptung „Mit unbiologischem Design wird (…) letztlich immer am Bedürfnis und damit am Markt vorbeiproduziert.“ sucht man vergebens nach einem überzeugenden Beleg.  Aufgrund der vielen von Vester zuvor angeführten Beispiele für das unsystemische Handeln des Menschen, durch das Vester das längerfristige Überleben des Menschen für gefährdet ansieht \cite[vgl. S.72ff]{KunstVester}, erwartet der Leser statt der bloßen Behauptung hier zumindest eine kurze Begründung.
 Weiterhin erinnert Vesters Forderung, dass alles dem Überleben unserer Spezies zu dienen habe, auf eigentümliche Weise an den alttestamentarischen Spruch  „Macht Euch die Erde untertan!“ (Genesis 1, 28) Bedenkt man, dass diese Aussage allzu oft als Legitimation für die Ausbeutung genutzt wurden. Die Forderungen des Club of Rome, dessen Mitglied Vester schließlich war, stehen diesem Gedankengut diametral gegenüber. Es wäre  zu überlegen, ob diese Regel nicht aus dem Katalog der biokybernetischen Regeln gestrichen werden sollte.\\\\
Statt des meist üblichen Beispiels aus der Biologie zur Veranschaulichung seiner Thesen bringt Vester hier eine Art Negativbeispiel, das jedoch nicht schlüssig überzeugt: Vester warnt vor dem Ausbau des Internets, das seiner Ansicht nach kein ‚biologisches Design‘ besäße. Er zieht einen Vergleich mit der Natur, in der (ihm zufolge)  keine direkte Vernetzung verschiedener Organismen stattfände: „Weder Blutkreisläufe noch Nervensysteme sind über den individuellen Organismus hinaus miteinander verbunden, (…); denn Störungen und Fehler an einer Stelle sollen gerade nicht gleichzeitig auf alle anderen übertragen werden. Nicht umsonst hat die Natur auf eine internet-ähnliche Infrastruktur verzichtet“  \cite[S.140f]{KunstVester}.  Offensichtlich berücksichtigt Vester hier nicht das in der Natur häufig zu findende Phänomen der Schwarmintelligenz, wie sie z.B. bei Bienen- oder Ameisenvölkern anzutreffen ist.\\
Aufgrund des fehlenden Bezugs zur Frage, inwiefern diese Regel eine Aussage zur Überlebensfähigkeit eines Systems zulässt, wird darauf verzichtet, sie auf ihre Anwendbarkeit bei sozio-technischen Systemen zu prüfen.\\\\

\textbf{Fazit:}\\
Es hat sich gezeigt, dass die Allgemeingültigkeit, die Vester seinen Regeln zuschreibt, nicht zutreffend ist. Vier Regeln, also 50\%, sind auf sozio-technische Systeme nicht anwendbar, sodass die kybernetischen Regeln als Klassifizierungskatalog für lebensfähige Systeme im sozio-technischen Bereich nicht einsetzbar sind.

%% file: 7_kap7.tex
\newpage
\section{Fazit}

Entsprechend der eingangs aufgestellten Zielbeschreibungen lassen sich zusammenfassend  aus den Ergebnissen der sozioinformatischen Analyse folgende Resultate und Schlussfolgerungen ableiten:\\
\textbf{‚Sensitivitätsmodell‘}\\
Das Sensitivitätsmodell zeigte sich mit gewissen Einschränkungen auf sozio-technische Systeme übertragbar, mit dem entscheidenden Vorteil, durch die Erstellung des Wirkungsgefüges  Einblicke und  Erkenntnisse über innere Systemzusammenhänge zu erhalten, die bei einer  unsystemischen Analysen nicht erfasst würden: So entstanden während der Erarbeitung des Wirkungsgefüges durch die Integration einzelner Beziehungen zwischen Variablen auch Rückkopplungszyklen, die nicht gezielt eingezeichnet wurden und somit auch für mit dem System vertrauten Personen neue Erkenntnisse über Abhängigkeiten  verschaffen können, die ohne ein solches Wirkungsgefüge unbedacht geblieben wären. \\
Genauso offenbarte die abschließende Regelkreisanalyse einen durch das Verfahren belegbaren Satz an für das System wichtigen Einflüssen, welche das System stabilisieren bzw. als Hebel oder Stütze fungieren können und zu einem differenzierterem Systemverständnis führen.\\\\
Allerdings konnten während der sozioinformatischen Analyse manche von Vester vorgesehenen Teilschritte für UBER als komplexes sozio-technisches Systeme nicht direkt umgesetzt werden, da Vester in seiner Vorgehensweise kaum Abweichungen vom vorgeschriebenen Plan aufzeigt (Kapitel \ref{kapitel5}).  Das Vorgehen im Sensitivitätsmodell wird von ihm sehr idealtypisch beschrieben, obwohl von einer Vielfalt an existierenden komplexen Systemen  auszugehen ist, wenn eine universale Anwendbarkeit angeben wird.  Es fehlt also trotz des postulierten breiten Anwendungsspektrums das nötige Reaktionspotential auf systemspezifische Charakteristika. \\
So besteht der involvierte Personenkreis vielleicht nur aus wenigen Menschen, wenn zum Beispiel die Technikfolgenabschätzung eines unbekannten Produkts angestrebt wird, und somit eine breit gefächerte  Diskussion über einzelne Einflussfaktoren kaum zu realisieren ist. \\
Wäre das System dann jedoch bereits aufgebaut und etabliert, so hätte die Analyse mit einem entsprechend anders gelagertem Problem zu kämpfen, wenn nach Vester sämtliche mit dem System korrelierende Personen zur Analyse des Variablensatzes und der weiteren Betrachtung herangezogen werden. Auf das eigentlich betrachtete System UBER angewandt, wäre es ein beträchtlicher Personenkreis, wenn alle involvierten  Personen bis hin zu Personengruppen kämen, die für eine objektive Beurteilung gehört oder zumindest vertreten sein sollten. Aus diesem Anstieg an Personen,  mit welchen über das System gesprochen werden muss, wäre es wichtig einen einheitlichen Katalog an Wortbedeutungen sowie Bewertungskriterien festzulegen, um überhaupt die gesammelten Meinungen strukturiert verwerten zu können. \\\\
\textbf{Biokybernetische Grundregeln} \\
Wie die ausführliche Prüfung der einzelnen biokybernetischen Regeln (Kapitel 6) nachgewiesen hat, ist Vesters  Behauptung einer Allgemeingültigkeit der acht Regeln für alle komplexen Systeme unzutreffend, da vier Regeln, also genau 50\% auf  sozio-technische Systeme nicht anwendbar sind. 
Insofern sind nur die verbleibenden vier kybernetischen Regeln als praktischer ‚Klassifizierungskatalog‘ zur Bestimmung lebensfähiger Systeme im sozio-technischen Bereich nicht einsetzbar.\\
Alternativ könnten jedoch die vier anwendbaren Regeln zwei, drei, fünf und sieben als eine Art ‚Minimalkatalog zur Bestimmung funktionierender sozio-technischer Systeme‘ genutzt werden.
Es wäre sicherlich auch aufgrund der zusätzlichen Erweiterungen und Auslegungen in der Anwendung der Regeln, die sich bei der Analyse abgezeichnet haben, zu überlegen, ob  nicht dieser ‚Minimalkatalog‘ hinsichtlich der Bestimmung sozio-technischer Systeme zu systematisieren und auszubauen wäre.  Dies würde die Analyse und Entwicklung neuer sozio-technischer Systeme zukünftiger Forschungsprojekte erleichtern, wenn eine angepasste Begrifflichkeit und geeignete Bestimmungskriterien zur Verfügung ständen.  \\\\
\textbf{Sozioinformatischer Ansatz:}\\
Das Unternehmen konnte nach der Definition von Andrea Kienle und Gabriele Kunau eindeutig als sozio-technisches System bestimmt werden.\\
Darüber hinaus erwies sich die Definition als sehr hilfreich und effizient bei der Prüfung der biokybernetischen Regeln hinsichtlich ihrer Anwendbarkeit bei sozio-technischen Systemen. \\\\
Im Gegensatz zu Vesters Ansatz für sein Sensitivitätsmodell wird die in ein sozio-technisches System stark integrierte technische Komponente nicht apriori als deterministisch angesehen,  man bedenke an dieser Stelle z.B. menschliches Versagen sowie Materialfehler, was den Vorteil der vorgestellten sozioinformatischen Methode deutlich herausstellt.\\
Durch die zugrunde gelegte Definition sozio-technischer Systeme konnten wichtige Einflussnahmen auf das Gesamtsystem aufgezeigt werden, die bei einer einseitig ausgerichteten Analyse technischer oder sozialer Komponenten sich sonst dem Betrachtenden \\
nicht erschlossen hätten, sodass Reaktionen des Systems weder vorhersehbar noch im Nachhinein erklärbar wären.\newpage
\textbf{Transdisziplinäre Weiterentwicklung}\\
Der Begriff der ‚transdiziplinären‘ Forschung und Betrachtungsweise ist noch nicht einheitlich geregelt, befasst sich jedoch mit einer fächerübergreifenden Sicht und Arbeitsweise. Nach Thomas Jahn vom Institut für sozial-ökologische Forschung ISOE müssen für eine transdisziplinäre Forschungsarbeit „Experten/innen aus verschiedenen Fächern bzw. Disziplinen und aus der Praxis zusammenwirken, um die komplexen Problematik umfassend behandeln zu können“ \cite{Jahn}.\\
Dennoch wurde Vesters Herangehensweise bereits als „zwischen (interdisziplinär) und über (transdisziplinär)“ den Disziplinen beschrieben\cite{Streich}, was dem transdisziplinären Charakter dieser Methodik entspricht. Das Vorgehen für solche Weiterentwicklungen von transdisziplinären Methoden wird von  Dusseldorp als Lernprozess bezeichnet, denn hier „wird nach geeigneten Konstellationen gesucht, wie Wissenschaft in gesellschaftliche Prozesse eingebunden sein kann“ \cite[S.18]{Dussel}. \\
Insofern kann der in dieser Arbeit beschriebene sozioinformatische Analyseansatz durchaus als Weiterentwicklung des Sensitivitätsmodells als ‚transdisziplinär‘ bezeichnet werden.\\\\
\textbf{Ausblick:}\\
Abschließend sei als positiver Ausblick noch auf die ‚Ecopolicyade‘ 2005 \cite{ecoo} verwiesen, bei der durch zwei Hauptschullehrer aus Schleswig-Holstein der Nachweis erbracht wurde, dass vernetztes Denken  lernbar ist.\\
 Die beiden Lehrer forderten erst den Kreistag, dann Abgeordnete des Landtags zu einer Partie Ecopolicy mit ihren SchülerInnen heraus. Die SchülerInnen gewannen problemlos beide Partien und spielten dann auch noch erfolgreich gegen Mitglieder des Bundestags in Berlin, was mitten in der Pisa-Debatte einiges Aufsehen erregte.   \\
Wenn Schüler bei einem solchen Planspiel Politiker in ihrem eigenen Wirkungsbereich, in dem es um die miteinander vernetzten gesellschaftspolitischen Fragen geht,  besiegen, so zeigt dies Beispiel, dass wir mit einiger Übung zu einem vernetzten Denken und Planen fähig und insofern in der Lage sind, die Probleme unserer komplexen Welt zu lösen.

%% file: erklaerung.tex
\section*{Eidesstattliche Erklärung}
\thispagestyle{empty}

\begin{verbatim}

\end{verbatim}

\begin{LARGE}Eidesstattliche Erklärung zur Bachelorarbeit\end{LARGE}
\begin{verbatim}

\end{verbatim}
Ich versichere, die von mir vorgelegte Arbeit selbstständig verfasst zu haben. Alle Stellen, die wörtlich oder sinngemäß aus veröffentlichten oder nicht veröffentlichten Arbeiten anderer entnommen sind, habe ich als entnommen kenntlich gemacht. Sämtliche Quellen und Hilfsmittel, die ich für die Arbeit benutzt habe, sind angegeben. Die Arbeit hat mit gleichem Inhalt bzw. in wesentlichen Teilen noch keiner anderen Prüfungsbehörde vorgelegen.

\begin{displaymath}
\begin{array}{ll}
Unterschrift:~~~~~~~~~~~~~~~~~~~~~~~~~~~~~~~~~~~~~~~~~~
& Ort, Datum:~~~~~~~~~~~~~~~~~~~~~~~~~~~~~~~~~~~~~~~~~~
\end{array}
\end{displaymath}